%% file: main.tex
\documentclass[fleqn,10pt]{article}
\usepackage{latexsym, graphicx, epsfig, amsmath, amssymb,amsfonts}
\usepackage{natbib,amsthm,version}
\usepackage{amsbsy,bm,multirow,enumerate}
\usepackage[titletoc,page]{appendix}
\usepackage[mathscr]{eucal}
\usepackage{mathtools}
\usepackage{color}
\usepackage[utf8]{inputenc}
\usepackage[english]{babel}
\usepackage{amsthm}
\usepackage{enumerate}
\usepackage[hidelinks]{hyperref}
\usepackage{url}
\usepackage{subfigure}
\usepackage[lined,boxed,linesnumbered,ruled]{algorithm2e}
\usepackage{float}
\usepackage{dsfont,bbm}

\DeclareMathOperator{\sech}{sech}

\newcommand{\mbs}[1]{\mathbf{#1}}

\newtheorem{remark}{Remark}[section]

\theoremstyle{definition}

\title{
  A Modified Batch Intrinsic Plasticity Method for Pre-training
  the Random Coefficients of Extreme Learning Machines
} 
\author{
  Suchuan Dong$^1$\thanks{Author of correpondence.
    Email: sdong@purdue.edu},\ \ Zongwei Li$^2$
  \\
  $^1$Center for Computational and Applied Mathematics \\
  Department of Mathematics,
  Purdue University\\
  West Lafayette, Indiana, USA \\
  $^2$ Department of Mathematics,
  Purdue University \\
  Fort Wayne, Indiana, USA
 } 

\date{(March 14, 2021)}

\begin{document}
\maketitle


\input Abstract


\vspace{0.05cm}



\input Introduction

\input Method

\input Test

\input Summary

\section*{Acknowledgement}
This work was partially supported by
NSF (DMS-2012415, DMS-1522537).

\bibliographystyle{plain}
\bibliography{elm,mypub,dnn,sem,obc}

\end{document}

%% file: Abstract.tex
\begin{abstract}

  In extreme learning machines (ELM) the hidden-layer coefficients
  are randomly set and fixed,
  while the output-layer coefficients of the neural network are computed
  by a least squares method. The randomly-assigned coefficients in ELM are
  known to influence its performance and accuracy significantly.
  In this paper we present a modified batch intrinsic plasticity (modBIP)
  method for pre-training the random coefficients
  in the ELM neural networks. The current method is devised based on
  the same principle as the batch intrinsic plasticity (BIP) method,
  namely, by enhancing the information transmission in every node
  of the neural network. It differs from BIP in two prominent aspects. First,
  modBIP does not involve the activation function in its algorithm,
  and it can be applied with any activation function in the neural network.
  In contrast, BIP employs the inverse of the activation function in its construction,
  and requires the activation function to be invertible (or monotonic).
  The modBIP method can work with the often-used non-monotonic
  activation functions (e.g.~Gaussian, swish, Gaussian error linear unit, and
  radial-basis type functions), with which BIP breaks down.
  Second, modBIP generates target samples on random intervals with a minimum size,
  which leads to highly accurate computation results when combined with ELM.
  The combined ELM/modBIP method is markedly more accurate than 
  ELM/BIP in numerical simulations. Ample numerical experiments are presented
  with shallow and deep neural networks
  for function approximation and boundary/initial value problems with partial
  differential equations. They demonstrate that the combined ELM/modBIP method
  produces highly accurate simulation results, and that its accuracy is insensitive
  to the random-coefficient initializations in the neural network. This is in sharp
  contrast with the ELM results without pre-training of the random coefficients.

\end{abstract}

%% file: Introduction.tex
\section{Introduction}
\label{sec:intro}


This work concerns the use of extreme learning machines (ELM)
for scientific computing, chiefly~for solving ordinary and
partial differential equations (ODE/PDE).
ELM is proposed in~\cite{HuangZS2006} for single
hidden-layer feed-forward networks (SLFN),
and consists of two main ideas: (i) the weights/biases
in the hidden layer are randomly set and
fixed, and (ii) the weights of the linear
output layer are computed/trained by a linear least squares method
or by using the pseudo-inverse (Moore-Penrose inverse) of
the coefficient matrix.
In the context of the current paper we will broadly
refer to neural network-based methods
adopting these strategies as ELM methods,
including those that employ multiple hidden layers
in the neural network and those that train the output-layer
coefficients by nonlinear least squares computations
(see e.g.~\cite{DongL2020}).


ELM is one type of random-weight neural
networks~\cite{ScardapaneW2017,FreireRB2020},
which randomly assign and fix a subset of the network's weights 
so that the resultant optimization task of training the neural network
can be simpler, and often linear, for example,
formulated as a linear least squares problem.
Randomization can be applied to both feed-forward and recurrent
networks, leading to  methodologies such as
the random vector functional link (RVFL)
networks~\cite{PaoPS1994,IgelnikP1995}, the extreme
learning machine~\cite{HuangZS2006,HuangHSY2015},
the no-propagation network~\cite{WidrowGKP2013},
the echo-state network~\cite{JaegerLPS2007,LukoseviciusJ2009},
and the  liquid state machine~\cite{MaasM2004}.
The universal approximation property of
the random-weight feed-forward neural networks
has been studied and proven
in e.g.~\cite{IgelnikP1995,LiCIP1997,HuangZS2006}.
Randomized neural networks can be traced to the un-organized
machines by Turing~\cite{Webster2012} and the perceptron by
Rosenblatt~\cite{Rosenblatt1958} in the 1950s.
After a  hiatus of several decades, contributions
started to appear in the 1990s, and in recent years such methods
have witnessed a strong revival.
We refer to e.g.~\cite{ScardapaneW2017}
for a historical overview of randomized neural networks.


With ELM one employs the least squares method, either
linear or nonlinear least squares~\cite{DongL2020}, to compute/train
the training parameters, which consist of only
the weights in the linear output layer of the neural network.
This training method is different from the back propagation
(or gradient descent) type algorithms~\cite{Werbos1974,Haykin1999},
which have been widely used in the deep neural network (DNN) based PDE solvers
in recent years (see e.g.~\cite{SirignanoS2018,RaissiPK2019} and
related approaches \cite{LagarisLF1998,RuddF2015,EY2018,WinovichRL2019,HeX2019,LuJK2019,XingKZ2019,ZangBYZ2020,WangL2020,JagtapKK2020,Samaniegoetal2020,Xu2020,DongN2020}).
This is the primary factor that accounts for ELM's lower
computational cost observed in numerical simulations~\cite{DongL2020}.

\begin{figure}
  \centerline{
    \includegraphics[width=2.2in]{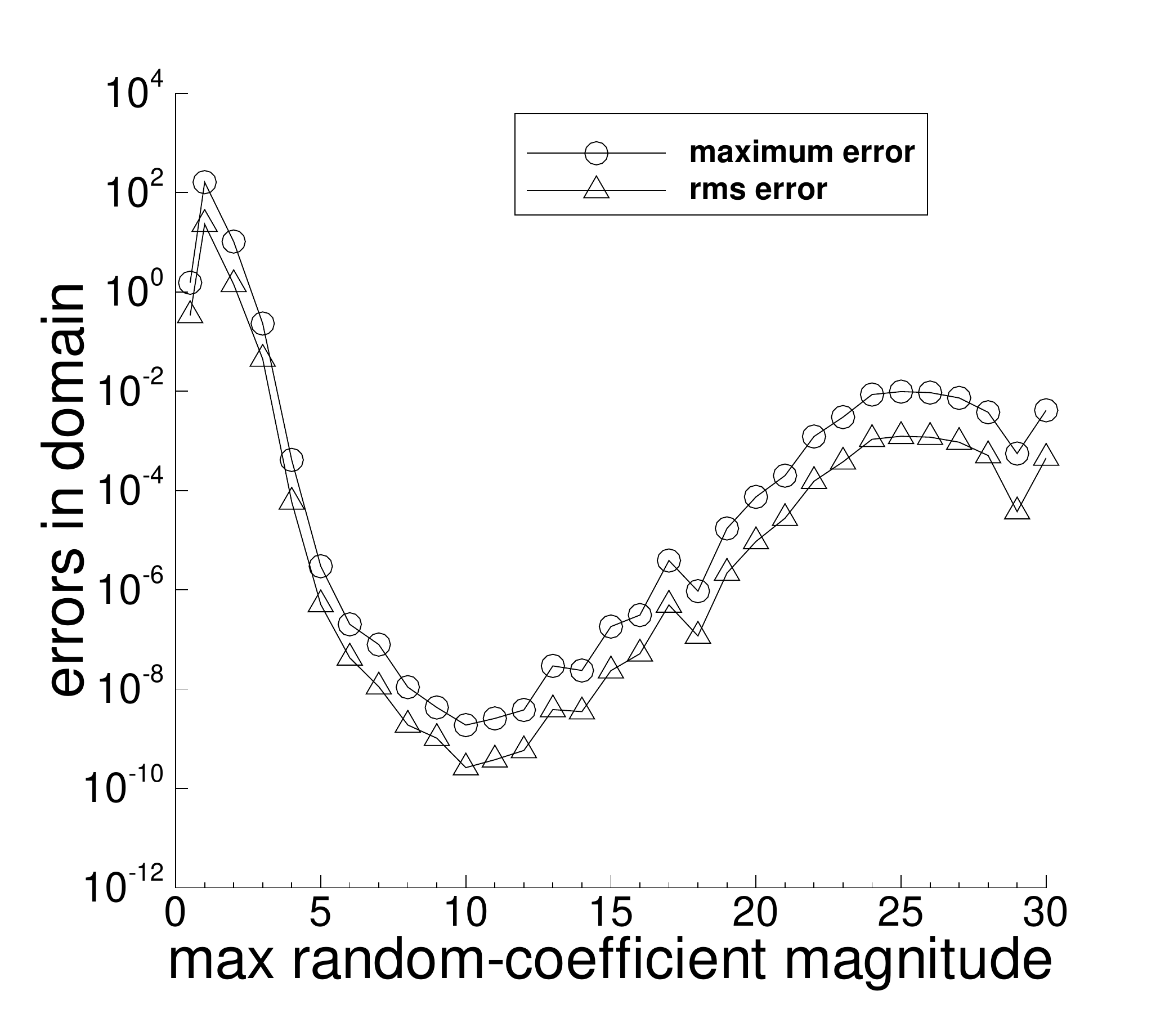}
  }
  \caption{Illustration of the random-coefficient effect on
    ELM accuracy: the maximum/rms errors of the ELM solution versus $R_m$
    (maximum magnitude of the random coefficients), for solving
    1D Helmholtz equation.
  }
  \label{fg_rm}
\end{figure}


The randomly-assigned coefficients in the neural network are
crucial to the performance of ELM,
and strongly influence its accuracy.
As an illustration, Figure \ref{fg_rm}
shows a typical plot of the $L^{\infty}$ (maximum) and
$L^2$ (root-mean-squares or rms) norms of the absolute error of the ELM solution
as a function of $R_m$, which denotes the maximum magnitude
of the random coefficients, for solving the one-dimensional (1D)
Helmholtz equation with Dirichlet boundary conditions.
Here the hidden-layer coefficients of the network
are assigned to uniform random values generated on the interval $[-R_m,R_m]$.
It is evident that the random coefficients  are critical to the ELM performance.
We refer to~\cite{DongL2020} for a recent fairly detailed study of
the random-coefficient effects on the ELM simulation results in
solving linear and nonlinear partial differential equations.
The effects of the random coefficients on the performance
of ELM and other random-weight neural networks
have also been recognized in regression and classification
problems other than scientific computing
(see e.g.~\cite{NeumannS2013,McDonnellTVST2015,YangW2016,WangL2017,Dudek2019,FreireRB2020}, among others).


How to choose, or perhaps pre-train, the random coefficients in the
ELM (or related random-weight) neural networks is an
important issue, and this issue is the focus of the current work.
Several studies in this regard are available from the literature
in the past few years.
In \cite{NeumannS2013} the batch intrinsic plasticity (BIP) method
is proposed to pre-train and adapt the activation function of the
hidden-layer neurons by a pseudo-inverse technique to achieve a desired
output distribution, so that the information transmission
of the neural network can be improved.
BIP is inspired by the biological intrinsic-plasticity mechanism~\cite{Triesch2005},
which, when applied to recurrent networks, can enhance the encoding and
improve the information transmission of the network~\cite{Steil2007}.
In \cite{McDonnellTVST2015} the authors employ the ELM method in handwritten
digit classification, and investigate ways to set the input weights
as a function of the input data by aiming to e.g.~increase the inner product
between the weights and the training data samples, constrain the input
weights to a set of difference vectors, or make the input weights sparse.
A combination of such ideas is also studied therein.
In \cite{YangW2016} the authors present an algorithm to grow the single
hidden-layer feed-forward network incrementally, by adding a
macro node each time, which consists of several hidden nodes and is called
a subnetwork hidden node. The method calculates
the subnetwork hidden nodes by pulling back the network error
into the hidden layer for invertible activation functions, and by aiming to
reduce the norms of the weights.
In \cite{WangL2017} the authors present a technique
to constructively build single hidden-layer
feed-forward networks by stochastic configuration algorithms
(called stochastic configuration networks or SCN).
The constructive process starts with a small network, and
the hidden nodes are added incrementally until an acceptable tolerance is achieved.
The added weights/biases are assigned by a supervisory mechanism
to satisfy certain inequality constraints guided by the universal approximation
property.
In addition to the above works, other researchers have aimed to
utilize the relationship between the input-data rank and
the performance of randomized neural networks, or to pick
the weights/bias based on the input data range and the activation function
type, or to consider the numerical
stabilities (see e.g.~\cite{CaoHGWM2020,Dudek2019,FreireRB2020}, among others).


In the current paper we present a modified batch intrinsic plasticity
(modBIP) method 
for pre-training the random coefficients of the ELM neural networks,
which can be shallow (single hidden layer) or deep (multiple hidden layers).
By random coefficient pre-training we mean that, after
the weight/bias coefficients of the hidden layers are initialized
to random values, we update these coefficients systematically by a well-defined
procedure. The updated coefficients are then fixed, and employed
in ELM for computing/training the weights in
the output layer (i.e.~the training parameters) by the least squares method.

The current modBIP method is devised based on the same principle
as the batch intrinsic plasticity (BIP) method~\cite{NeumannS2013}, namely, by
enhancing the information transmission in every node of the neural
network. The current method differs from BIP~\cite{NeumannS2013}
in two key aspects.
First, modBIP does not
involve the activation function in its algorithm, and it
can work with any activation function in the ELM neural network.
In contrast, BIP~\cite{NeumannS2013}
employs the inverse of the activation
function in its algorithm, and requires the activation function
to be invertible (i.e.~monotonic). BIP can only work with those
activation functions that are monotonic. This excludes many
often-used activation functions that are non-invertible,
such as the Gaussian function, the swish function~\cite{ElfwingUD2018},
the Gaussian error linear unit (GELU)~\cite{HendrycksG2020}, and other radial-basis
type activation functions.
Second, the modBIP method generates the target samples
on random intervals with some minimum size, which leads to highly accurate
simulation results when combined with ELM.
The combined ELM/modBIP method is observed
to be markedly more accurate than
the combined ELM/BIP method in numerical simulations.


We present a number of numerical examples of boundary-value
and boundary/initial-value problems with linear partial
differential equations to evaluate the performance
of modBIP and the combined ELM/modBIP method.
We compare their performance with those of the combined ELM/BIP method
and the ELM method without pre-training of the random coefficients.
These numerical experiments show that the combined ELM/modBIP method produces
highly accurate simulation results with both shallow and
deep neural networks, and that the accuracy of the ELM/modBIP
solution is insensitive to the initial random coefficients
in the neural network. More precisely, with the hidden-layer
coefficients initialized as uniform random values generated on $[-R_m,R_m]$,
for an arbitrary $R_m$, the combined ELM/modBIP method results in
very accurate results. This is in sharp contrast with the ELM method
without pre-training of the random coefficients (see e.g.~Figure \ref{fg_rm}).
The numerical results demonstrate that the combined ELM/modBIP method,
with non-invertible activation functions such as
the Gaussian/swish/GELU functions in the neural network,
exhibits the same properties of high accuracy and insensitivity to
the random coefficient initialization.
This is in sharp contrast with the ELM/BIP method, which breaks down
with the class of non-invertible activation functions.
The simulation results also signify the exponential decrease
in the numerical errors of the ELM/modBIP method as
the number of degrees of freedom (e.g.~number of training collocation points,
number of training parameters) in the system increases,
analogous to the observations of~\cite{DongL2020}.


We have also looked into the computational cost of the modBIP pre-training
of the random coefficients, as compared to that of the ELM training of
the neural networks. For every hidden-layer node, the primary operations with modBIP
consist of (i) the computation of the total input to the current node induced by
the input samples to the network, and (ii) the solution
of a small linear system consisting of two unknown
variables by the linear least squares method.
The modBIP pre-training cost increases linearly
or nearly linearly with increasing number of training parameters and
collocation points.
The pre-training cost is insignificant, and
it is only a fraction of the ELM training cost 
for the neural network. In typical numerical simulations,
the modBIP pre-training cost is within $10\%$ of the ELM
network training cost.


The contribution of this paper lies in the development of the modBIP
algorithm for pre-training the random coefficients of
shallow and deep ELM neural networks.
The algorithm has been shown to
be effective, efficient, and highly accurate. The combined ELM/modBIP method
is observed to be a promising  technique for
scientific computing.


The rest of this paper is structured as follows.
In Section \ref{sec:method} we present the modBIP algorithm for
pre-training the random coefficients of ELM neural networks,
and outline how to employ the combined ELM/modBIP method to solve
linear differential equations.
In Section \ref{sec:tests} we test the performance of the modBIP
algorithm and the combined ELM/modBIP method with function approximation
and several PDEs commonly encountered in computational
science/engineering~\cite{DongS2015,Dong2015clesobc,Dong2017}.
We compare the performance of the current modBIP algorithm, the
BIP algorithm,
and the case with no pre-training of the random coefficients.
The effectiveness of modBIP for shallow and deep neural networks, and
with invertible and non-invertible activation functions is demonstrated.
Section \ref{sec:summary} then concludes the discussions with
some closing remarks.


%% file: Method.tex
\section{Pre-training Random Coefficients of Extreme Learning Machines}
\label{sec:method}

\subsection{Extreme Learning Machine and Random Coefficients}


We consider a feed-forward
neural network~\cite{GoodfellowBC2016} with one
or multiple hidden layers, and the function representation using
this network. We assume the following in its configuration and settings:
\begin{itemize}
\item
  The weight/bias coefficients in all the hidden layers are pre-set to
  random values, and are fixed throughout the computation once they are set.
  In this work we follow~\cite{DongL2020}
  and set the hidden-layer coefficients to uniform random values
  generated on $[-R_m,R_m]$, where $R_m>0$ is a user-provided
  parameter.

\item
  The last hidden layer, i.e.~the layer before the output layer,
  can be wide. It may contain a large number of nodes.

\item
  The output layer is linear (i.e.~no activation function applied) and has
  zero bias.
  The training parameters consist of the weights of the output layer,
  and will be adjusted by the training computation.

\item
  The network is to be trained, and 
  the training parameters are to be determined by a least squares computation.

\end{itemize}
In the current work we concentrate on function approximation and
linear partial differential equations, and so the network
training is via a linear least squares computation. We refer to~\cite{DongL2020}
for the network training by a nonlinear least squares method
for solving nonlinear partial differential equations.

A feed-forward neural network with the above settings, when containing
a single hidden layer, is known as an extreme learning machine
(ELM)~\cite{HuangZS2006,HuangHSY2015}.
In the current work we consider neural networks with both a single
and multiple hidden layers, and we follow
this terminology and will refer to
them as shallow and deep extreme learning machines, respectively.


The random coefficients in the hidden layers of the neural network are crucial
to the performance and accuracy of
ELM~\cite{NeumannS2013,McDonnellTVST2015,FreireRB2020,DongL2020}. 
It has been observed from the numerical experiments in \cite{DongL2020} that
the ELM accuracy can be influenced strongly by
the maximum magnitude of the random coefficients (i.e.~$R_m$),
where uniform random coefficients generated on $[-R_m,R_m]$ are employed.
When $R_m$ is very large or very small,
ELM tends to produce results with poor accuracy.
More accurate results tend to be attained with $R_m$ in a
range of moderate values. This ``optimal'' range for $R_m$ is 
problem dependent and is also affected by the simulation resolution
(e.g.~the number of training
parameters, number of training data points)~\cite{DongL2020}.
For many problems the optimal range for $R_m$
appears to reside somewhere between $1$
and $15$.
As discussed in the Introduction section,
Figure \ref{fg_rm} is 
an illustration of the effect of $R_m$
on the ELM accuracy.


Our goal here is to devise a method for pre-training
the random coefficients once they are initialized,
so that accurate ELM results
can be obtained with random coefficients initialized by
essentially an arbitrary $R_m$. 
Once the hidden-layer coefficients in 
the neural network are initialized to uniform random values
from $[-R_m,R_m]$, for some given $R_m$, our method can be applied to
update or adjust these random coefficients.
The updated hidden-layer coefficients are then fixed, and the usual
ELM method and the least squares procedure can be employed to
determine the training parameters (i.e.~the output-layer coefficients).

\subsection{Modified Batch Intrinsic Plasticity (modBIP) Algorithm}
\label{modbip}

Consider a feed-forward neural network~\cite{GoodfellowBC2016}
with $(L+1)$ layers, and let $M_l$ denote the number of nodes in layer $l$
for $0\leqslant l\leqslant L$.
The layer zero represents the input to the neural network,
and let the matrix $\mbs X$ of dimension $N_s \times M_0$ denote the input data,
where $N_s$ is the number of samples in the input data.
The layer $L$ represents the output of the neural network,
denoted by the matrix $\mbs U$ of dimension $N_s\times M_L$.
The layers in between are the hidden layers.
Let the matrix $\bm \Phi_l$, with dimension $N_s\times M_l$,
denote the output data of layer $l$
for $0\leqslant l\leqslant L$,
with $\bm \Phi_0=\mbs X$ and $\bm\Phi_L=\mbs U$.
Then the logic of the hidden layer $l$ ($1\leqslant l\leqslant L-1$) is given by,
\begin{equation}\label{eq_1}
  \bm\Phi_l = \sigma\left(\bm\Phi_{l-1} \mbs W_l + \mbs b_l \right), \quad
  1\leqslant l \leqslant L-1,
\end{equation}
where $\sigma(\cdot)$ denotes the activation function, 
the $M_{l-1}\times M_l$ matrix $\mbs W_l$ denotes the weights
of layer $l$, and
the row vector $\mbs b_l$ (with dimension $1\times M_l$)
denotes the biases of this layer.
Note that here we have adopted the convention (as in the computer language Python)
that when computing
$(\bm\Phi_{l-1} \mbs W_l + \mbs b_l)$, the data in the vector $\mbs b_l$ will first be
propagated along the first dimension to form a
$N_s\times M_l$ matrix.
In equation ~\eqref{eq_1} we have also assumed for simplicity
that the same activation function
is employed for different hidden layers.
The logic of the output layer is given by
\begin{equation}\label{eq_2}
  \mbs U = \bm\Phi_{L-1} \mbs W_L
\end{equation}
where the $M_{L-1}\times M_L$ matrix $\mbs W_L$ denotes the weights
of the output layer, and they are the training parameters of
the neural network. As discussed before, the output layer is assumed to contain no
bias and no activation function.
The weight and bias coefficients of the hidden layers, $\mbs W_l$ and
$\mbs b_l$ ($1\leqslant l\leqslant L-1$), are initialized to uniform
random values generated on the interval $[-R_m,R_m]$ for some prescribed $R_m$.
Once the specific problem is given,
the training parameters $\mbs W_L$ can be determined by
a least squares computation based on the ELM procedure~\cite{DongL2020}.
The parameter value $R_m$, and hence
the random coefficients $\mbs W_l$ and $\mbs b_l$,
strongly influence
the ELM accuracy, as discussed in the previous subsection.

Given the input data $\mbs X$ and the initial random coefficients
$\mbs W_l$ and $\mbs b_l$ ($1\leqslant l\leqslant L-1$),
we will compute a set of new coefficients $\mbs W'_l$ and $\mbs b'_l$
for $1\leqslant l\leqslant L-1$ based on a procedure presented below,
and replace $\mbs W_l$ and $\mbs b_l$  by
the newly computed values, so that the resultant neural
network will give rise to results that are more accurate
and less sensitive or insensitive to  $R_m$.
We refer to this process as the pre-training of the random
coefficients. Once the random hidden-layer coefficients are
pre-trained, they will be fixed throughout the rest of the computation, when 
the training parameters are determined by the least squares method
in the usual ELM algorithm.


To pre-train the random hidden-layer coefficients,
we follow the philosophy as advocated in~\cite{NeumannS2013}. In other words,
these coefficients should assume values that will
facilitate the information transmission within each neuron. 
Consider a particular node (or neuron) in a particular hidden
layer of the neural network.
Let $s$ denote the total input (synaptic input) to this neuron from the previous layer,
and $\phi$ denote the output signal of this neuron.
Then $\phi = \sigma(s)$, where $\sigma$ is the activation function
of this neuron. For commonly used activation
functions (e.g.~$\tanh$, sigmoid, Gaussian),
if the magnitude of the input $s$ is very large, the output $\phi$ of
the neuron will reach the level of saturation, which is unfavorable
for the information transmission in this neuron. Therefore, the magnitude of
the synaptic input to the neuron should not be too large, in order to
facilitate the information transfer.
Let us further suppose that  the synaptic input to this neuron consists of $N_s$ independent
samples $s_i$ ($1\leqslant i\leqslant N_s$), and
let $s_{\max}$ and $s_{\min}$ denote the maximum and the minimum of these
input samples. If $s_{\max}$ and $s_{\min}$ are very close to each other,
then this neuron will output essentially a constant value under these
$N_s$ input samples, which is unfavorable for the
information transmission. Therefore, the samples of
the synaptic input to a neuron should maintain
a reasonable spread in their values, in order
to facilitate the information transfer.
In light of these considerations, in order to facilitate the
information transmission, we will impose the following
requirements on the synaptic input to any neuron:
\begin{itemize}
\item
  The synaptic input to the neuron should fall within a range $[-S_b, S_b]$,
  where $S_b>0$ is a user-provided hyper-parameter.
  The larger the $S_b$ parameter, the more likely the input will cause a saturation
  in the neuron response.

\item
  The samples of the synaptic input to the neuron should be such that
  $s_{\max}-s_{\min}>S_c$, where $S_c$ (with $0\leqslant S_c<2S_b$)
  is a user-provided hyper-parameter.
  A non-zero $S_c$ ensures that the input samples to the neuron have a spread
  of at least $S_c$ in their values.

\end{itemize}
These requirements provide the basis for the algorithm described below
for pre-training the random hidden-layer coefficients in
the neural network.

Given $S_b$ and $S_c$,
we pre-train the random coefficients as follows.
We start with the first hidden layer, and pre-train the random coefficients
in each layer individually and {\em successively}, until the last hidden layer
is pre-trained. It should be noted that
pre-training a later hidden layer depends on
the updated weight/bias coefficients in previous layers that
are already pre-trained.
Within each hidden layer, we pre-train the random coefficients
associated with each node individually and independently. We start with
the first node and proceed until all the nodes in the layer are pre-trained.

The general idea for pre-training a node is as follows. 
For any particular node in a layer, we first compute the total input to this node
corresponding to all the $N_s$ input data samples to the neural network.
This produces the input samples $s_i$ ($1\leqslant i\leqslant N_s$) to this node.
We generate a {\em random} sub-interval $[t_{\min}, t_{\max}]\subset [-S_b,S_b]$,
satisfying the condition $t_{\max} - t_{\min}>S_c$.
Then we generate $N_s$ random numbers $t_i$ ($1\leqslant i\leqslant N_s$)
on the interval $[t_{\min}, t_{\max}]$, which will be referred to as
the target samples.
We sort the input samples $s_i$ and the target samples $t_i$ in the ascending order,
respectively. Then we compute an affine  mapping between $s_i$ and $t_i$
by a linear least squares method. The weight/bias coefficients associated with
this node are then updated by the computed affine mapping coefficients
to complete the pre-training for this node.

Let us now expand on the general idea to provide more details for pre-training
a node. We consider pre-training the random coefficients associated
with node $k$ in the hidden layer $l$, for $1\leqslant l\leqslant L-1$
and $1\leqslant k\leqslant M_l$.
At this point, all the previous hidden layers have been pre-trained and
their weight/bias coefficients have been updated.
We evaluate the neural network against the input data $\mbs X$
to attain the output of the layer $l-1$, which is
denoted by the matrix $\bm \Phi_{l-1}$
of dimension $N_s\times M_{l-1}$. Note that $\bm\Phi_{l-1}=\mbs X$
if this is the first hidden layer (i.e.~$l=1$).
The current weights for layer $l$ are given by
the $M_{l-1}\times M_l$ matrix $\mbs W_l$, and the current biases
for this layer are given by
the row vector $\mbs b_l$ with dimension $M_l$. Let $\mbs w$ denote
the $k$-th column of $\mbs W_l$, and $b_k$ denote the $k$-th component of
$\mbs b_l$. Then the synaptic input, $\mbs s$, to the current node in consideration is
given by
\begin{equation}\label{eq_3}
  (s_1,s_2,\dots,s_{N_s})^T = \mbs s = \mbs \Phi_{l-1}\mbs w + \mathds{1} b_k,
\end{equation}
where $s_i$ ($1\leqslant i\leqslant N_s$) are the components of $\mbs s$,
and $\mathds{1}$ denotes the vector of all ones.
The values of $\mbs w$ and $b_k$ will be updated when this node is pre-trained.

Next we generate two uniform random numbers $t_{\min}$ and $t_{\max}$ on $[-S_b,S_b]$
that satisfy the condition $t_{\max}-t_{\min}>S_c$.
Then we generate $N_s$ random numbers $t_i$ ($1\leqslant i\leqslant N_s$) on
the interval $[t_{\min},t_{\max}]$ as the target samples.
In the current implementation we have considered two distributions when generating
the target samples:
\begin{itemize}

\item
  $t_i$ ($1\leqslant i\leqslant N_s$) are generated on $[t_{\min},t_{\max}]$
  from a normal distribution with a mean $\frac12(t_{\min}+t_{\max})$ and
  a standard deviation $\frac14(t_{\max}-t_{\min})$.
  When drawing from the normal distribution, if the generated random number
  is out of the range $[t_{\min},t_{\max}]$, a simple sub-iteration can produce
  a random number on $[t_{\min},t_{\max}]$.

\item
  $t_i$ ($1\leqslant i\leqslant N_s$) are uniform random numbers on $[t_{\min},t_{\max}]$.
  
\end{itemize}

We sort the input samples $s_i$ ($1\leqslant i\leqslant N_s$) in the ascending order,
and also sort the target samples $t_i$ ($1\leqslant i\leqslant N_s$)
in the ascending order.
Then we solve for two scalar numbers $\xi$ and $\eta$ from the following
linear system by the linear least squares method,
\begin{equation}\label{eq_4}
  s_i \xi + \eta = t_i, \quad 1\leqslant i\leqslant N_s.
\end{equation}
Finally, we update the column $k$ of the weight-coefficient matrix
$\mbs W_l$ and the $k$-th component of the bias vector $\mbs b_l$
by the following relations:
\begin{equation}\label{eq_5}
  \mbs w \longleftarrow \xi\mbs w, \quad
  b_k \longleftarrow \xi b_k + \eta.
\end{equation}
This completes the pre-training of the node.

\begin{algorithm}[tb]
  \DontPrintSemicolon
  \SetKwInOut{Input}{input}\SetKwInOut{Output}{output}

  \Input{input data $\mbs X$; initial random weight coefficients $\mbs W_l$
    and initial random bias coefficients $\mbs b_l$,
    for $1\leqslant l\leqslant L-1$; constant
  $S_b>0$; constant $S_c$, with $0\leqslant S_c<2S_b$.}
  \Output{updated weight coefficients $\mbs W_l$ and updated bias coefficients $\mbs b_l$,
  for $1\leqslant l\leqslant L-1$.}
  \BlankLine\BlankLine
  \For{$l\leftarrow 1$ \KwTo $L-1$}{
    \eIf{$l$ equals $1$}{set $\bm\Phi_{l-1}=\mbs X$}{
      compute $\bm\Phi_{l-1}$ by evaluating the
      neural network (first $l-1$ layers) on the input data $\mbs X$\;
    }
    \BlankLine\BlankLine
    \For{$k\leftarrow 1$ \KwTo $M_l$}{
      set $\mbs w$ to point to the column $k$ of $\mbs W_l$\;
      set $b_k$ to point to the $k$-th component of $\mbs b_l$\;
      \BlankLine
      compute the input samples $s_i$ ($1\leqslant i\leqslant N_s$) by equation \eqref{eq_3}\;
      sort $s_i$ ($1\leqslant i\leqslant N_s$)\;
      \BlankLine
      generate uniform random numbers $t_{\min}$ and $t_{\max}$ on $[-S_b,S_b]$ satisfying
      $t_{\max}-t_{\min}>S_c$\;
      generate random numbers $t_i$ ($1\leqslant i\leqslant N_s$)
      on $[t_{\min},t_{\max}]$ by a normal distribution
      (mean: $(t_{\min}+t_{\max})/2$, stddev: $(t_{\max}-t_{\min})/4$) or a uniform distribution\;
      sort $t_i$ ($1\leqslant i\leqslant N_s$)\;
      \BlankLine
      solve equation \eqref{eq_4} for $\xi$ and $\eta$ by the linear least squares method\;
      update $\mbs w$ and $b_k$ by equation \eqref{eq_5}
    } 
  } 
  \caption{modBIP algorithm}
  \label{alg_1}
\end{algorithm}

The overall pre-training procedure by the modBIP method is summarized
in Algorithm~\ref{alg_1}.
%
A key construction in the algorithm that enables high accuracy of
this method is the adoption of random sub-intervals $[t_{\min},t_{\max}]$
with a minimum size $S_c$ when generating the target samples.
If this interval is taken to be $[-S_b,S_b]$ or some fixed sub-interval of
$[-S_b,S_b]$, numerical experiments show that the method will be
much less accurate. 
Target samples generated on $[t_{\min},t_{\max}]$ from a uniform distribution
and from a normal distribution seem to produce results with comparable accuracy.
For different problems one type of distribution may lead to results
with slightly better accuracy than the other, but their error levels are
largely the same. In the numerical tests of Section \ref{sec:tests},
we employ the normal distribution when generating
random target samples on $[t_{\min},t_{\max}]$.

\begin{figure}
  \centerline{
    \includegraphics[width=2.2in]{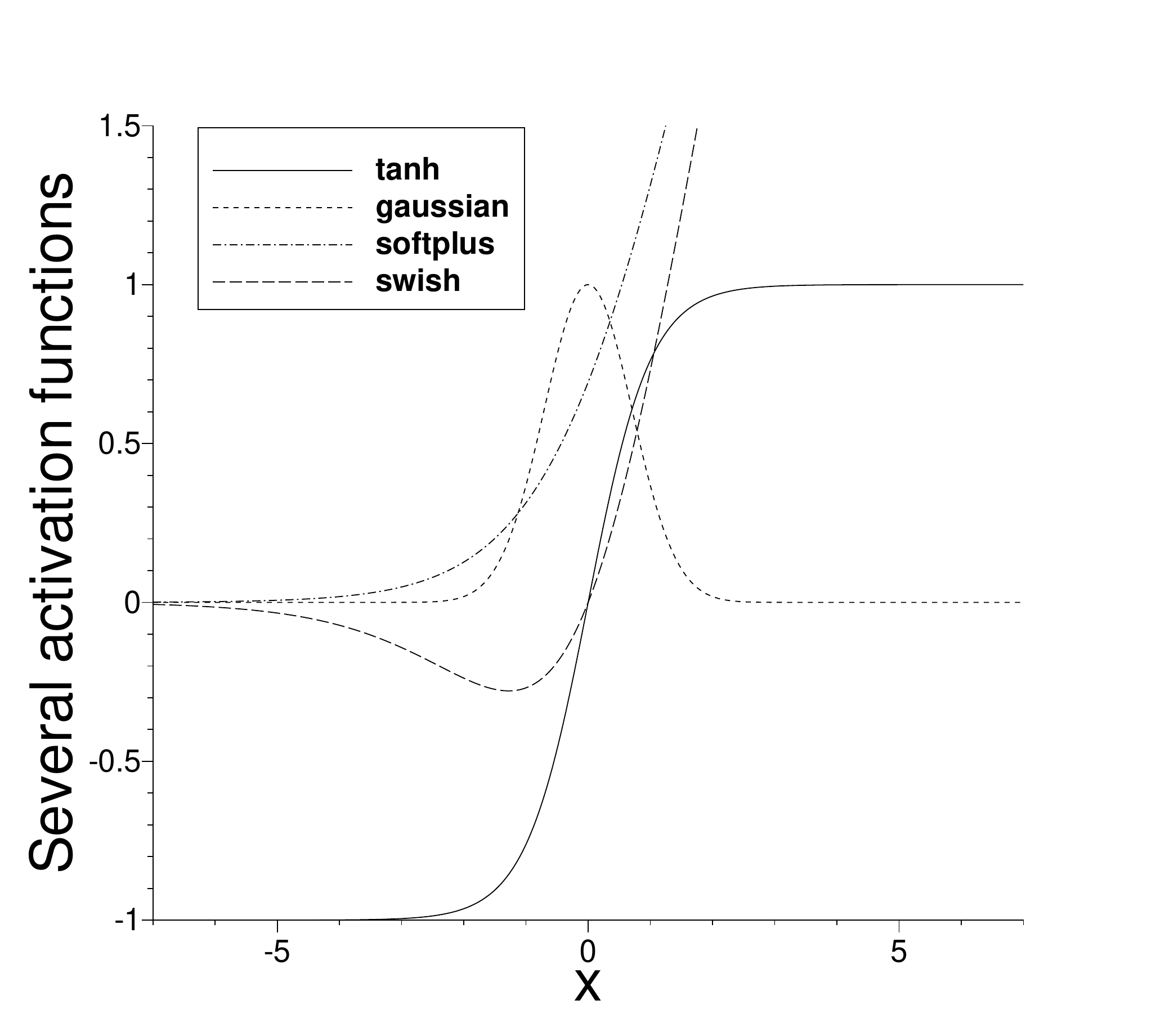}
  }
  \caption{Profiles of several commonly-used activation functions.}
  \label{fg_activ}
\end{figure}


\begin{remark}\label{rem_2}
  The parameter $S_c$ controls the minimum size of the random sub-interval
  $[t_{\min},t_{\max}]$. As $S_c\rightarrow 0$, random intervals with a near-zero size
  may be generated. This will cause the mapped synaptic input (and also the neuron
  response) to cluster
  around a constant level, which will affect the accuracy adversely.
  On the other hand, as $S_c\rightarrow 2S_b$, all the random sub-intervals
  $[t_{\min},t_{\max}]\rightarrow [-S_b,S_b]$. So the randomness in the sub-interval
  will be lost, and the accuracy will deteriorate as mentioned before.
  We observe from numerical experiments that 
  a value around $S_c=S_b/2$  
  produces results with very good (and oftentimes the best) accuracy.
  So in the current paper we will employ $S_c=S_b/2$ with modBIP
  in the numerical simulations.

\end{remark}

\begin{remark}\label{rem_1}

  The parameter $S_b$ controls which regime the pre-training algorithm
  generally maps the synaptic input data into.
  If $S_b$ is very small, the mapped synaptic input
  will be close to zero,
  and many activation functions are close to
  a linear function in this regime. Since a linear function reduces the approximation
  capability of a neuron, the accuracy in this case will be limited.
  If $S_b$ is very large, the magnitude of the mapped
  synaptic input to the neuron can be
  large. It is thus more likely to cause saturation in the neuron response,
  which is unfavorable for the information transmission and
  can have an adverse effect on the accuracy.
  Figure \ref{fg_activ} shows the profiles of several commonly-used
  activation functions, including the $\tanh$, Gaussian ($\sigma(x)=e^{-x^2}$),
  softplus ($\sigma(x)=\log(1+e^x)$), and swish ($\sigma(x)=x/(1+e^{-x})$)
  functions. They suggest that a reasonable range for
  the input to the activation function
  seems to be somewhere around $[-3,3]$, and perhaps even a little larger
  with the swish and softplus functions.
  We observe from numerical experiments that,
  with the $\tanh$ (and Gaussian) activation function and a single hidden layer in the
  neural network,
  a value around $S_b=2\sim 3$
  will produce results with good accuracy.
  For certain problems, the method achieves even better
  accuracy if $S_b$ is adjusted from this base value.
  When the neural network contains more hidden layers,
  it is observed that
  $S_b$ should typically be decreased from this reference range
  to achieve a better accuracy.

  While the above reference range for $S_b$ is useful, for a given problem
  can we estimate the $S_b$ value that provides the best or close
  to the best accuracy? The answer is positive, and we next outline a procedure
  for estimating the optimal $S_b$ using simple numerical experiments.
  Let us use the problem of solving linear PDEs for illustration.
  When training the ELM neural network,
  suppose the linear system resulting from
  the PDE that one needs to solve using the least squares method
  is given by the following (see Section \ref{sec:pde}),
  \begin{equation}\label{eq_6}
    \mbs A \mbs x = \mbs b
  \end{equation}
  where $\mbs A$, $\mbs x$, and $\mbs b$ denote the coefficient matrix (non-square),
  the vector of unknowns, and the right-hand-side vector, respectively.
  Let $\mbs r$ denote the residual vector associated with the least
  squares solution,
  \begin{equation}
    \mbs r = \mbs b - \mbs A \mbs x^{+},
  \end{equation}
  where $\mbs x^{+}$ denotes the least squares solution to equation \eqref{eq_6} (with
  minimum norm if rank-deficient).
  We use the residual norm $\| \mbs r\|$ as an indicator to the accuracy
  of the least squares solution. Since $\| \mbs r \|$ can be readily evaluated,
  we will use preliminary simulations to compute
  $\| \mbs r\|$ and estimate the best $S_b$.
  The main steps are as follows:
  \begin{itemize}
  \item
    Consider a set of points from a range for $S_b$ (e.g.~$S_b\in[0.5, 5]$).

  \item
    For each $S_b$ value, set $S_c=S_b/2$ and pre-train the random coefficients of
    the network using modBIP.

  \item
    Perform a preliminary ELM simulation using the pre-trained neural network, and
    compute  $\| \mbs r \|$.

  \item
    Collect $\| \mbs r  \|$ for the set of $S_b$ values.
    Find the $S_b$ corresponding
    to the smallest (or close to the smallest)
    $\| \mbs r  \|$. Use this value as an estimate
    for the best $S_b$.

    
  \end{itemize}
  With the estimate for $S_b$ available,
  we can then use it in modBIP and perform actual simulations
  for the given problem with ELM.
  It should be noted that in the simulations for estimating $S_b$,
  the number of training
  data points should be larger than the number of
  training parameters in the neural
  network to avoid the regime of rank deficiency in the least squares
  solution.

\end{remark}

\begin{remark}
  \label{rem_2a}
  In the current work we have considered a symmetric interval $[-S_b,S_b]$
  in the modBIP algorithm. For activation functions that are
  not symmetric or anti-symmetric
  (e.g.~swish, softplus), one can imagine that the use of a non-symmetric
  interval such as $[S_{b1}, S_{b2}]$ in the algorithm
  might be more favorable.
  This aspect is not considered here, and we employ a symmetric
  interval in the current work.

\end{remark}


\begin{remark}\label{rem_3}
  We observe that the ELM method, with the random coefficients pre-trained
  by the current modBIP algorithm, produces highly accurate simulation results,
  and that the accuracy of the combined ELM/modBIP method is
  insensitive to the random coefficient initializations.
  More specifically, with the hidden-layer 
  coefficients initialized as random values
  generated on $[-R_m,R_m]$, for an arbitrary $R_m$,
  the combined ELM/modBIP method produces accurate results and the accuracy
  is insensitive to $R_m$.
  This is very different from the behavior of ELM without pre-training of
  the random coefficients (see e.g.~Figure \ref{fg_rm}).
  We will demonstrate this point with numerical experiments
  in Section \ref{sec:tests}.

\end{remark}


\begin{remark}\label{rem_4}
  It is evident that the  modBIP algorithm does not involve
  the activation function in its construction. Therefore,
  essentially any activation function can be used in the neural network together
  with the current method. This is in sharp contrast with
  the BIP algorithm~\cite{NeumannS2013}, which employs the inverse
  of the activation function in its construction. BIP requires the
  activation functions in the neural network to be invertible (i.e.~monotonic).
  This precludes many often-used activation
  functions such as the  Sigmoid
  weighted linear unit (SiLU or swish)~\cite{ElfwingUD2018},
  Gaussian error linear unit (GELU)~\cite{HendrycksG2020}, Gaussian,
  and other radial basis-type functions.

\end{remark}


\begin{remark}\label{rem_5}
  
  We briefly mention another method for generating target samples
  from a normal distribution, which is different from what has been discussed
  above. In Algorithm~\ref{alg_1} we replace the constant $S_c$ by
  two constants $S_{c_1}$ and $S_{c_2}$, with $0<S_{c_1}\leqslant S_{c_2}$.
  So now there are three constant parameters in the input, $S_b$, $S_{c_1}$ and $S_{c_2}$.
  We replace the lines $12$ and $13$ of Algorithm~\ref{alg_1}
  by the following steps for generating the random target
  samples $t_i$ ($1\leqslant i\leqslant N_s$):
    \begin{align*}
      &
      \textnormal{generate a uniform random number}\ \mu\
      \textnormal{on}\ [-S_b,S_b];\\
      &
      \textnormal{generate a uniform random number}\ \delta\
      \textnormal{on}\ [S_{c_1},S_{c_2}];\\
      &
      \textnormal{generate random numbers}\ t_i\ (1\leqslant i\leqslant N_s)\
      \textnormal{from a normal distribution with mean}=\mu,\ \textnormal{stddev}=\delta.
  \end{align*}
    Here we use a random mean $\mu$ from $[-S_b,S_b]$ and
    a random standard deviation $\delta$ from $[S_{c_1},S_{c_2}]$ for generating
    the target samples $t_i$.
    We observe that 
    a value around $S_b=2\sim 2.5$, $S_{c_1}=0.2$ and
    $S_{c_2}=S_b/2$ generally produce results with good accuracy.

\end{remark}


\subsection{Solving Linear Differential Equations with Combined ELM/modBIP}
\label{sec:pde}

We will test the modBIP pre-training algorithm by combining
it with ELM for solving linear partial differential equations.
We first initialize
the hidden-layer coefficients in the neural network
by uniform random values from 
$[-R_m,R_m]$, for some prescribed $R_m$.
Then we pre-train these random coefficients by modBIP, and afterwards fix
the updated hidden-layer coefficients.
At this point, we can use the ELM method and the pre-trained neural network
in the usual fashion.
We can compute the training parameters (output-layer coefficients)
by a linear least squares method for solving linear partial differential
equations. This will be discussed in this subsection.

The use of ELM for solving linear partial differential equations has been
discussed in a number of previous works;
see e.g.~\cite{PanghalK2020,DwivediS2020,DongL2020} and the references therein.
For the sake of completeness, we summarize the main procedure below, and
we refer the reader to e.g.~\cite{DongL2020}
for more detailed discussions of related aspects.
Here we assume that the hidden-layer coefficients of
the neural network
have been pre-trained by modBIP
as discussed above. So the weight/bias coefficients in the hidden layers
are fixed throughout the computations to be discussed below.


For illustration we consider a rectangular domain in two dimensions (2D),
$\Omega = \{(x,y)\ |\ x\in[a_1,a_2],\ y\in[b_1,b_2] \}$.
If the problem is time-dependent, we will treat the time $t$ in
the same way as the spatial variables, and use the last independent variable
to denote the time $t$. In the case with two independent variables,
for time-dependent
problems, the last independent variable (i.e.~$y$)
denotes the time $t$. With this notation, we can treat time-dependent and
time-independent problems in a unified fashion.
So the following discussions
also apply to time-dependent problems.

Consider a generic linear partial differential equation on $\Omega$,
\begin{subequations}
\begin{align}
  &
  \mbs L u = f(x,y), \label{eq_6a} \\
  &
  \mbs B u = g(x,y), \quad \text{on}\ \partial\Omega, \label{eq_6b}
\end{align}
\end{subequations}
where $u(x,y)$ is the field function to be solved for,
$\mbs L$ is a linear differential operator, $\mbs B$ is a linear operator,
$f(x,y)$ is a prescribed source term on the domain, and $g(x,y)$
is a prescribed source term defined on the domain boundary $\partial\Omega$.
We assume that the system as given by \eqref{eq_6a}--\eqref{eq_6b} is well-posed.
Note that, depending on the order of $\mbs L$, the boundary condition~\eqref{eq_6b}
may be imposed only on  a part of the domain boundary, and that it should include
the initial condition(s) if this is a time-dependent problem.

To solve equations \eqref{eq_6a}--\eqref{eq_6b}, we use an extreme learning
machine (feed-forward neural network), with its
random hidden-layer coefficients pre-trained by modBIP,
to represent the solution field $u(x,y)$.
The input layer of the neural network consists of two nodes, representing
$x$ and $y$, respectively. The output layer of the neural network consists of
one node, representing the solution $u$. The neural network contains
one or multiple hidden layers in between. Let $M$ denote the number of
nodes in the last hidden layer of the neural network, and
let $V_j(x,y)$ ($1\leqslant j\leqslant M$) denote the output fields of
the last hidden layer. Then the logic of the output layer is given by
\begin{equation}\label{eq_7}
  u(x,y) = \sum_{j=1}^M \beta_j V_j(x,y),
\end{equation}
where $\beta_j$ ($1\leqslant j\leqslant M$) denote the weight coefficients
in the output layer, which are the training parameters of the neural network.

We employ $(Q_x+1)$ and $(Q_y+1)$ uniform grid points 
in the $x$ and $y$ directions,
respectively, with their coordinates given by
\begin{equation}\label{eq_8}
  x_p = a_1+\frac{a_2-a_1}{Q_x}p, \quad
  y_q = b_1 + \frac{b_2-b_1}{Q_y}q, \quad
  0\leqslant p\leqslant Q_x, \quad
  0\leqslant q\leqslant Q_y.
\end{equation}
We enforce the equation \eqref{eq_6a} on all the grid points $(x_p,y_q)$,
which will be refer to as the collocation points, and arrive at
\begin{equation}\label{eq_9}
  \sum_{j=1}^M \left[ \mbs LV_j(x_p,y_q)\right] \beta_j = f(x_p,y_q), \quad
  0\leqslant p\leqslant Q_x, \quad 0\leqslant q\leqslant Q_y,
\end{equation}
where equation \eqref{eq_7} has been employed.
Let $X_b$ denote the set of collocation points, among $(x_p,y_q)$,
that reside on the
domain boundary $\partial\Omega$ where the boundary condition~\eqref{eq_6b}
is imposed on. Let $Q_b$ denote the number of points in $X_b$.
We enforce the boundary condition \eqref{eq_6b} on each $(x'_m,y'_m)\in X_b$,
and arrive at
\begin{equation}\label{eq_10}
  \sum_{j=1}^M [\mbs BV_j(x'_m,y'_m)]\beta_j = g(x'_m,y'_m), \quad
  0\leqslant m\leqslant Q_b-1.
\end{equation}

Equations \eqref{eq_9} and \eqref{eq_10} form a linear algebraic
system about the training parameters $\beta_j$ ($1\leqslant j\leqslant M$).
This system consists of $[(Q_x+1)(Q_y+1)+Q_b]$ equations and $M$ unknowns.
The terms involved in the coefficient matrix,
such as $V_j(x_p,y_q)$, $\mbs LV_j(x_p,y_q)$, $V_j(x'_m,y'_m)$
and $\mbs BV_j(x'_m,y'_m)$, can be computed
by a forward evaluation of the neural network or by auto-differentiation.
We seek a least squares solution (with minimum norm if the problem
is rank deficient) to this system, and solve it by
the linear least squares method~\cite{GolubL1996}.
In the current implementation, we have employed the linear least squares routine from
LAPACK, available through the wrapper function in the scipy package
in Python (function scipy.linalg.lstsq).


In the current paper, we implement the neural network  in Python
employing the Tensorflow and Keras libraries. The neural-network layers
are implemented as the ``Dense'' layers in Keras. The input (training) data
to the neural network consist of the coordinates of all
the collocation points $(x_p,y_q)$ ($0\leqslant p\leqslant Q_x$ and
$0\leqslant q\leqslant Q_y$) in the domain.
In our implementation, we have incorporated an affine mapping
between the input layer and the first hidden layer to normalize
the input $(x,y)$ data from $\Omega=[a_1,a_2]\times[b_1,b_2]$
to the domain $[-1,1]\times[-1,1]$.
This mapping is implemented by a ``lambda'' layer in Keras, which contains
no weight/bias coefficients. This lambda layer does not need to be pre-trained
by modBIP. With this lambda layer incorporated, in line $3$ of 
Algorithm~\ref{alg_1}, $\bm\Phi_{l-1}$ should be the output of the lambda
layer, i.e.~the normalized data, instead of the original
input data $\mbs X$.

After the linear system consisting of \eqref{eq_9} and \eqref{eq_10}
is solved by the linear least squares method, the weight coefficients
of the output layer  will be set to the computed solution.
Then the neural network is evaluated on a set of finer grid points,
which is different from the training data points, to attain the field solution
data $u(x,y)$. The solution data is then
compared with e.g.~the exact solution to compute
the errors and other useful quantities.
These steps have been followed in the numerical experiments
of Section \ref{sec:tests}.


\begin{remark}\label{rem_6}
  For longer-time simulations of time-dependent partial differential equations,
  we employ the block time-marching scheme developed in~\cite{DongL2020}.
  The spatial-temporal domain is first divided into a number of windows
  along the time (time blocks).
  The equations~\eqref{eq_9} and \eqref{eq_10} are solved on each time block,
  individually and successively, using the method discussed in this sub-section.
  After one time block is computed, the solution at the last time instant
  will be evaluated, and used as the initial condition for the computation
  of the time block that follows.
  We refer to \cite{DongL2020} for more detailed discussions of this scheme.
  Block time marching has been employed for simulations of time-dependent
  problems in Section \ref{sec:tests}.

\end{remark}


%% file: Test.tex
\section{Representative Numerical Examples}
\label{sec:tests}


In this section we evaluate the performance of the modBIP
algorithm using function approximation and
linear partial differential equations
in one or two dimensions (1D/2D) in space, and plus time if the
problem is time-dependent.
We solve these equations numerically by the combined ELM/modBIP method
as discussed in Section \ref{sec:pde}.
The random hidden-layer coefficients in the neural network
are pre-trained by modBIP first, and then they are fixed and used
in ELM for finding the solutions to the differential equations.
In the numerical experiments reported below,
with modBIP we employ $S_c=S_b/2$, and estimate $S_b$ 
using the procedure outlined in Remark \ref{rem_1} by computing
the residual norm of the linear least squares (LLSQ) problem in ELM.
When generating the target samples on the random interval
$[t_{\min},t_{\max}]$, we have employed the normal distribution
with a mean $(t_{\min}+t_{\max})/2$ and a standard deviation
$(t_{\max}-t_{\min})/4$ (see line $13$ of Algorithm \ref{alg_1})
in all the numerical tests.
The numerical experiments are conducted on a MAC computer
(3.2GHz Quad-Core Intel Core i5 CPU, 24GB memory) in the authors' institution.
The wall clock time is collected by using the ``timeit'' module in Python.

In the current implementation, the initial random coefficients
in the hidden layers are generated using the random number generator
from the Tensorflow library (invoked by the initialization routines in
the Keras library), while the random values in the modBIP algorithm are generated
by the random number generator in the numpy package in Python.
In order to make all the simulation results reported here fully
repeatable, we have employed the same seed value for the random number
generators in both Tensorflow and numpy, and the seed value is fixed
for all the numerical experiments reported within a subsection.
More specifically, the seed value
is $1$ for the numerical experiments presented in Sections \ref{sec:func}
and \ref{sec:poisson}, $12$ for those in Section \ref{sec:wave},
and $22$ for those in Section \ref{sec:diffu}, respectively.

Hereafter we employ the vector $[M_0, M_1,\dots,M_L]$ ($L\geqslant 2$)
to represent the architecture
of the feed-forward neural network in ELM, where
the vector length $(L+1)$ denotes the number of layers in the network
and $M_i$ is the number of nodes in layer $i$ for $0\leqslant i\leqslant L$.
Note that $M_0$ and $M_L$ are
the numbers of nodes in the input and output layers,
respectively. The number of training parameters in ELM is
$M_{L-1}$, i.e.~the number of nodes in the last hidden layer, as discussed in
Section \ref{sec:method}.

\subsection{Function Approximation}
\label{sec:func}

\begin{figure}
  \centerline{
    \includegraphics[width=2in]{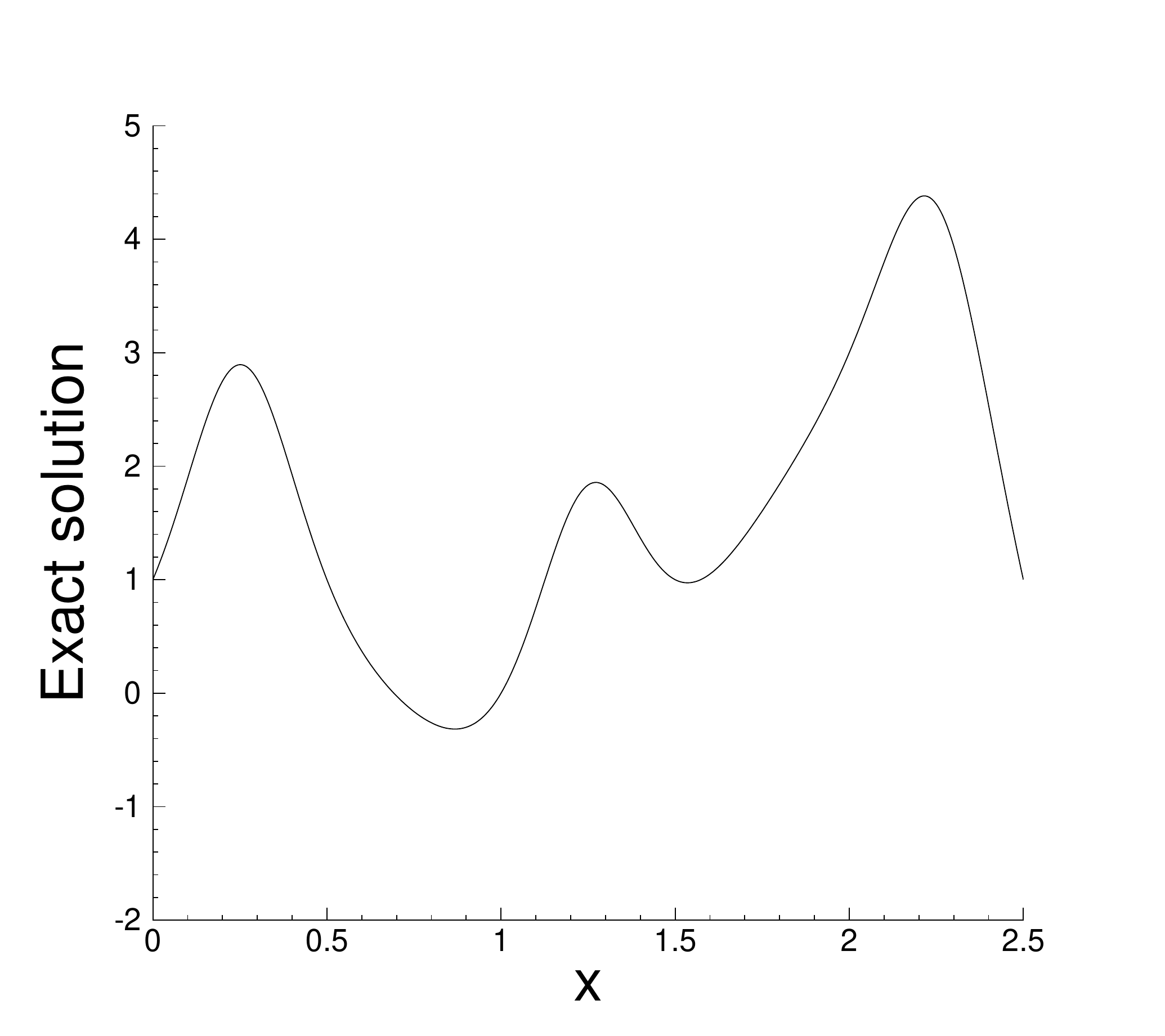}(a)
    \includegraphics[width=2in]{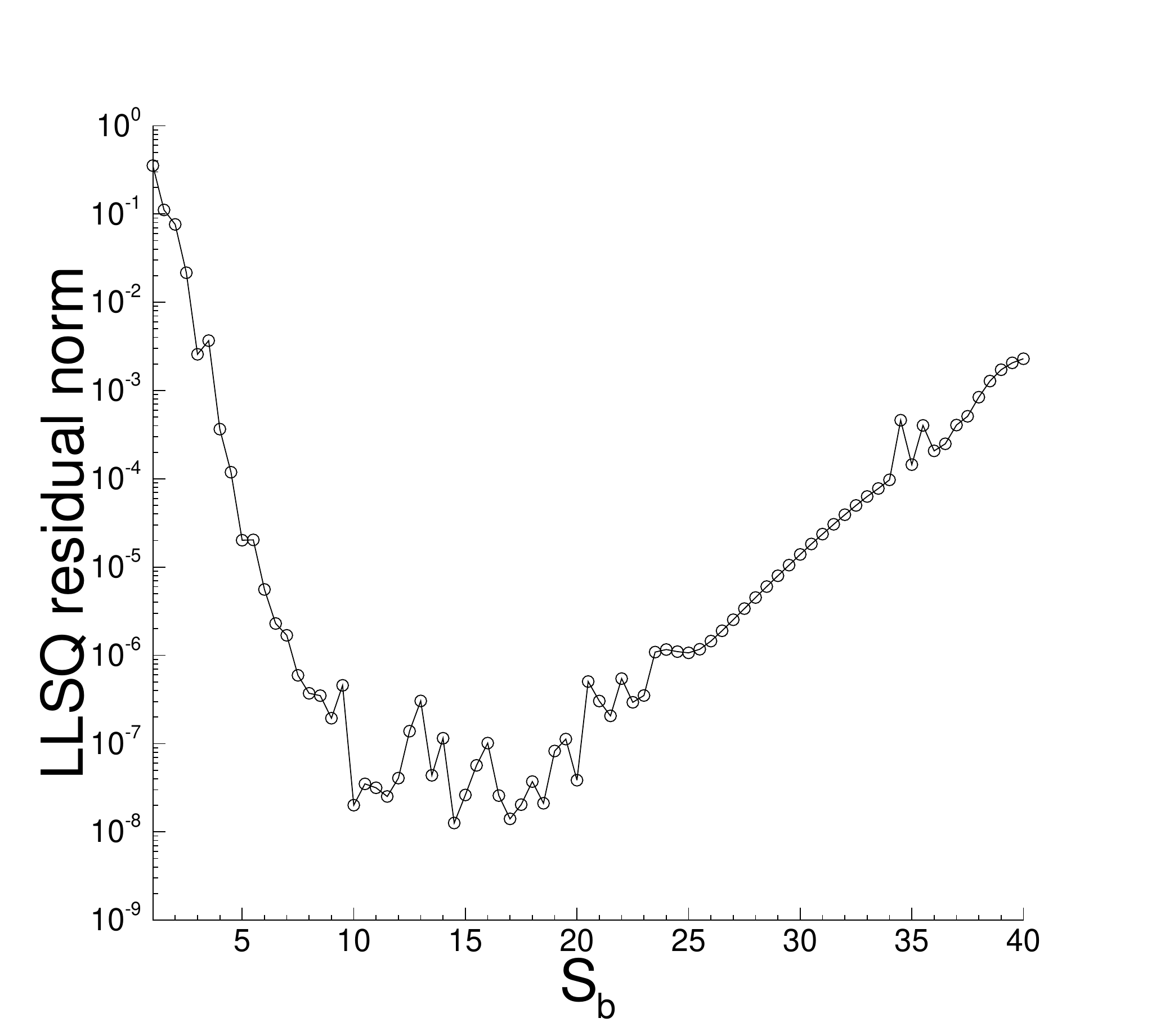}(b)
    \includegraphics[width=2in]{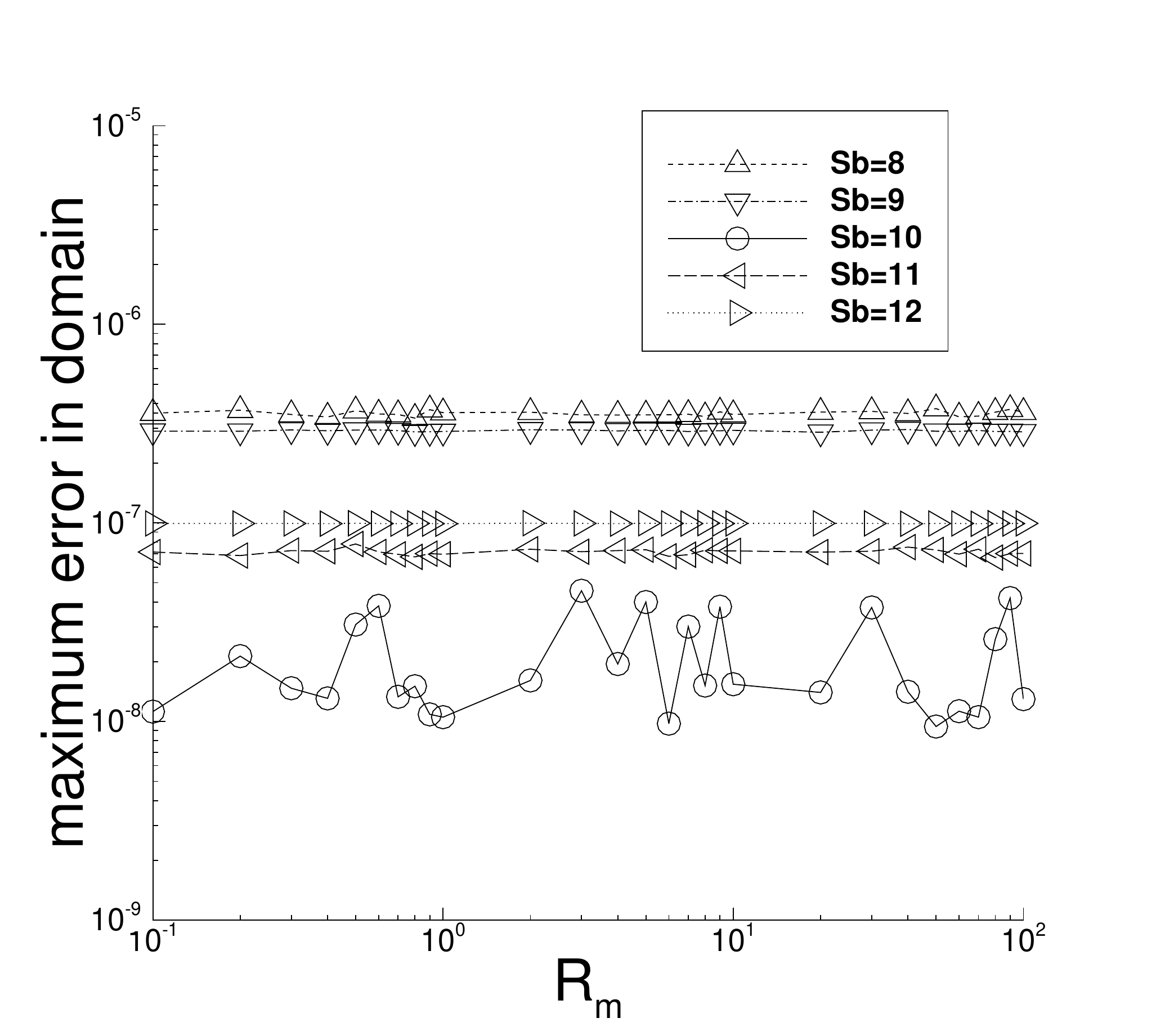}(c)
  }
  \caption{Function approximation: (a) Distribution of the exact function.
    (b) Residual norm of the linear least squares (LLSQ) problem versus $S_b$,
    for estimating the best $S_b$ in modBIP.
    (c) The maximum error of the ELM/modBIP solution as a function
    of $R_m$, corresponding to several $S_b$ in modBIP.
  }
  \label{fg_3}
\end{figure}

We approximate the following function $u(x)$ by the combined ELM/modBIP method,
\begin{equation}\label{eq_13}
  u = e^{\sin(2\pi x)} + x\cos(\pi x), \quad x\in \left[0, 2.5 \right].
\end{equation}
Figure \ref{fg_3}(a) shows the distribution of this function on
the domain. Note that the function approximation problem is equivalent
to solving the linear PDE \eqref{eq_6a}, with no boundary condition,
in which $\mbs L$ is given by
the identity operator and $f$ is given by the function to be approximated.

Let us first consider a single hidden layer in the neural network,
with the network architecture given by $[1, 100, 1]$ and $\tanh$ as the activation
function in the hidden layer (the output layer is linear).
The input node represents $x$, and the output node represents the function $u(x)$.
The input data to the neural network consists of $Q=121$ uniform grid (collocation) points
on $[0,2.5]$; see equation \eqref{eq_8} when restricted to 1D.
The function values on these collocation points are provided in
equation \eqref{eq_6a} as the data for the source term.
The hidden-layer coefficients are initialized by uniform random values
generated on $[-R_m,R_m]$, with $R_m$ specified below.


We first estimate the $S_b$ in modBIP using
the procedure from Remark \ref{rem_1}. 
Figure \ref{fg_3}(b) shows the residual norm of the linear least squares (LLSQ)
problem as a function of $S_b$ in a set of preliminary simulations.
Here the initial random coefficients in the neural network are generated
with $R_m=50$. They are pre-trained by modBIP,
with $S_c=S_b/2$ and $S_b$ from a range of values.
The residual norm of the LLSQ problem in ELM is collected corresponding
to these $S_b$ values and plotted in Figure \ref{fg_3}(b).
This plot indicates that,
while the residual norm at times fluctuates with
respect to $S_b$, it achieves a relatively low level for
a range of $S_b\approx 10 \sim 20$ for this problem.
We employ $S_b=10$ in modBIP in the majority of subsequent tests for this problem.

Figure \ref{fg_3}(c) illustrates the general behavior of the ELM
approximation error, with the random coefficients pre-trained by modBIP.
It shows the maximum error in the domain of the ELM approximant as
a function of $R_m$,
the maximum magnitude of the initial random coefficients,
corresponding to several $S_b$ values around $S_b=10$ in the modBIP algorithm.
Here for a given $S_b$ value, we vary $R_m$ systematically in the
range $0.1\leqslant R_m\leqslant 100$, and for each $R_m$ we initialize
the hidden-layer coefficients by uniform random values
generated on $[-R_m,R_m]$ and pre-train the random coefficients by modBIP
with the given $S_b$ and $S_c=S_b/2$. The pre-trained random coefficients are then
used in ELM to compute the training parameters (i.e.~the output-layer coefficients)
by the linear least squares method
for approximating the function \eqref{eq_13}.
So the approximation function is now represented by the fully trained
neural network.
We then evaluate the trained neural network  on a set of $401$ (finer)
uniform grid points to compute the approximant values, which are then
compared with the exact function \eqref{eq_13} to attain the errors.
We can observe from Figure \ref{fg_3}(c) that, with the modBIP pre-training of
the random coefficients, the ELM error is essentially independent of $R_m$,
although some fluctuations with respect to $R_m$ can be
observed in certain cases.
This insensitivity to $R_m$ is a common
characteristic of the combined ELM/modBIP method,
which will be observed repeatedly in subsequent numerical experiments.

\begin{figure}
  \centerline{
    \includegraphics[width=2.5in]{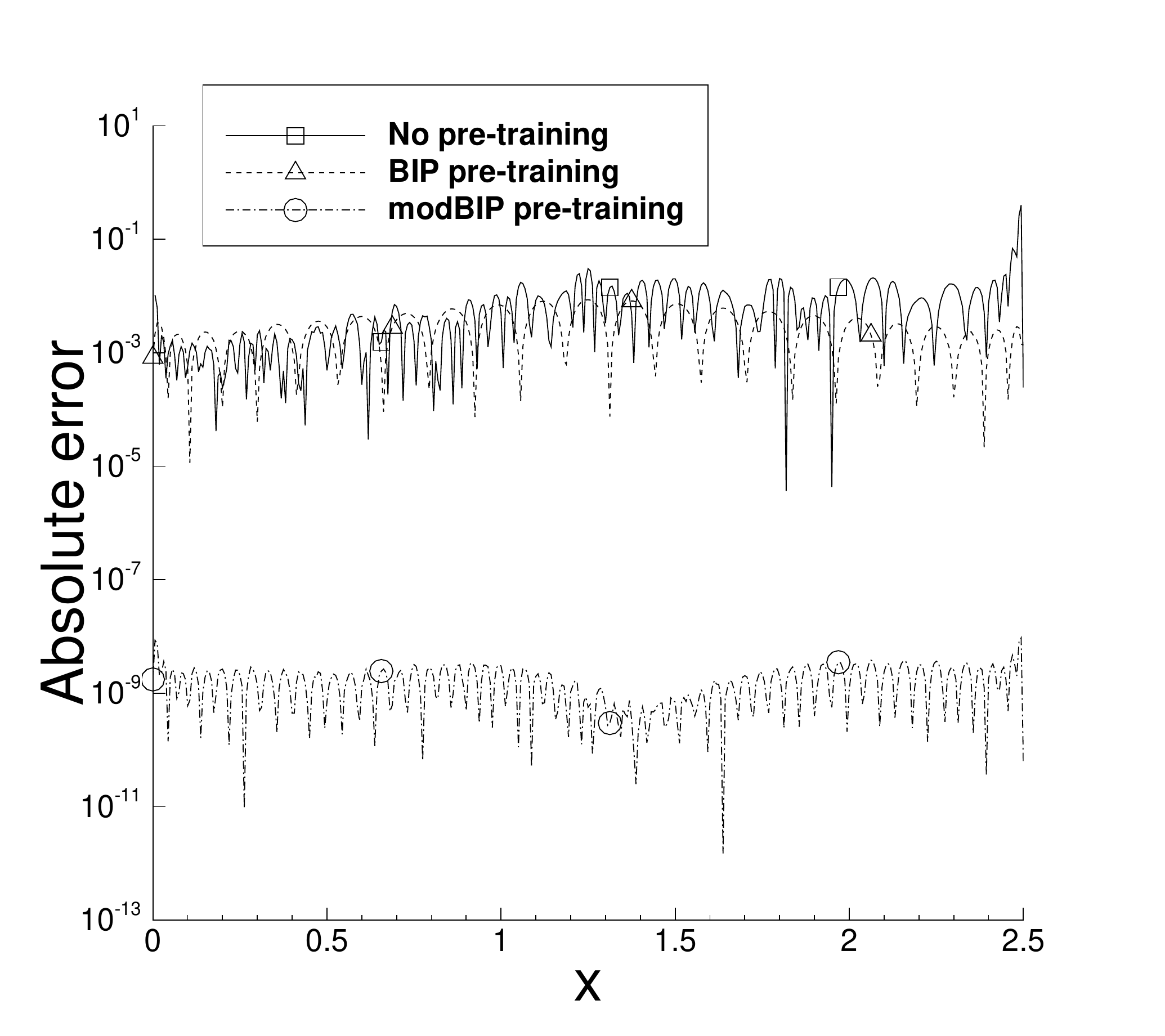}
  }
  \caption{Function approximation: Distributions of the absolute error of
    the ELM approximant obtained with (a) no pre-training, (b) BIP pre-training,
    and (c) modBIP pre-training of the random coefficients.
  }
  \label{fg_4}
\end{figure}

\begin{figure}
  \centerline{
    \includegraphics[width=2in]{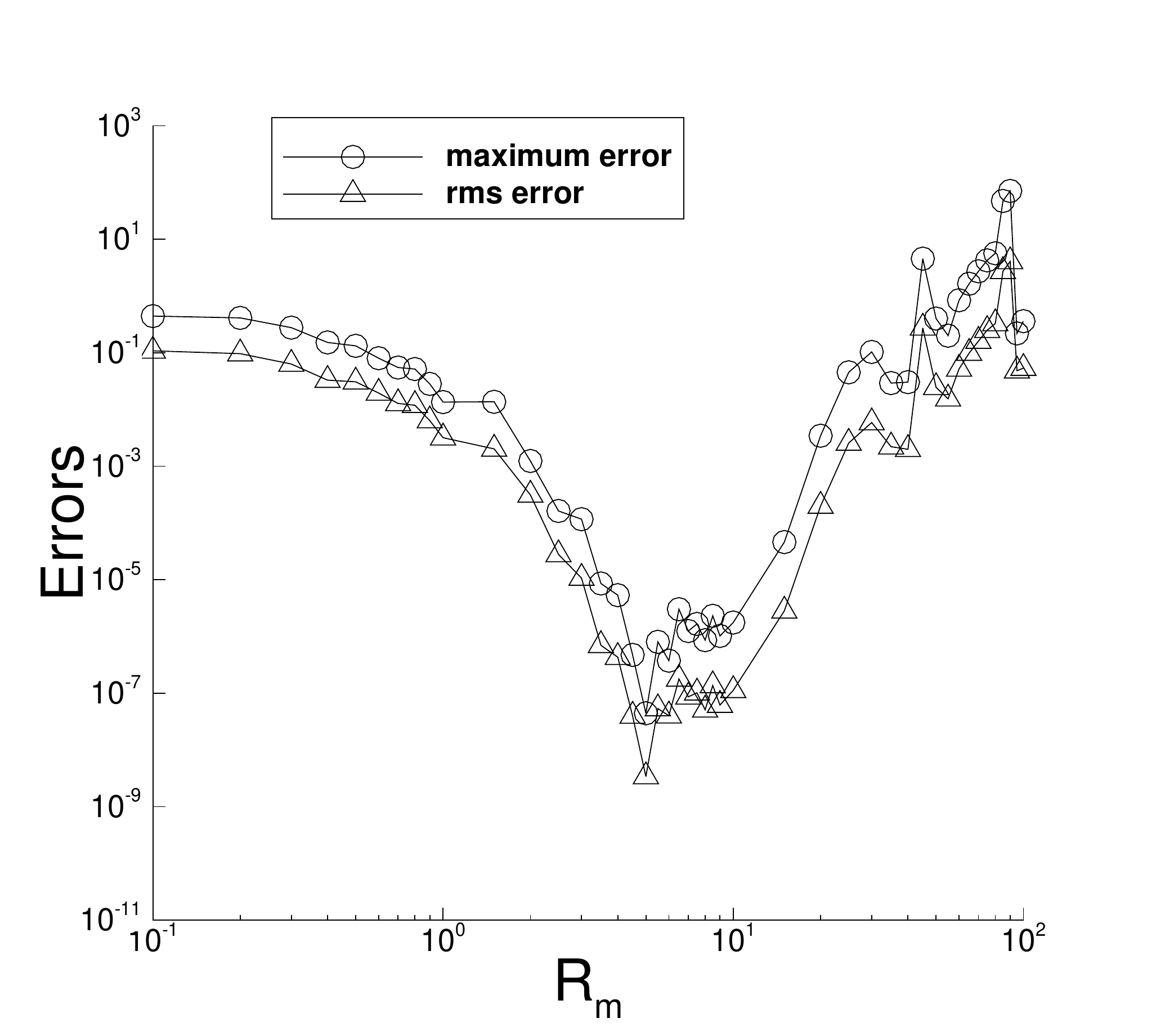}(a)
    \includegraphics[width=2in]{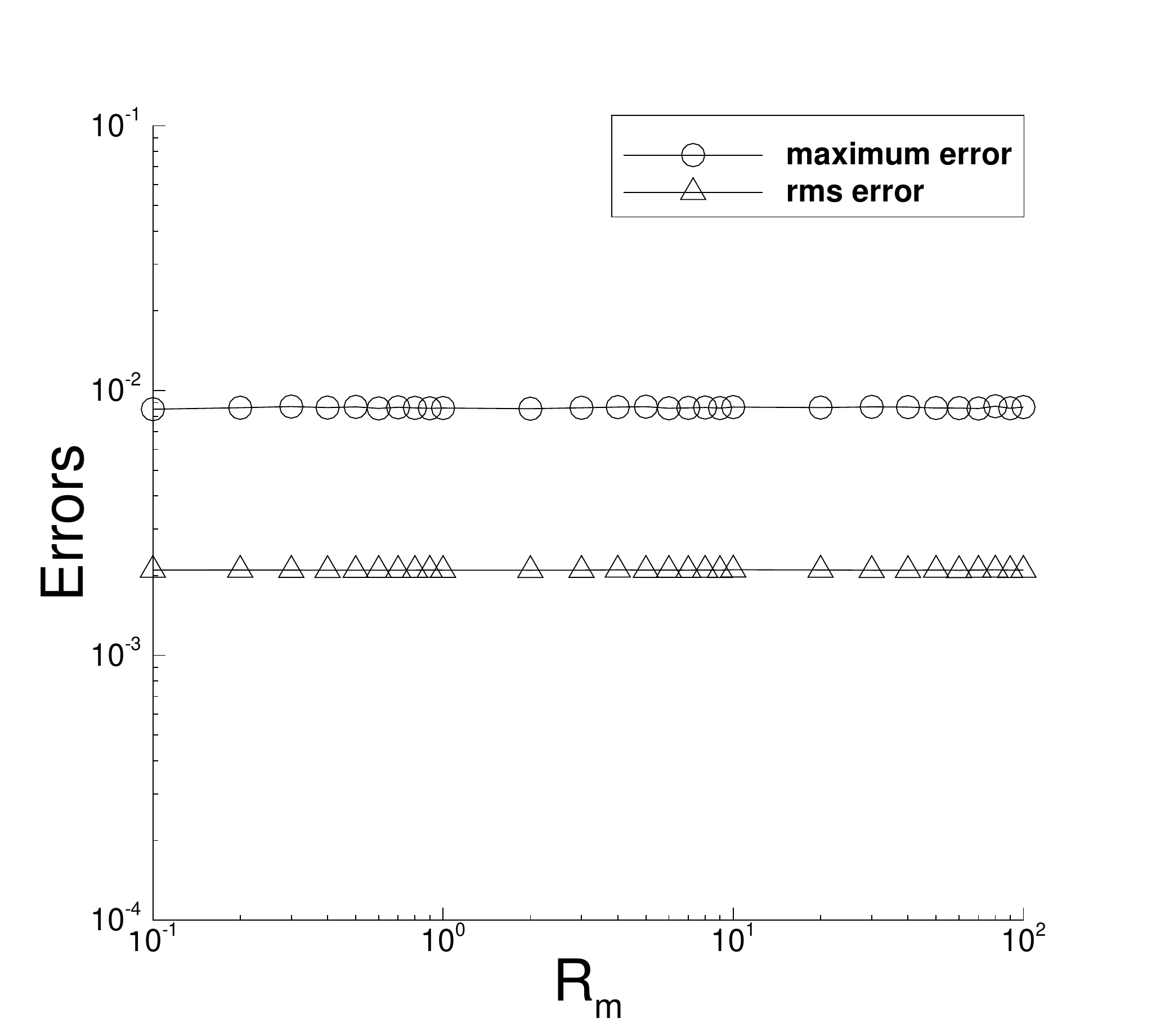}(b)
    \includegraphics[width=2in]{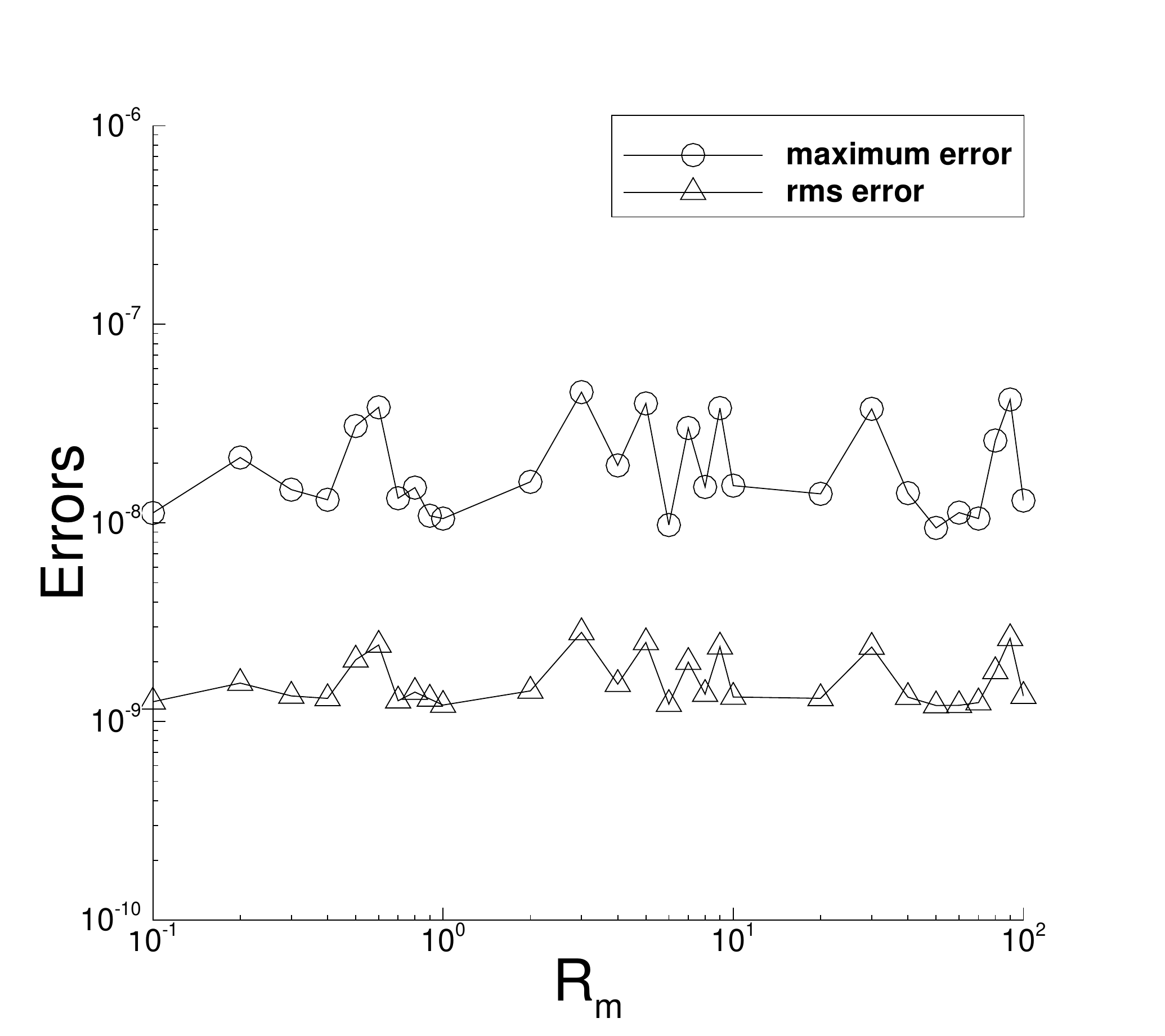}(c)
  }
  \caption{Function approximation: The maximum and rms errors in the domain
    of the ELM approximant as a function of $R_m$,
    attained with (a) no pre-training,
    (b) BIP pre-training, and (c) modBIP pre-training of the random coefficients.
  }
  \label{fg_5}
\end{figure}

Figures \ref{fg_4} and \ref{fg_5} are comparisons of the ELM errors
of the function approximation problem obtained with three configurations:
no pre-training of the random coefficients, and with pre-training of
the random coefficients by the BIP algorithm~\cite{NeumannS2013}
and by the current modBIP algorithm.
As mentioned before, the network architecture is
characterized by $[1, 100, 1]$, with the $\tanh$ activation function and
$Q=121$ uniform collocation points as the training data points.
In this set of tests the initial random coefficients  are
generated with $R_m$ either fixed at $R_m=50$ or varied systematically.
In the case without pre-training,
the initial random coefficients are directly used in ELM for computing
the training parameters and the approximation function.
In the cases with pre-training, the initial random coefficients are pre-trained
by BIP or modBIP first, and the pre-trained hidden-layer coefficients are then
used in ELM for computing the approximation function.
With BIP, we employ a normal distribution
for generating the target samples for each hidden-layer node,
with a random mean from $[-1,1]$
and a standard deviation $0.5$, as described in \cite{NeumannS2013}.
The inverse of $\tanh$ is then applied to the target samples,
which are then used to compute the mapping coefficients in
BIP~\cite{NeumannS2013}.
With modBIP pre-training, we employ $S_c=S_b/2$ and $S_b=10$ in
Algorithm~\ref{alg_1}.


Figure \ref{fg_4} compares profiles of the absolute error of
the ELM approximant obtained without pre-training,
with BIP pre-training, and with modBIP pre-training of the random
coefficients. The initial random coefficients are generated with $R_m=50$
in this set of tests. The error levels of the ELM result with BIP pre-training
and without pre-training are largely comparable, both on the order of $10^{-3}$.
In contrast, the error level of the ELM result with the modBIP pre-training is
considerably lower, on the order of $10^{-9}$.
This indicates that the combined ELM/modBIP method is markedly more
accurate than the ELM methods without pre-training and
with the BIP pre-training of the random coefficients.

Figure \ref{fg_5} shows the maximum and root-mean-squares (rms) errors
in the domain of the ELM approximants obtained without pre-training and
with BIP and modBIP pre-training of the random coefficients.
In this set of tests, $R_m$ is varied systematically between $0.1$ and $100$,
and the maximum/rms errors of the ELM solution
corresponding to the initial random coefficients generated on $[-R_m,R_m]$,
with and without pre-training, are computed and collected.
Without pre-training of the random coefficients, the ELM accuracy 
exhibits a strong dependence on $R_m$. It produces quite accurate
results in a range of moderate $R_m$ values, while outside this range
the accuracy can be quite poor; see Figure \ref{fg_5}(a).
With BIP and modBIP pre-training of the random coefficients,
the error of the ELM result is observed to be
largely independent of $R_m$. The ELM error level corresponding to
the modBIP pre-training
is considerably smaller than that of the BIP pre-training
(see Figures \ref{fg_5}(b,c)).


\begin{figure}
  \centerline{
    \includegraphics[width=2in]{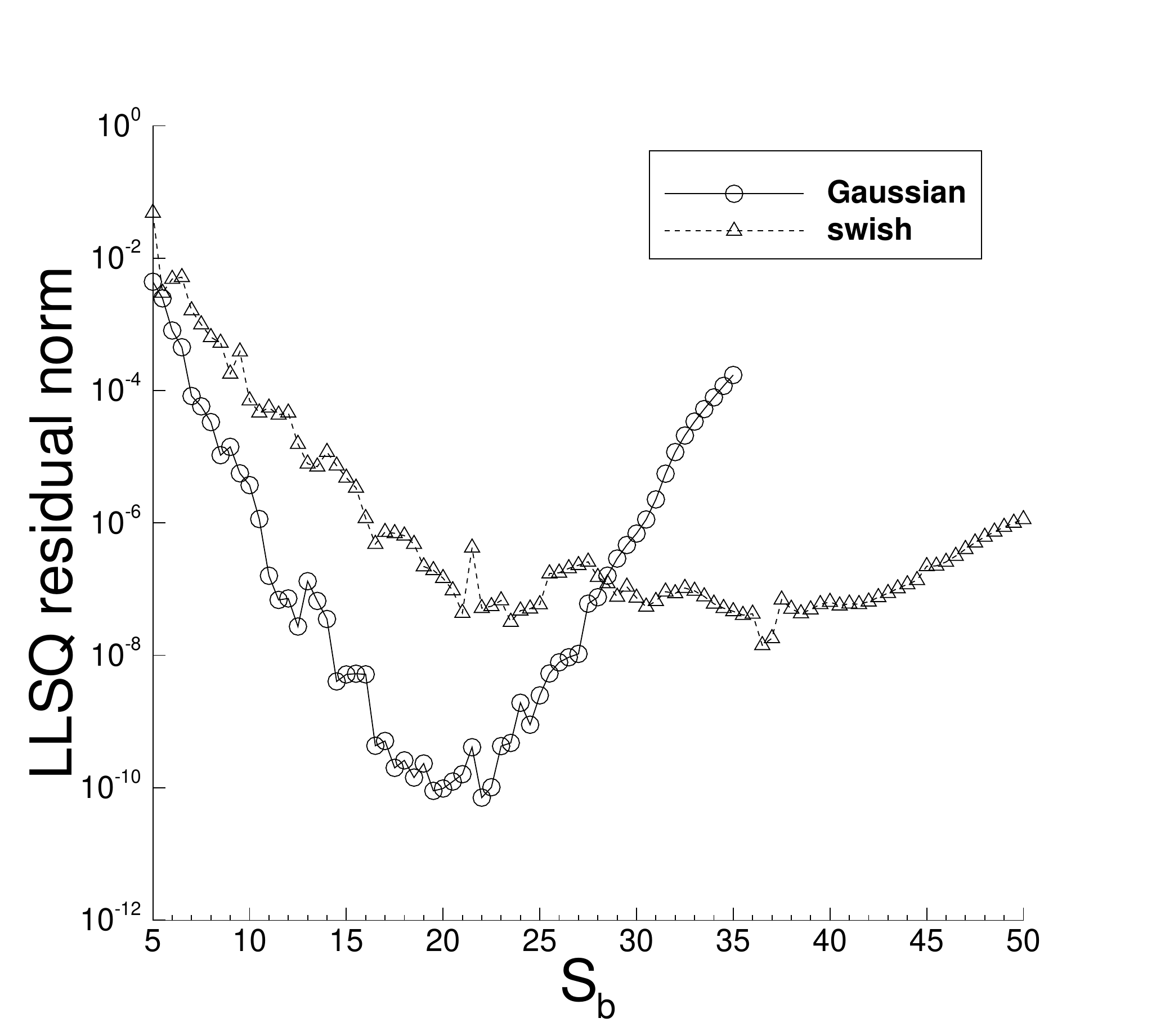}(a)
    \includegraphics[width=2in]{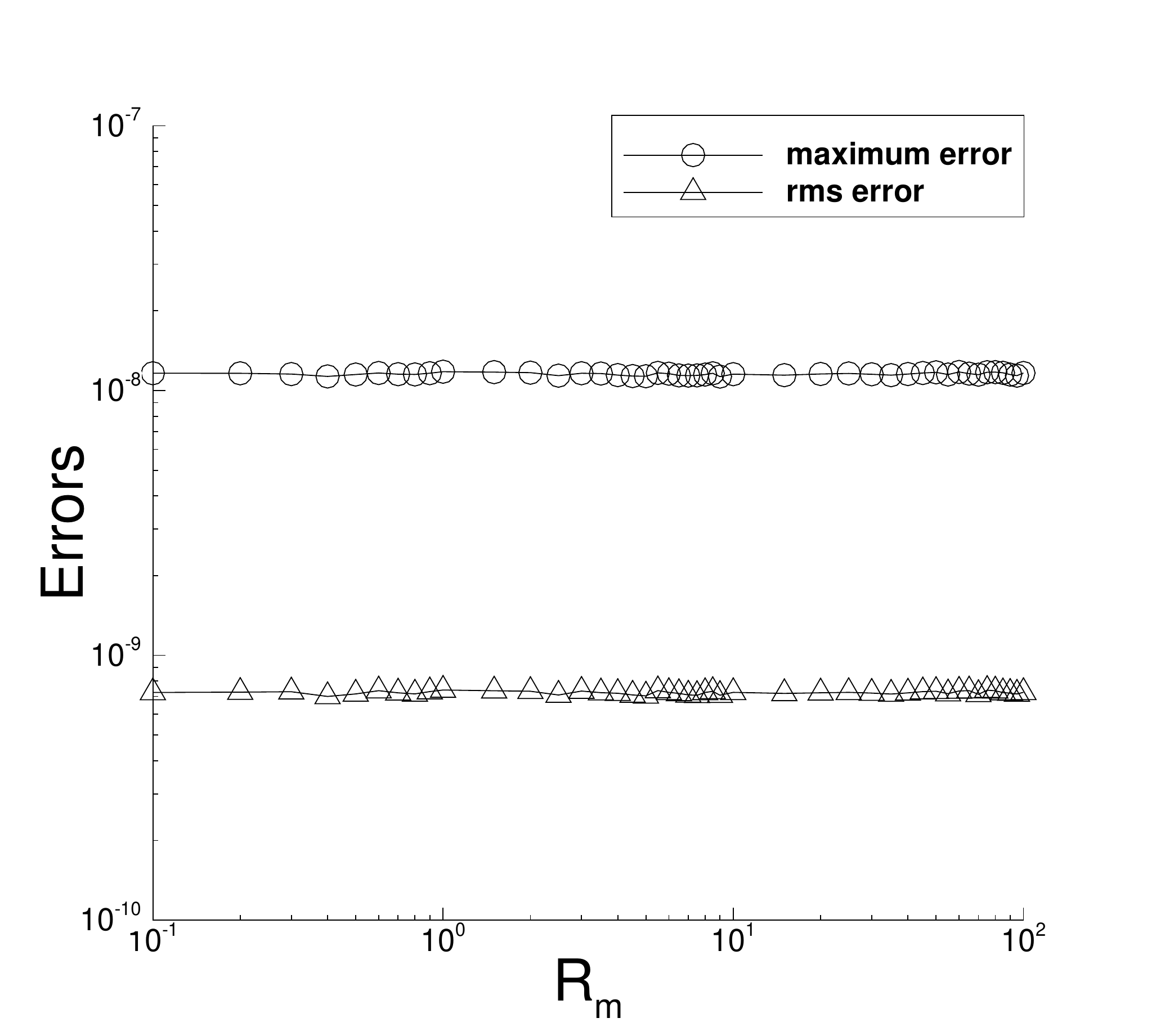}(b)
    \includegraphics[width=2in]{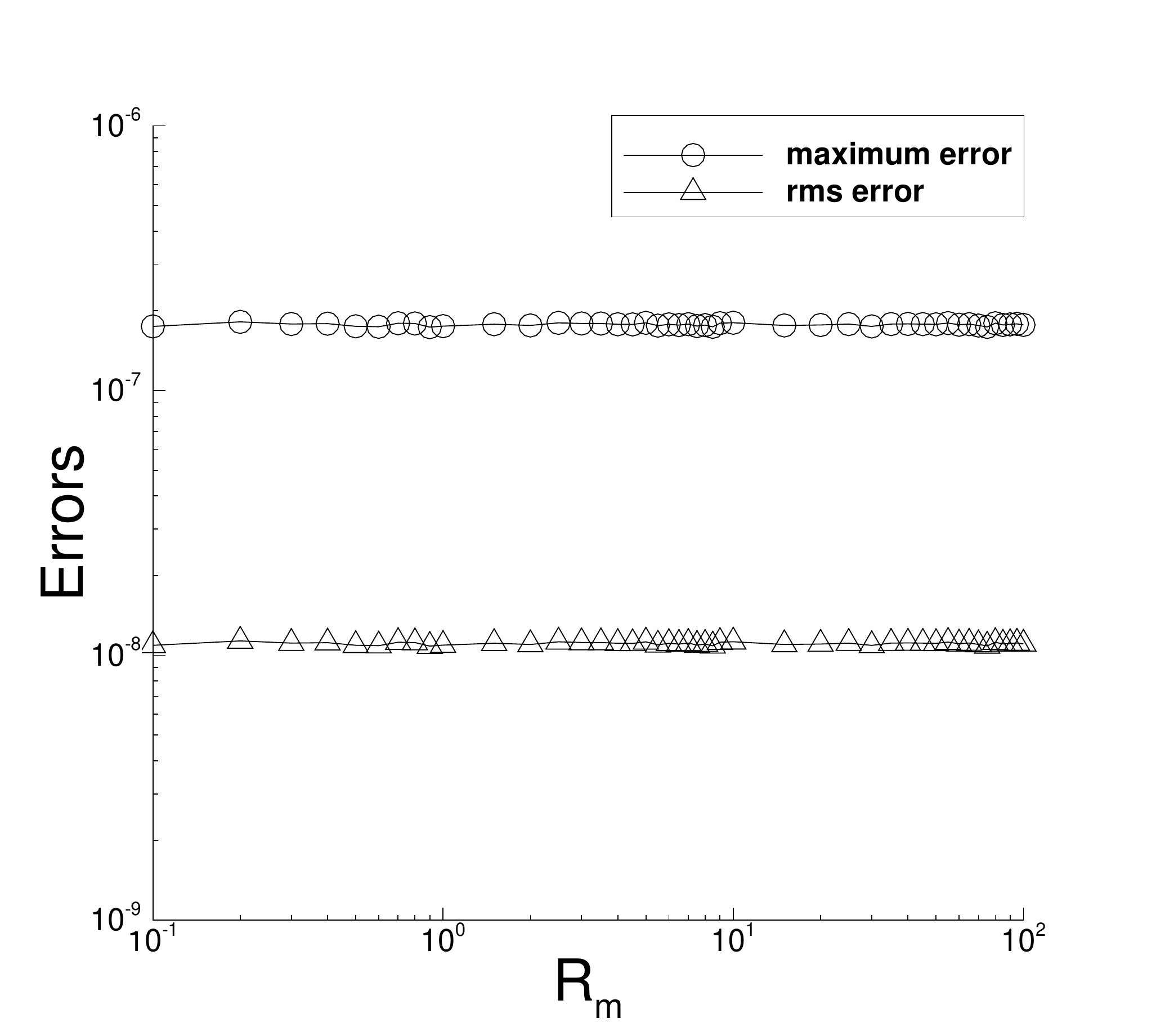}(c)
  }
  \caption{Function approximation (non-monotonic activation functions):
    (a) Residual norm of the LLSQ problem versus $S_b$, for estimating the best
    $S_b$ in modBIP.
    The maximum and rms errors of the ELM solution as a function of $R_m$
    obtained using (b) the Gaussian and (c) the swish activation functions,
    with the random coefficients pre-trained by modBIP.
  }
  \label{fg_6}
\end{figure}

A prominent advantage of modBIP over BIP lies in that modBIP does not
place any constraint on
the activation function, while BIP requires the activation function to be
invertible. So modBIP can be applied  with
many activation functions with which BIP breaks down.
Two such examples are provided in Figure \ref{fg_6},
with the Gaussian and the swish~\cite{ElfwingUD2018} activation functions.
Neither of these two functions has an inverse.
Here the neural network has the same architecture as before, but the activation function
for the hidden layer has been changed to the Gaussian function ($\sigma(x)=e^{-x^2}$)
or the swish function ($\sigma(x)=x/(1+e^{-x})$). The initial random
coefficients are generated with a fixed $R_m=50$ (plot (a))
or a varying $R_m$ (plots (b,c)),
and pre-trained by modBIP.
We employ the same training data points as before ($Q=121$), and $S_c=S_b/2$ in modBIP.
Figure \ref{fg_6}(a) shows the LLSQ residual norms for estimating the best $S_b$, suggesting
a value around $S_b\approx 20$ with the Gaussian function and around $S_b\approx 24$
with the swish function.
Figures \ref{fg_6}(b,c) show the maximum and rms errors in the domain of the ELM/modBIP
approximant as a function of $R_m$, corresponding to the Gaussian activation
function (with $S_b=20$) and to the swish activation function (with $S_b=24$).
The ELM/modBIP results exhibit a high accuracy (error level around $10^{-10}\sim 10^{-7}$),
which is insensitive to $R_m$ (or the initial random coefficients).
It should be noted that the BIP algorithm breaks down when used 
with these activation functions, because they do not have an inverse.

\begin{figure}
  \centerline{
    \includegraphics[width=1.8in]{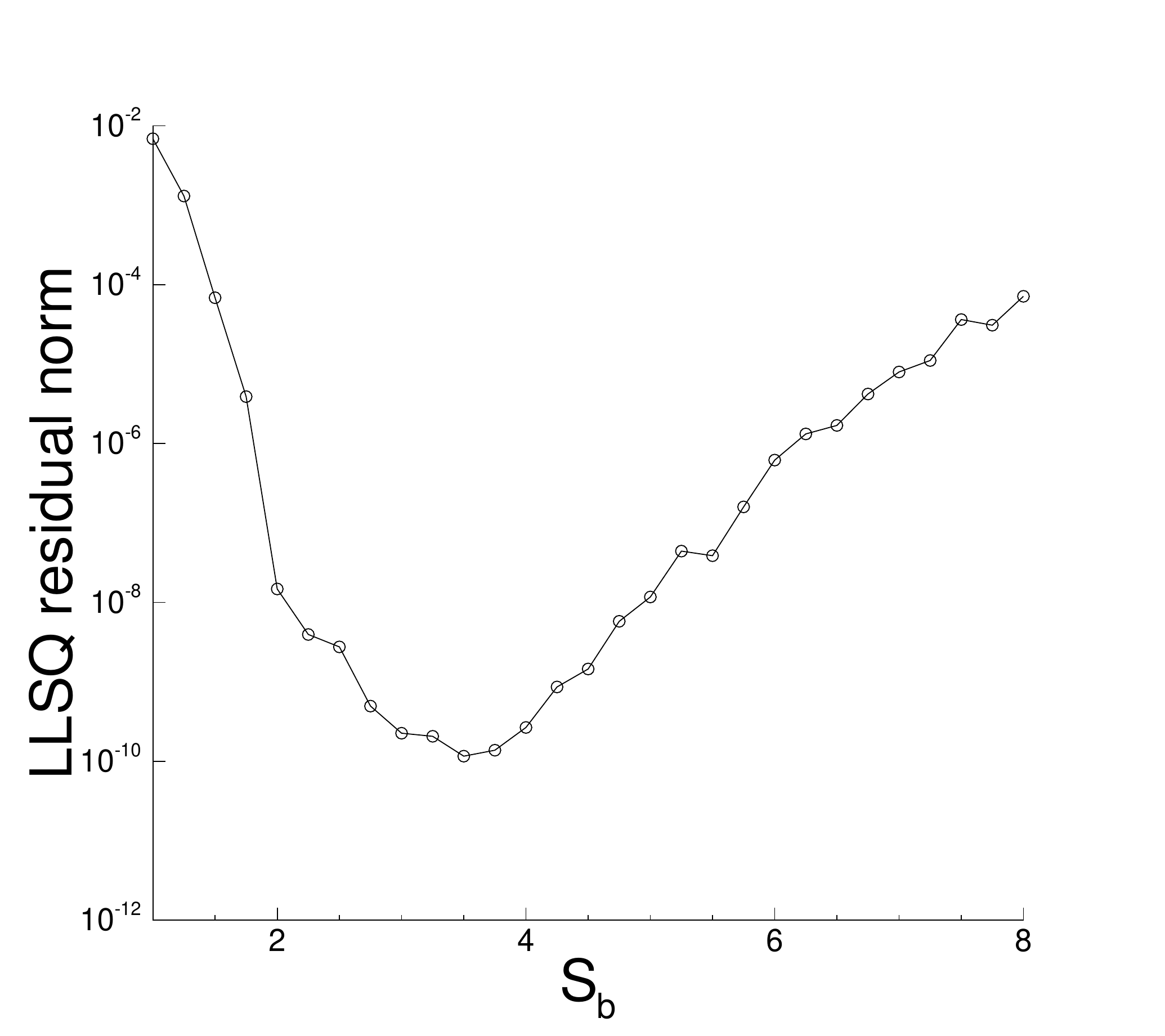}(a)
    \includegraphics[width=1.8in]{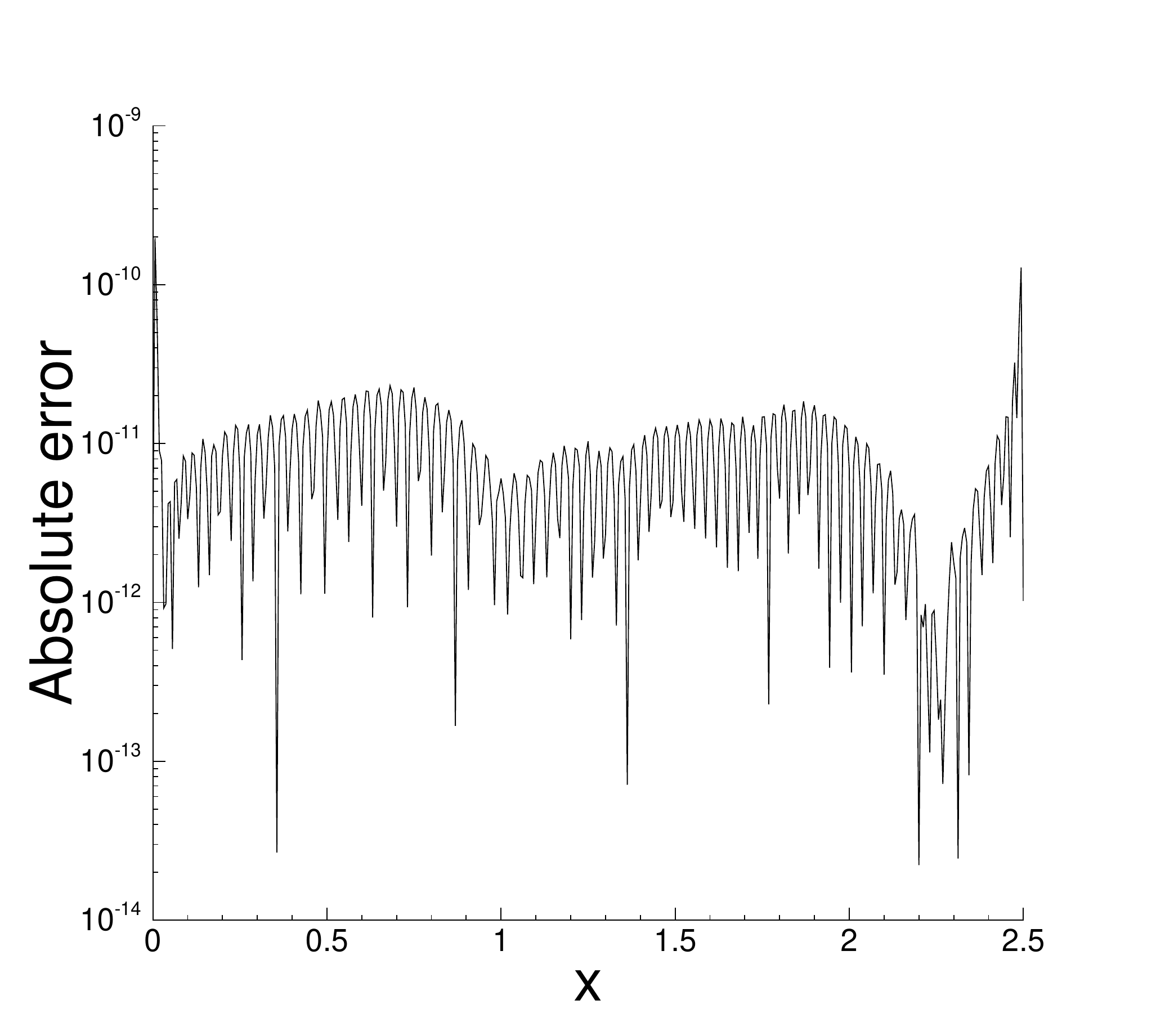}(b)
    \includegraphics[width=1.8in]{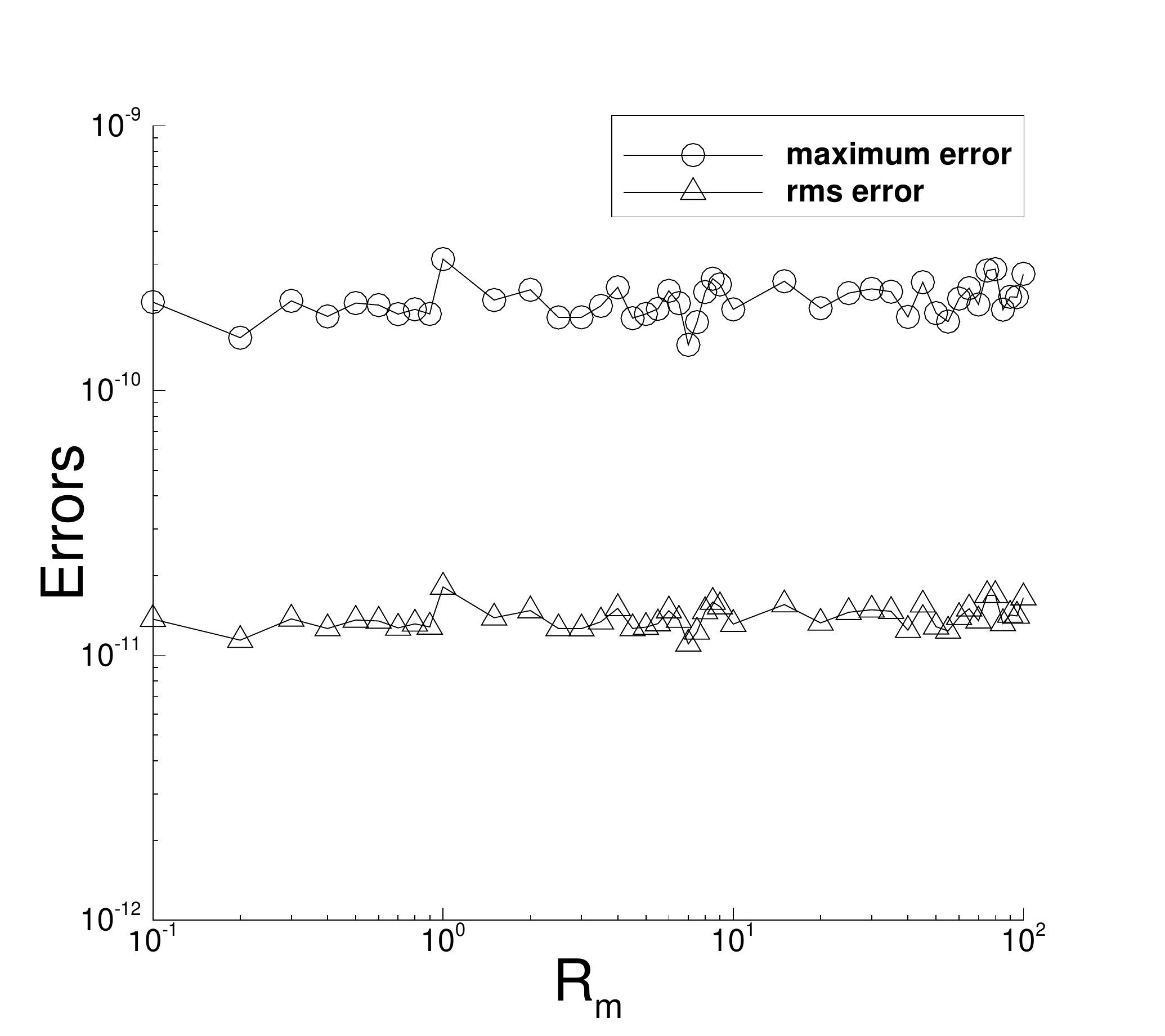}(c)
  }
  \centerline{
    \includegraphics[width=1.8in]{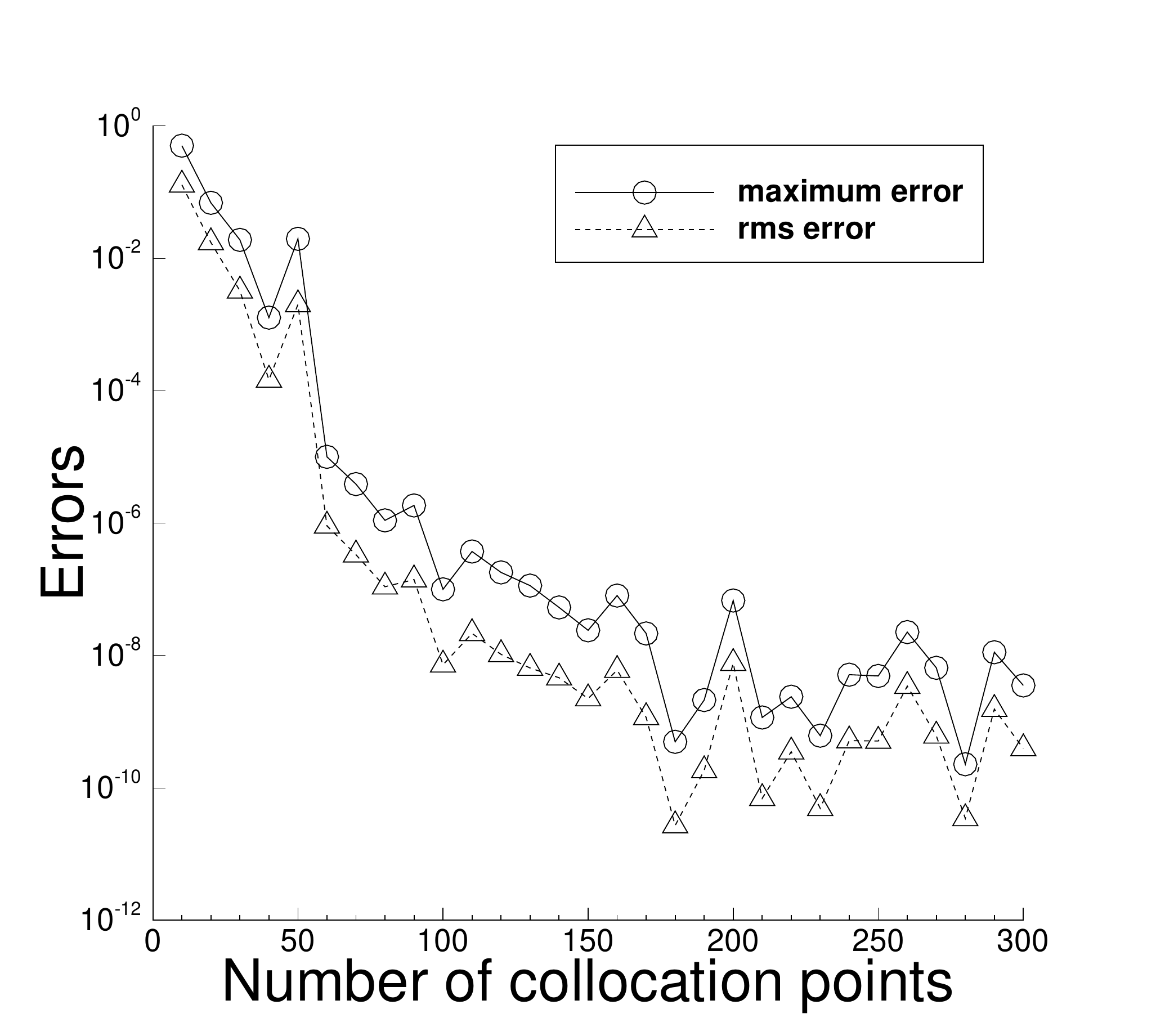}(d)
    \includegraphics[width=1.8in]{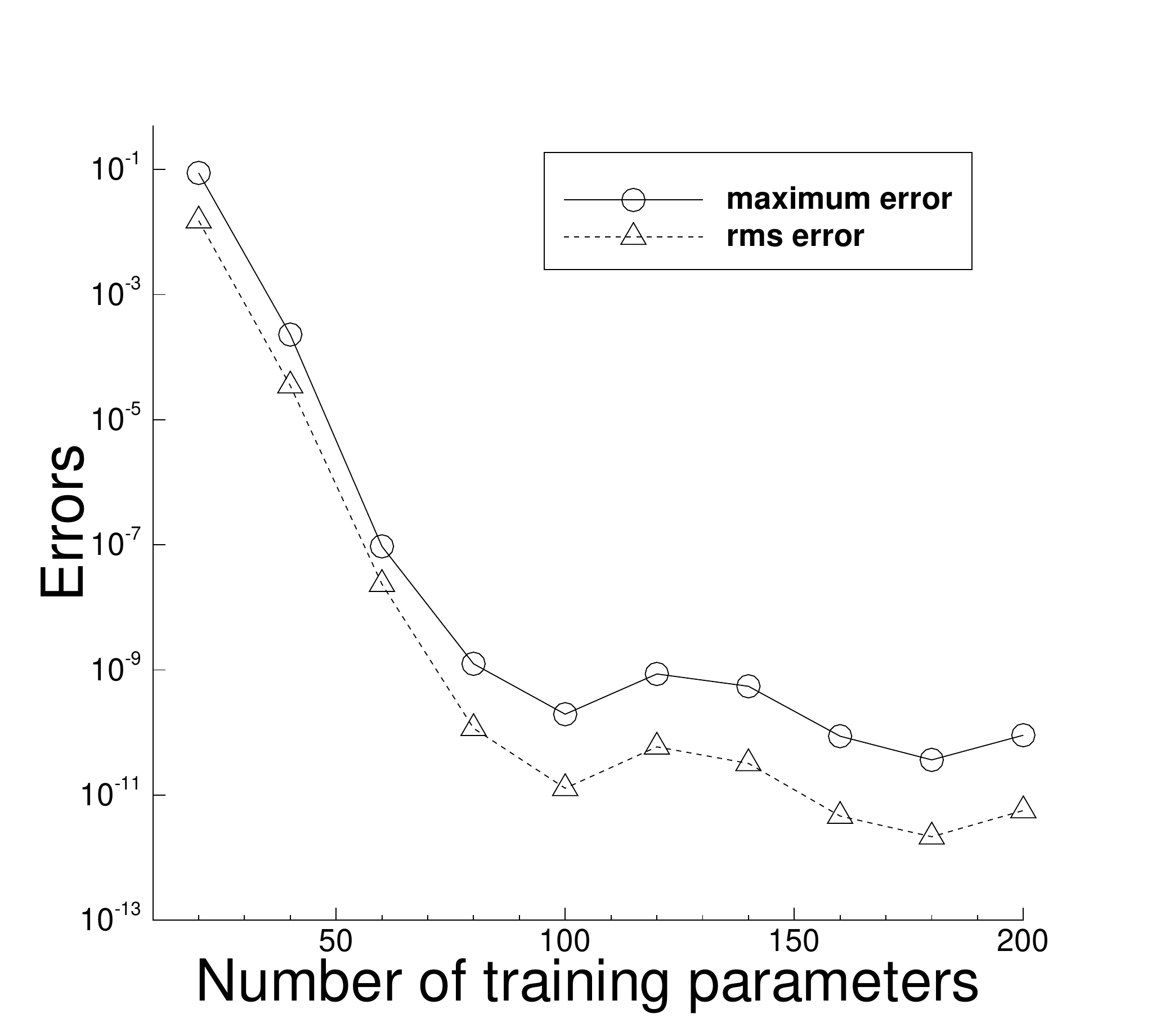}(e)
  }
  \caption{Function approximation (3 hidden layers in neural network):
    (a) The LLSQ residual norm versus $S_b$, for estimating
    the best $S_b$ in modBIP.
    (b) Distribution of the absolute error of the ELM/modBIP approximant.
    The maximum/rms errors of the ELM/modBIP approximant as a function of
    (c) $R_m$, (d) the number of collocation points $Q$, and (e) the number of
    training parameters $M$.
    $Q$ is fixed at $Q=150$ in (a,b,c,e) and varied in (d).
    $M$ is fixed at $M=100$ in (a,b,c,d) and varied in (e).
    $R_m$ is fixed at $R_m=50$ in (a,b,d,e) and varied in (c).
    $S_b$ is varied in (a) and fixed at $S_b=3.5$ in (b,c,d,e).
  }
  \label{fg_7}
\end{figure}

\begin{figure}
  \centerline{
    \includegraphics[width=1.8in]{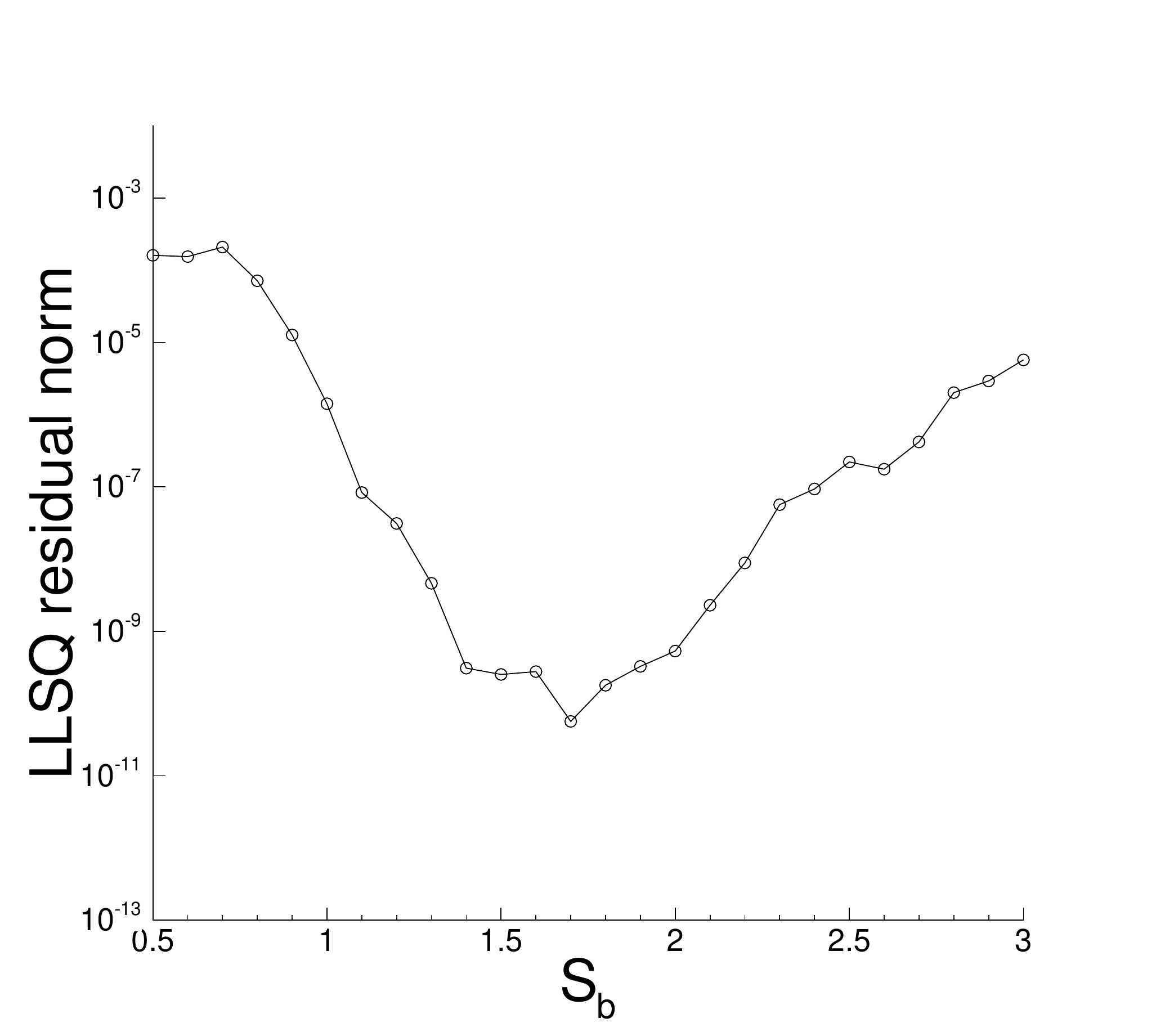}(a)
    \includegraphics[width=1.8in]{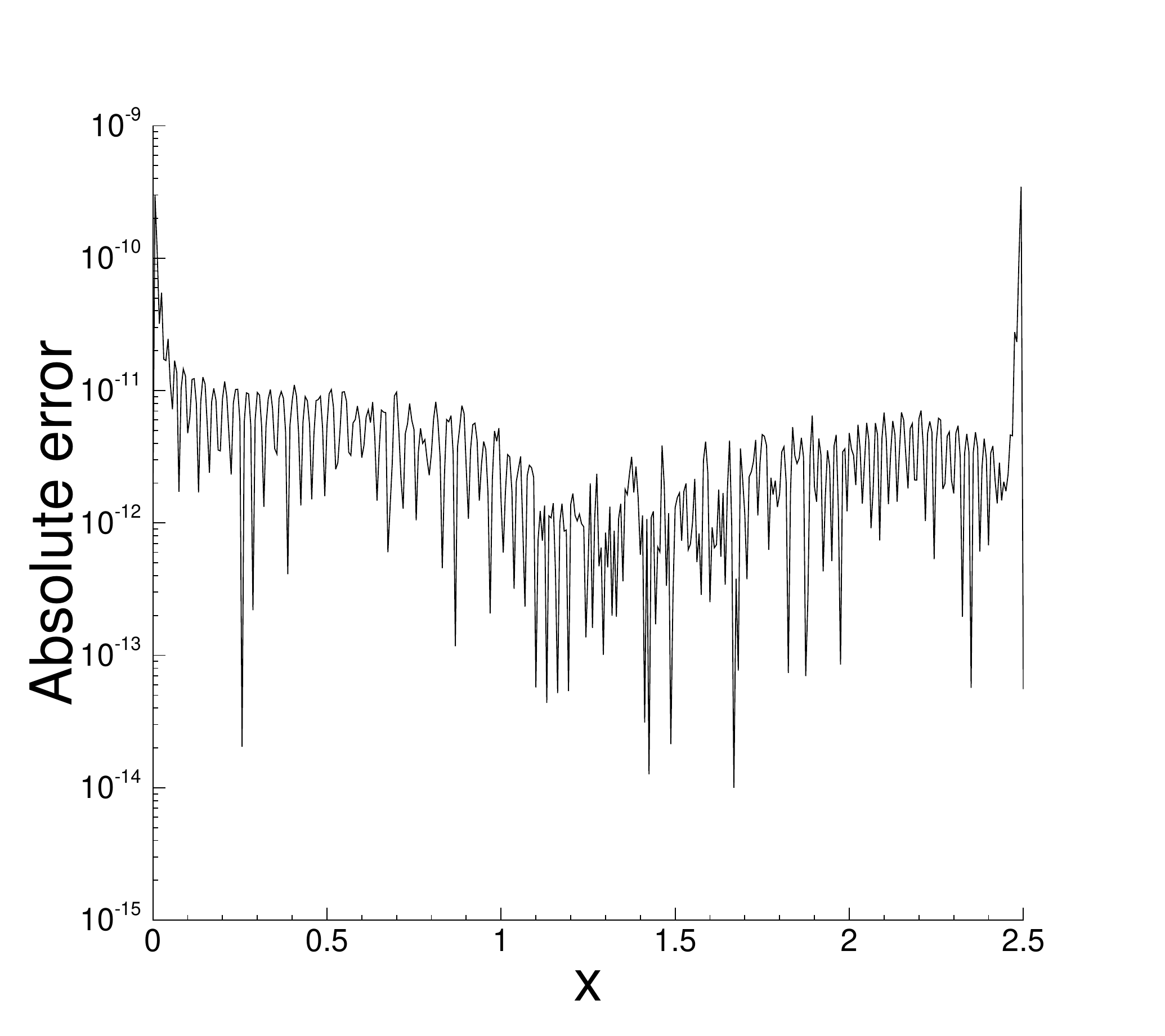}(b)
    \includegraphics[width=1.8in]{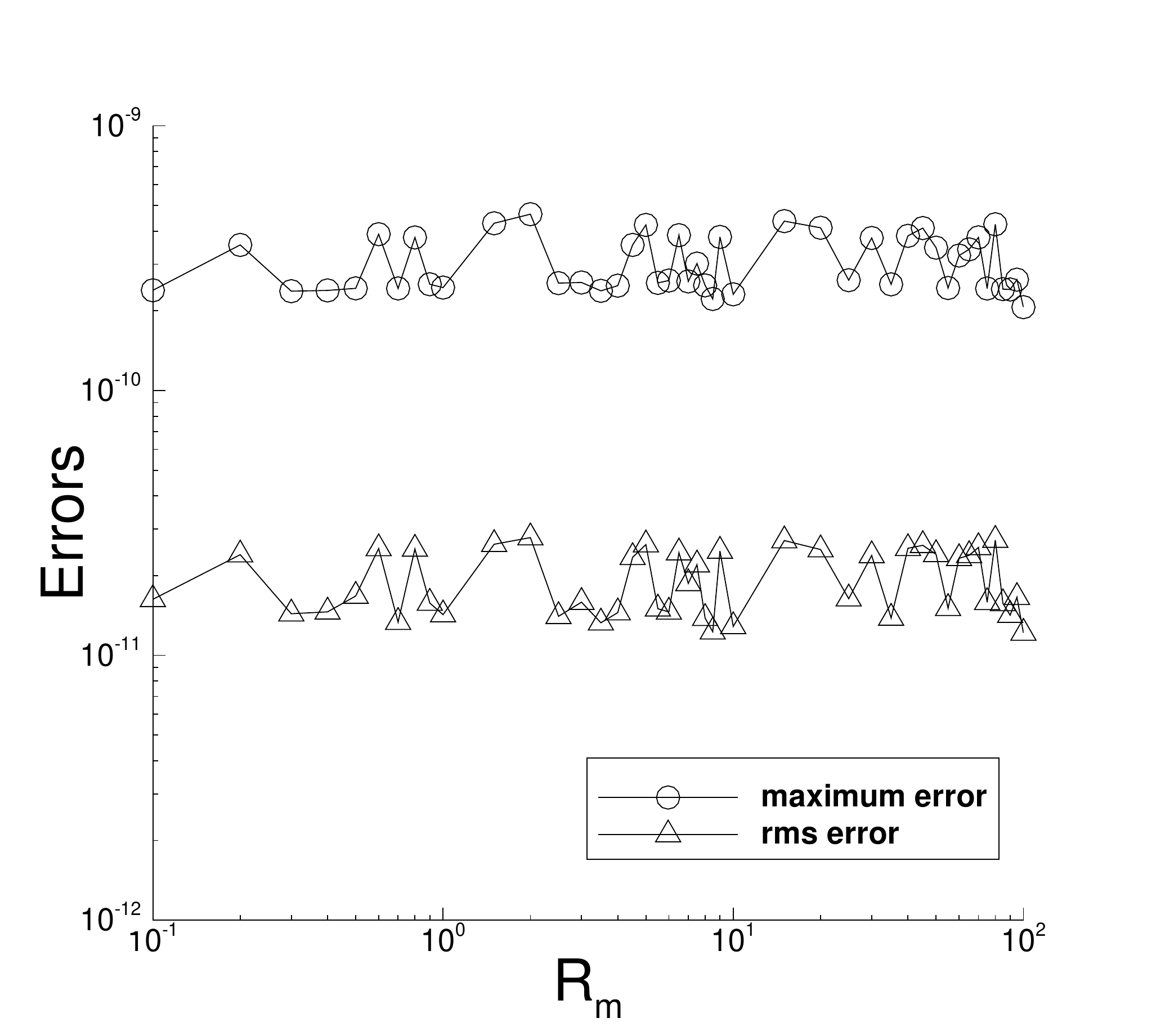}(c)
  }
  \centerline{
    \includegraphics[width=1.8in]{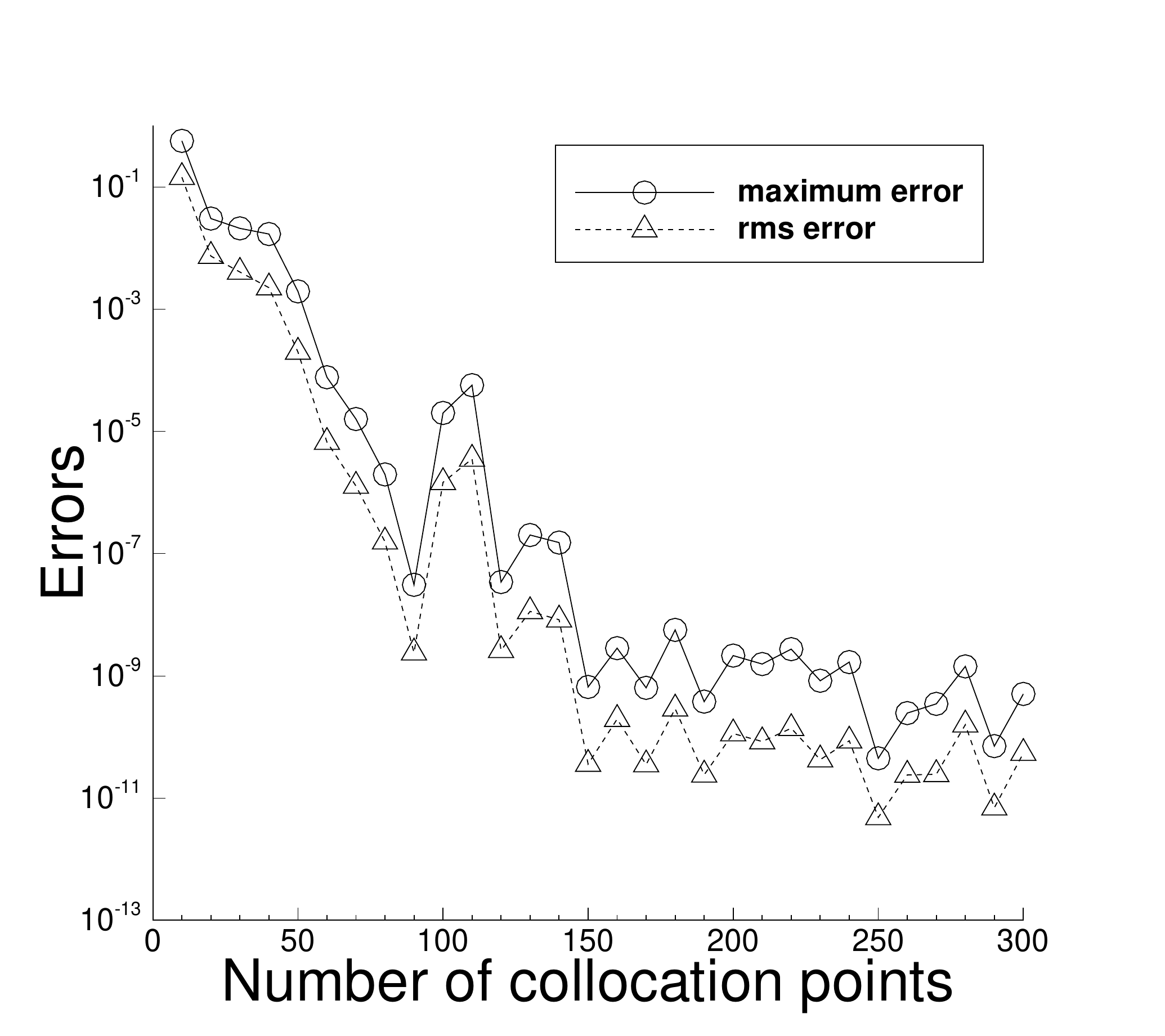}(d)
    \includegraphics[width=1.8in]{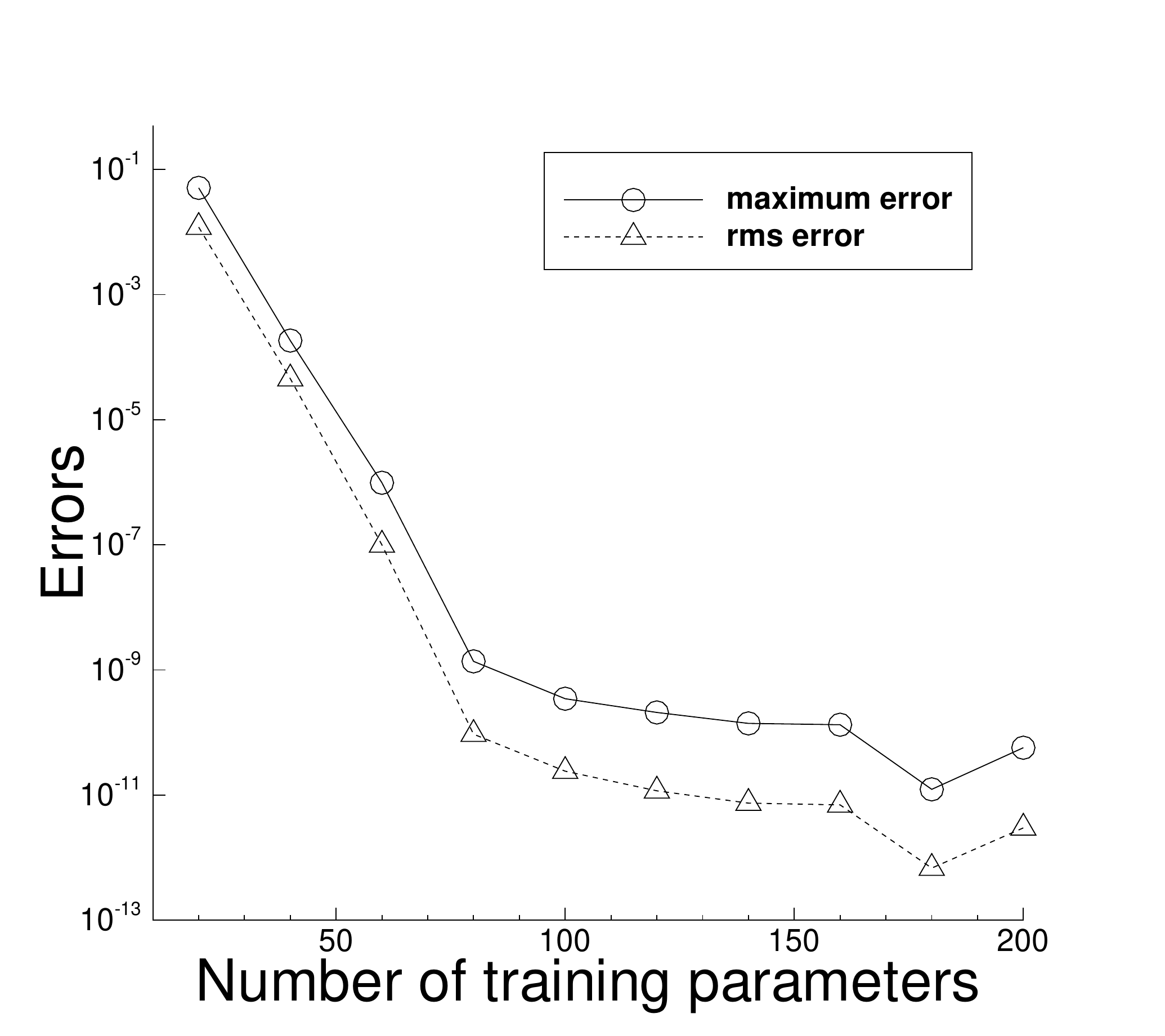}(e)
  }
  \caption{Function approximation (7 hidden layers in neural network):
    (a) The LLSQ residual norm versus $S_b$, for estimating
    the best $S_b$ in modBIP.
    (b) Distribution of the absolute error of the ELM/modBIP approximant.
    The maximum/rms errors of the ELM/modBIP approximant as a function of
    (c) $R_m$, (d) the number of collocation points $Q$, and (e) the number of
    training parameters $M$.
    $Q$ is fixed at $Q=150$ in (a,b,c,e) and varied in (d).
    $M$ is fixed at $M=100$ in (a,b,c,d) and varied in (e).
    $R_m$ is fixed at $R_m=50$ in (a,b,d,e) and varied in (c).
    $S_b$ is varied in (a) and fixed at $S_b=1.7$ in (b,c,d,e).
  }
  \label{fg_8}
\end{figure}

The results presented so far are obtained with a single hidden layer
in the neural network. Let us next investigate the performance
of the ELM/modBIP
method with neural networks containing multiple hidden layers.
Figures \ref{fg_7} and \ref{fg_8} display ELM/modBIP results obtained
with $3$ and $7$ hidden layers in the neural network, respectively.
The network architecture is characterized by the vectors
$[1, 40, 40, M, 1]$ and $[1, 40, 40, 40, 40, 40, 40, M, 1]$
in these two cases, respectively, where $M$ denotes the number of
training parameters and is either fixed at $M=100$ or varied between $M=20$
and $M=200$.
The activation function is $\tanh$ in all hidden layers,
and the output layer is linear.
We employ $Q$ uniform grid (collocation) points in the domain as
the training data points, with $Q$ either fixed at $Q=150$ or
varied between $Q=10$ and $Q=300$.
The initial hidden-layer coefficients  are
set to uniform random values generated on $[-R_m,R_m]$, with
$R_m$ either fixed at $R_m=50$ or varied between $R_m=0.1$ and $R_m=100$.
These initial random coefficients are pre-trained by modBIP
with $S_c=S_b/2$.

Figure \ref{fg_7} illustrates the ELM/modBIP results with three hidden layers in
the neural network. The plot (a) shows the LLSQ residual norms for estimating
the best $S_b$ in modBIP, suggesting a value around $S_b\approx 3.5$.
The plot (b) depicts the error distribution of the ELM/modBIP approximant
against the actual function \eqref{eq_13}.
The plots (c,d,e) show the maximum and rms errors in the domain of the ELM/modBIP
appximant as a function of $R_m$, the number of training collocation points $Q$,
and the number of training parameters $M$. The specific parameter values
employed for each plot are provided in the caption of Figure \ref{fg_7}.

Figure \ref{fg_8} shows the corresponding ELM/modBIP results obtained with seven
hidden layers in the neural network. The LLSQ residual norms in plot (a)
suggest a value around $S_b\approx 1.7$ for modBIP,
which has been employed to attain the error distribution and the maximum/rms errors
of the ELM/modBIP in the plots (b) to (e). The specific
parameter values for each case are provided in the caption of this figure.

The results in Figures \ref{fg_7} and \ref{fg_8} indicate that
the combined ELM/modBIP method produces highly accurate results with
multiple hidden layers in the neural network.
The best $S_b$ for modBIP (with $S_c=S_b/2$) appears to decrease with
increasing number of hidden layers in the neural network.
The ELM/modBIP errors are not sensitive to the initial random coefficients,
similar to what has been observed with a single hidden layer in the network.
These errors decrease approximately exponentially
as the number of collocation points or the
number of training parameters increases, until they essentially saturate 
when the number of collocation points or  training parameters
becomes sufficiently large.

\subsection{Poisson Equation}
\label{sec:poisson}

\begin{figure}
  \centerline{
    \includegraphics[width=2.in]{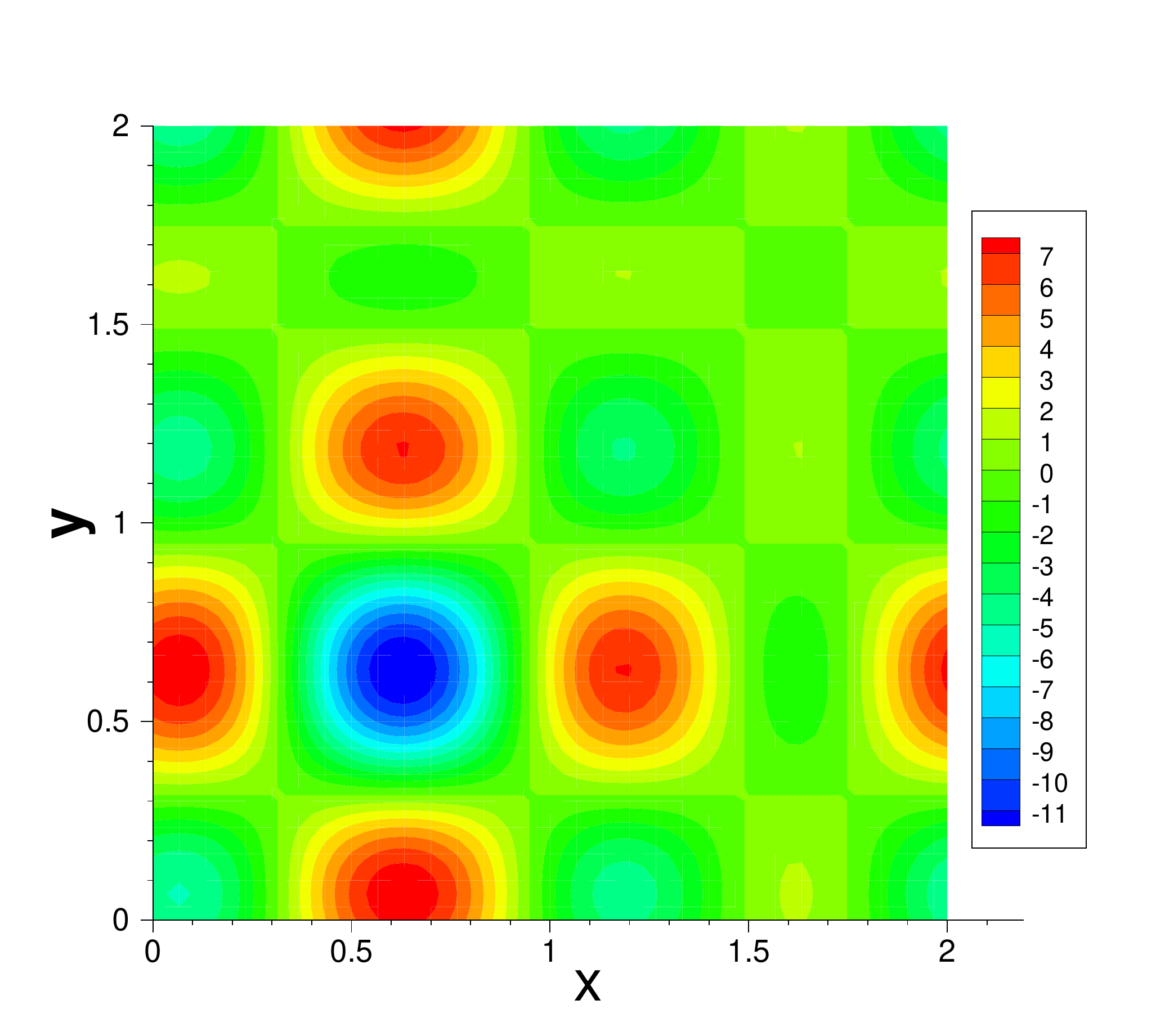}(a)
    \includegraphics[width=2.in]{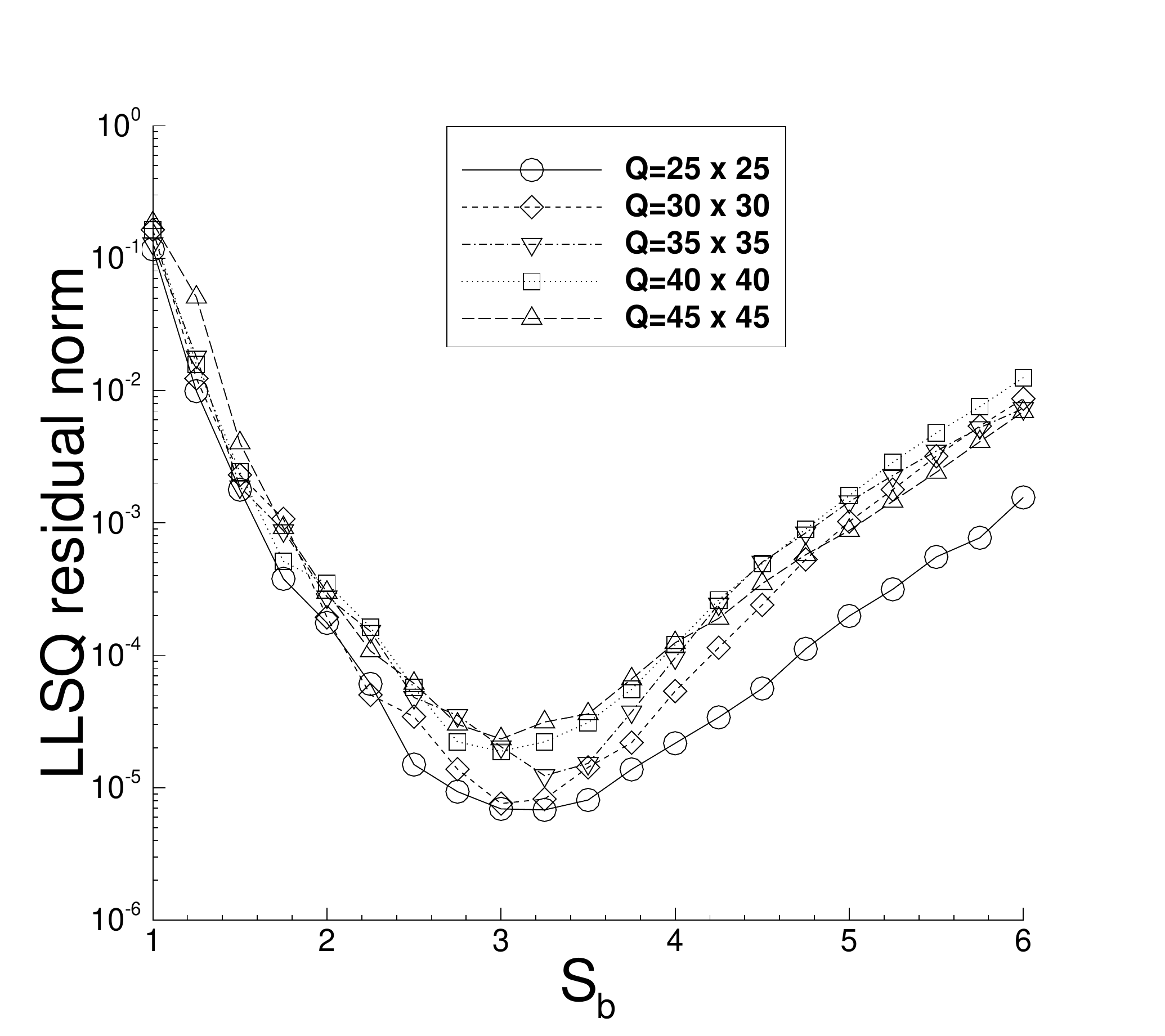}(b)
    \includegraphics[width=2.in]{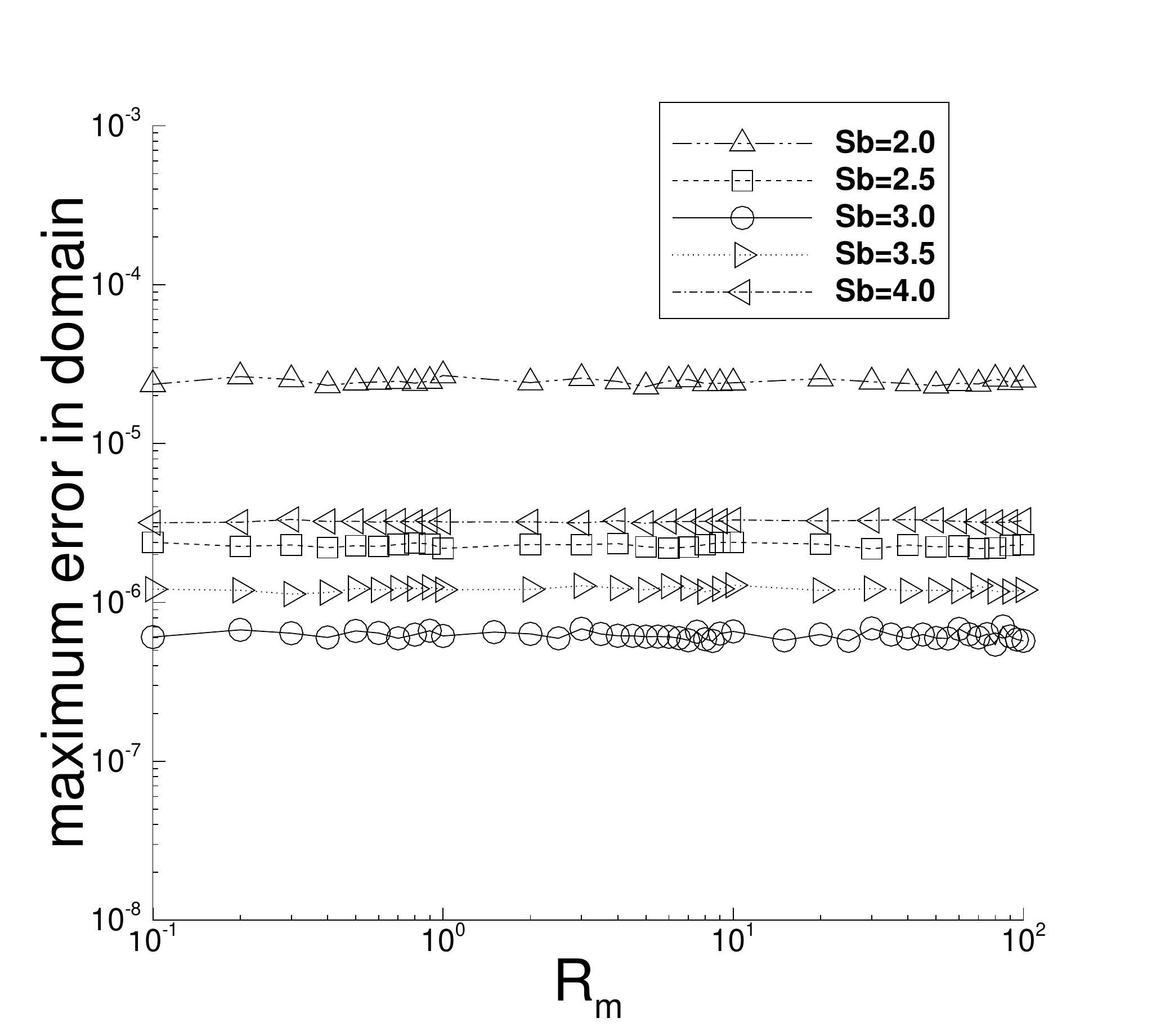}(c)
  }
  \caption{Poisson equation: (a) Distribution of the exact solution.
    (b) LLSQ residual norm 
    as a function of $S_b$, for estimating the best $S_b$ in modBIP.
    (c) The maximum error of the ELM/modBIP solution 
    as a function of $R_m$.
    $R_m=10$ in (b) and is varied in (c).
    $Q=25\times 25$ in (c), and takes several values in (b).
    $S_b$ takes several values in (c), and is varied in (b).
  }
  \label{fg_9}
\end{figure}

We next consider the two-dimensional (2D)
domain $\Omega=\{(x,y)\ |\ x\in[0,2],\ y\in[0,2] \}$, and 
test  the combined ELM/modBIP method using
the boundary value problem with the Poisson equation on $\Omega$:
\begin{subequations}\label{eq_14}
\begin{align}
  &
  \frac{\partial^2 u}{\partial x^2} + \frac{\partial^2 u}{\partial y^2} = f(x,y),
  \\
  &
  u(0,y) = g_1(y), \\
  &
  u(2,y) = g_2(y), \\
  &
  u(x,0) = h_1(x), \\
  &
  u(x,2) = h_2(x),
\end{align}
\end{subequations}
where $u(x,y)$ is the field function to be solved for, $f(x,y)$ is
a prescribed source term, and $g_1$, $g_2$, $h_1$ and $h_2$ are
the Dirichlet boundary distributions.
We consider the following manufactured solution to this problem,
\begin{equation}\label{eq_15}
  u(x,y) = -\left[\frac32\cos\left(\pi x + \frac{7\pi}{20}  \right)
    + 2\cos\left(2\pi x - \frac{\pi}{4}  \right) \right]
  \left[\frac32\cos\left(\pi y + \frac{7\pi}{20}  \right)
    + 2\cos\left(2\pi y - \frac{\pi}{4}  \right) \right].
\end{equation}
Accordingly, the source term $f$ and the boundary distributions
are chosen such that the expression \eqref{eq_15} satisfies
the system \eqref{eq_14}.
Figure \ref{fg_9}(a) illustrates the distribution of this analytic solution.

We employ the combined ELM/modBIP method to solve the system~\eqref{eq_14};
see Section \ref{sec:pde}.
We first consider a single hidden layer in the neural network,
with an architecture given by $[2, 500, 1]$, the $\tanh$ activation
function for the hidden layer, and a linear output layer.
The input layer ($2$ nodes) represents the coordinates $x$ and $y$,
and the output layer ($1$ node) represents the field solution $u(x,y)$.
We employ a set of $Q=25\times 25$ uniform grid (collocation) points
as the training data points,
i.e.~with $25$ points in both $x$ and $y$ directions
(see equation \eqref{eq_8}),  which constitute
the input data into the neural network.
The hidden-layer coefficients in the neural network are initialized to uniform
random values generated on $[-R_m,R_m]$, with $R_m$ either fixed at $R_m=10$
or varied between $R_m=0.1$ and $R_m=100$ in the following tests.
The initial random coefficients are pre-trained by modBIP with
$S_c=S_b/2$ and $S_b$ determined by the procedure
given in Remark~\ref{rem_1}.


Figure \ref{fg_9}(b) shows the residual norm of the linear least squares (LLSQ) problem
as a function of $S_b$ in modBIP, where the initial random coefficients are
generated with $R_m=10$. The results corresponding to $Q=25\times 25$ and several
other sets of collocation points are included,
which all suggest a value around $S_b\approx 3$
for modBIP. Figure \ref{fg_9}(c) shows the maximum and rms errors in the domain
of the ELM/modBIP solution as a function of $R_m$, corresponding to several $S_b$
values around $S_b=3$ in modBIP (with $S_c=S_b/2$), where $Q=25\times 25$
collocation points have been employed. The errors of the ELM/modBIP method
can be observed to be insensitive to $R_m$ (or the initial
random coefficients).


\begin{figure}
  \centerline{
    \includegraphics[width=2.in]{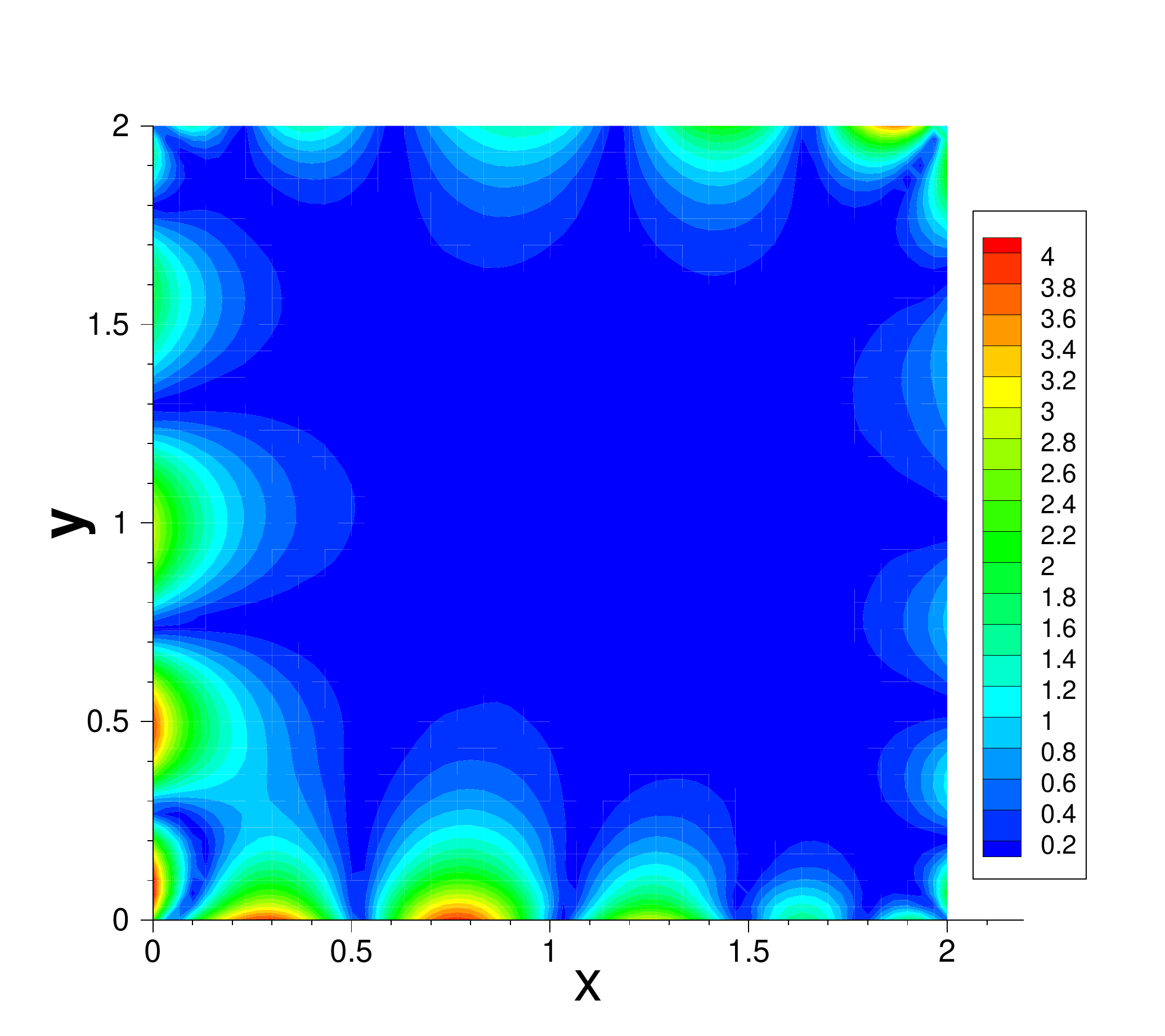}(b)
    \includegraphics[width=2.in]{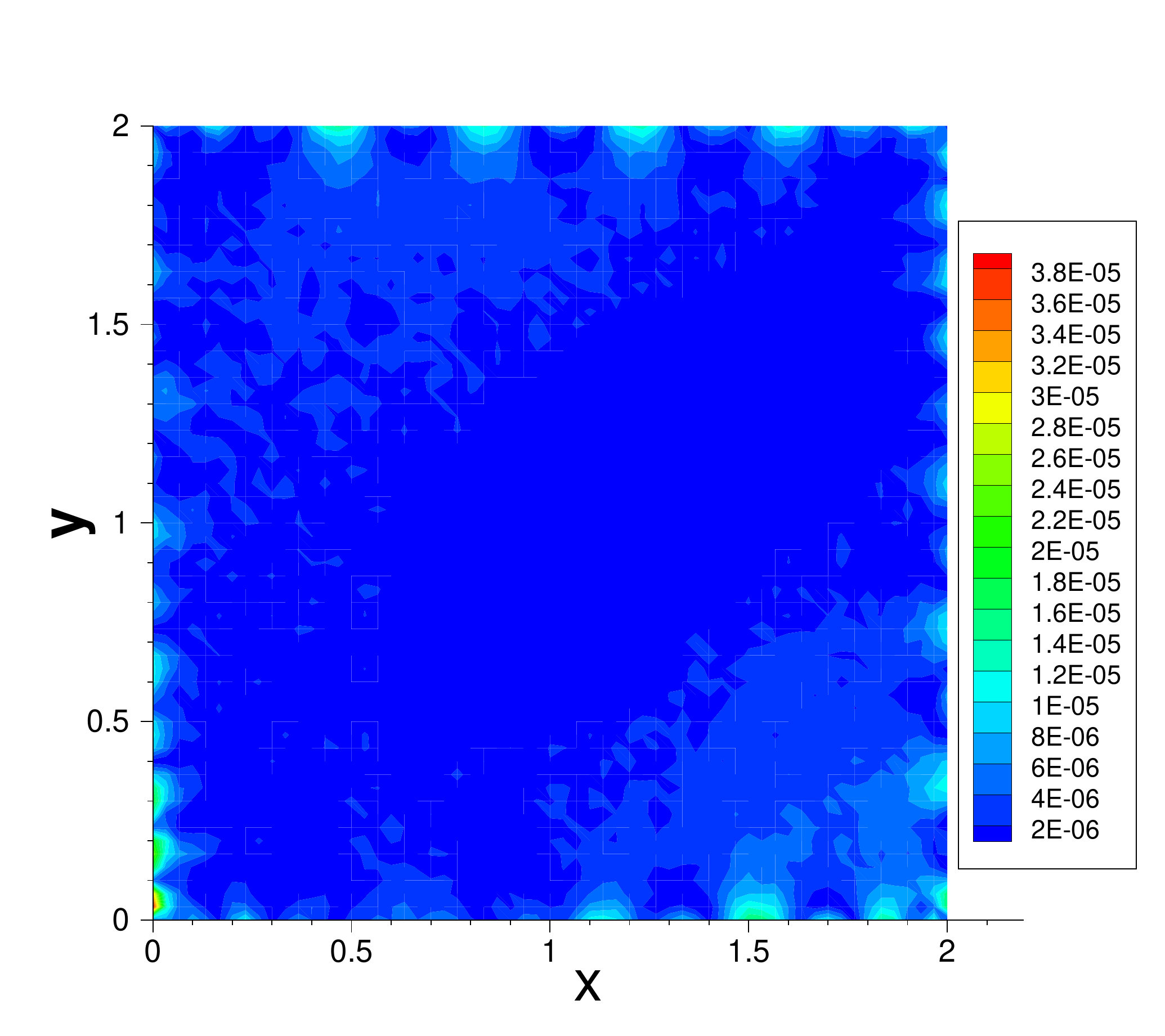}(c)
    \includegraphics[width=2.in]{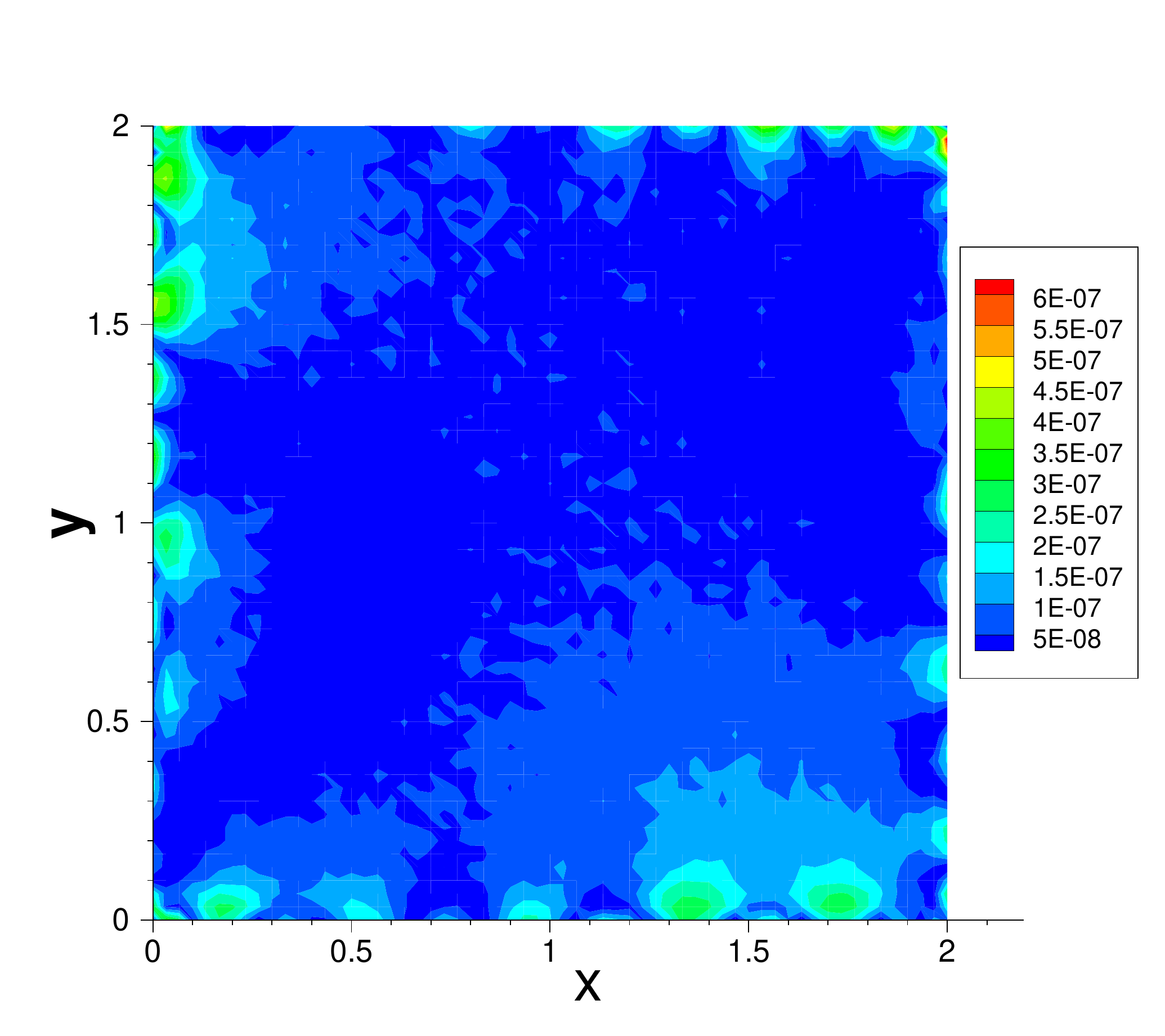}(d)
  }
  \caption{Poisson equation:
    Distributions of
     the absolute error of the ELM solution computed with (a) no pre-training,
    (b) BIP pre-training, and
    (c) modBIP pre-training of the random coefficients.
  }
  \label{fg_10}
\end{figure}

Figure \ref{fg_10} compares distributions of the absolute
error of the ELM solution obtained with no pre-training and with the BIP and
the modBIP pre-training of the random coefficients in the neural network.
Here we have employed a network
architecture given by $[2, 500, 1]$, the $\tanh$ activation function
for the hidden layer, $Q=25\times 25$ uniform collocation points, and
$R_m=10$ for generating the initial random coefficients.
With BIP, we employ a normal distribution for the target samples,
with a random mean generated on $[-1,1]$ from a uniform distribution and
a standard deviation $0.5$~\cite{NeumannS2013}.
With modBIP we employ $S_b=3$ and $S_c=S_b/2$ in the algorithm.
One can observe that the ELM result is inaccurate without pre-training of
the random coefficients.
With both BIP and modBIP pre-training of the random coefficients, the ELM method
produces accurate solutions to the Poisson equation. The ELM/modBIP
solution is markedly more accurate than that of ELM/BIP.


\begin{figure}
  \centerline{
    \includegraphics[width=2in]{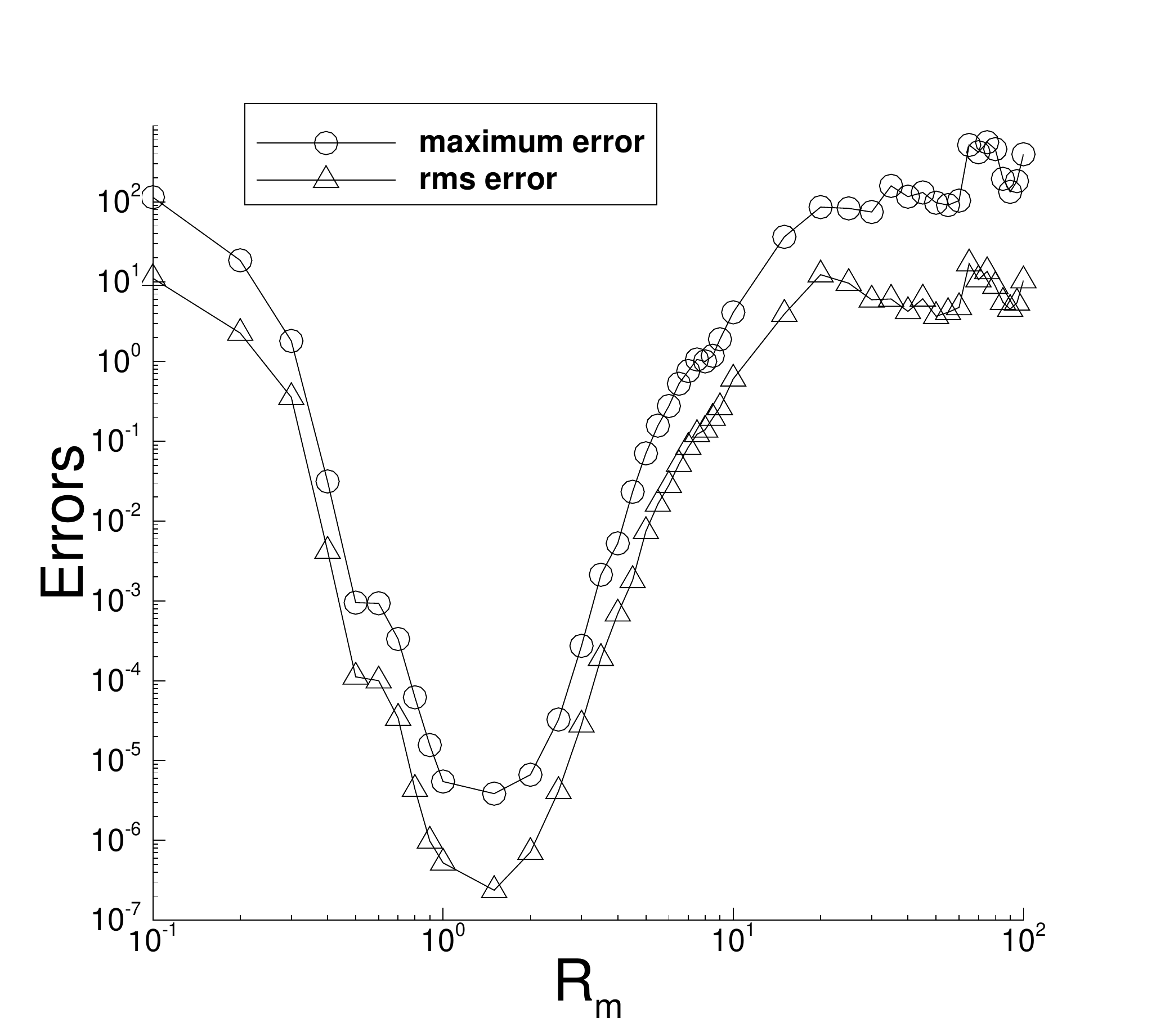}(a)
    \includegraphics[width=2in]{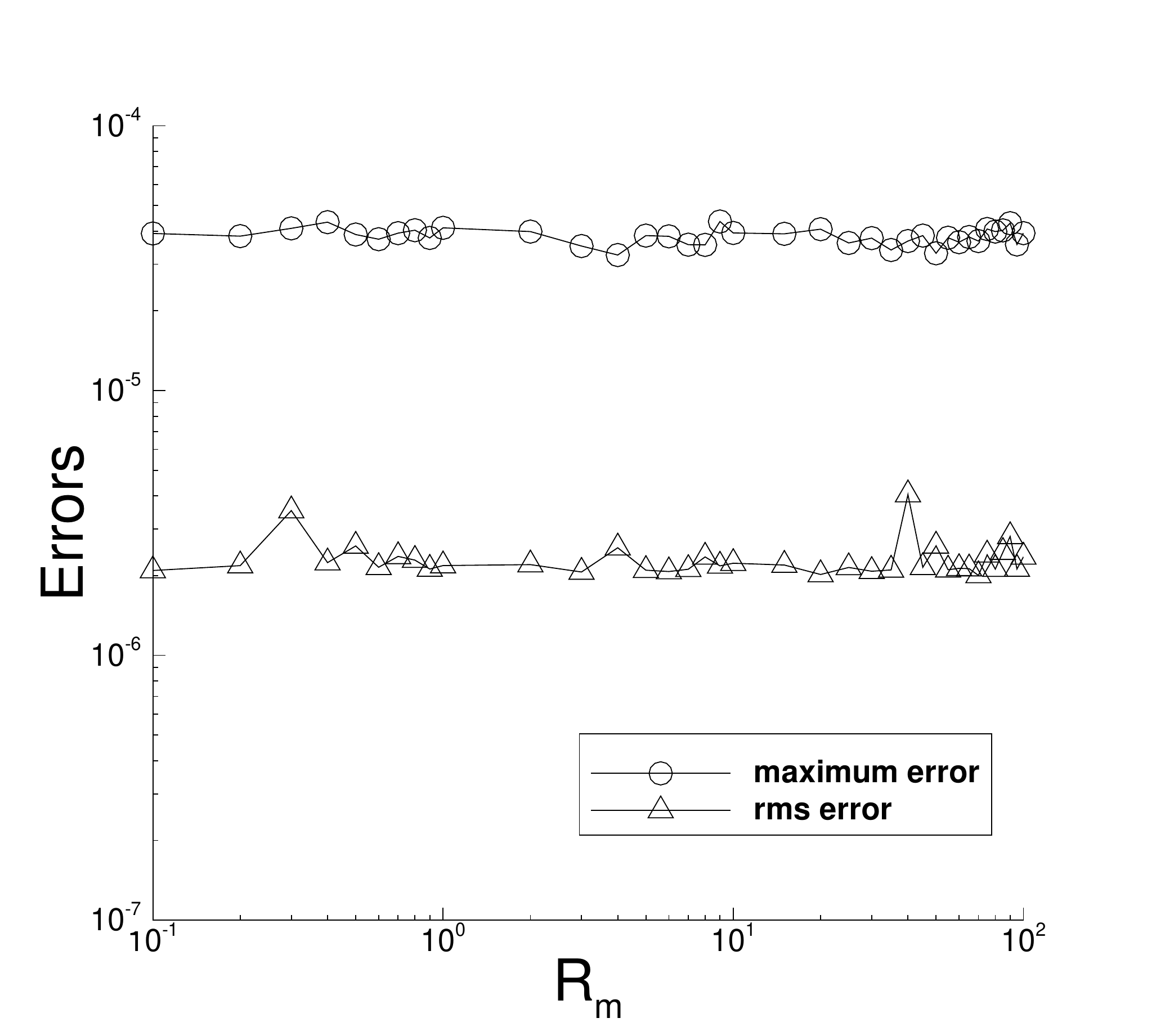}(b)
    \includegraphics[width=2in]{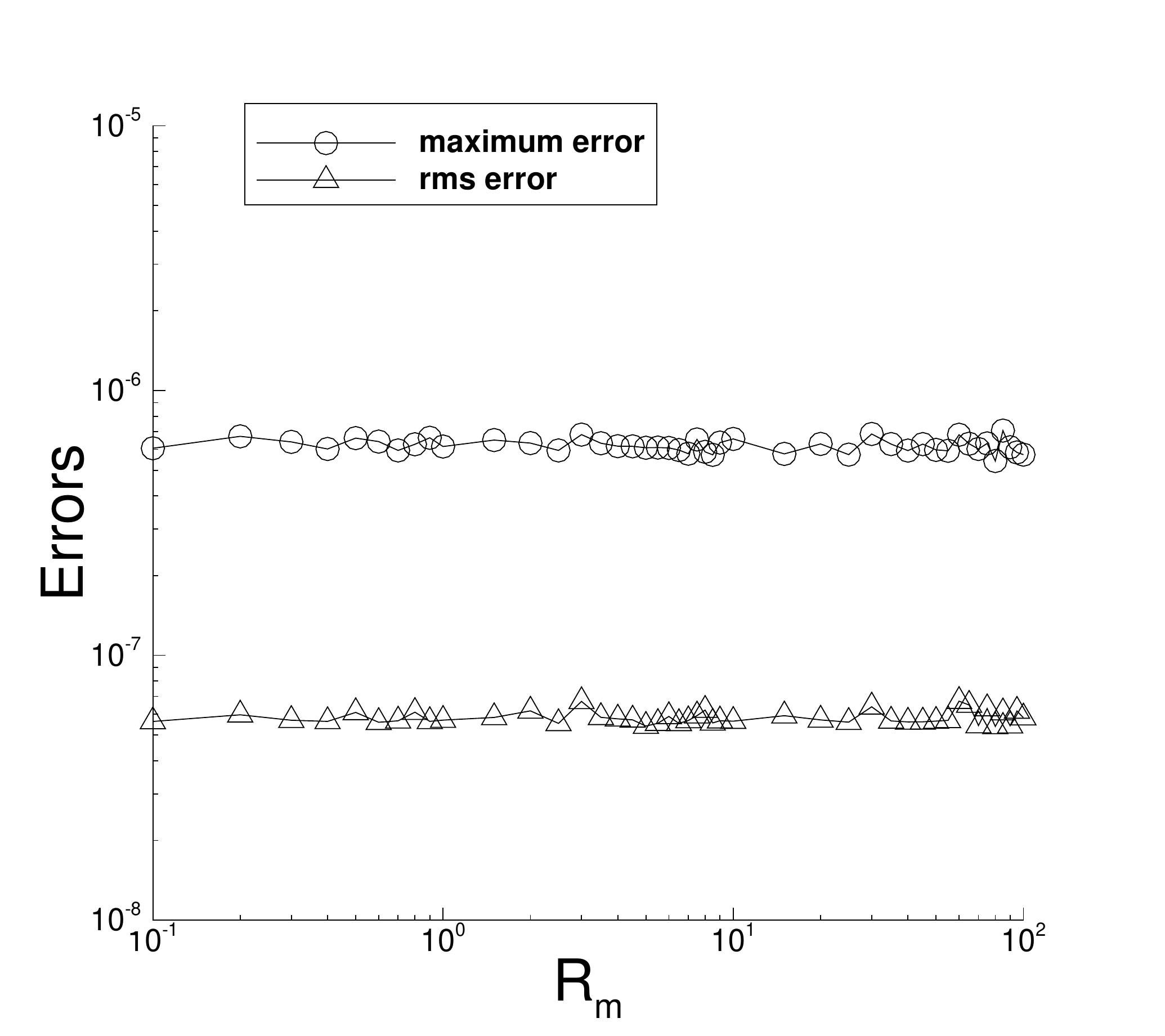}(c)
  }
  \caption{Poisson equation: the maximum and rms errors in the domain as a
    function of $R_m$, 
    obtained with (a) no pre-training, (b) BIP pre-training,
    and (c) modBIP pre-training of the random coefficients.
  }
  \label{fg_11}
\end{figure}

Figure \ref{fg_11} is a further comparison of the cases with no pre-training
and with BIP and modBIP pre-training of the random coefficients.
Here we vary $R_m$ systematically, and for each $R_m$ we initialize
the hidden-layer coefficients to uniform random values
from $[-R_m,R_m]$, which are then pre-trained by BIP
or modBIP and used in the ELM computations.
The other parameter values are identical to those for Figure \ref{fg_10}.
The three plots show the maximum and rms errors in the domain of the ELM
solution as a function of $R_m$,
obtained with no pre-training and with the BIP and modBIP pre-training
of the random coefficients.
With no pre-training, $R_m$ is observed to strongly influence the accuracy
of the ELM solution. With both BIP and modBIP pre-training of
the random coefficients, the accuracy of the ELM solution becomes
essentially independent of $R_m$. The error level of
the ELM/modBIP solution is markedly lower than that of the ELM/BIP solution.


\begin{figure}
  \centerline{
    \includegraphics[width=2in]{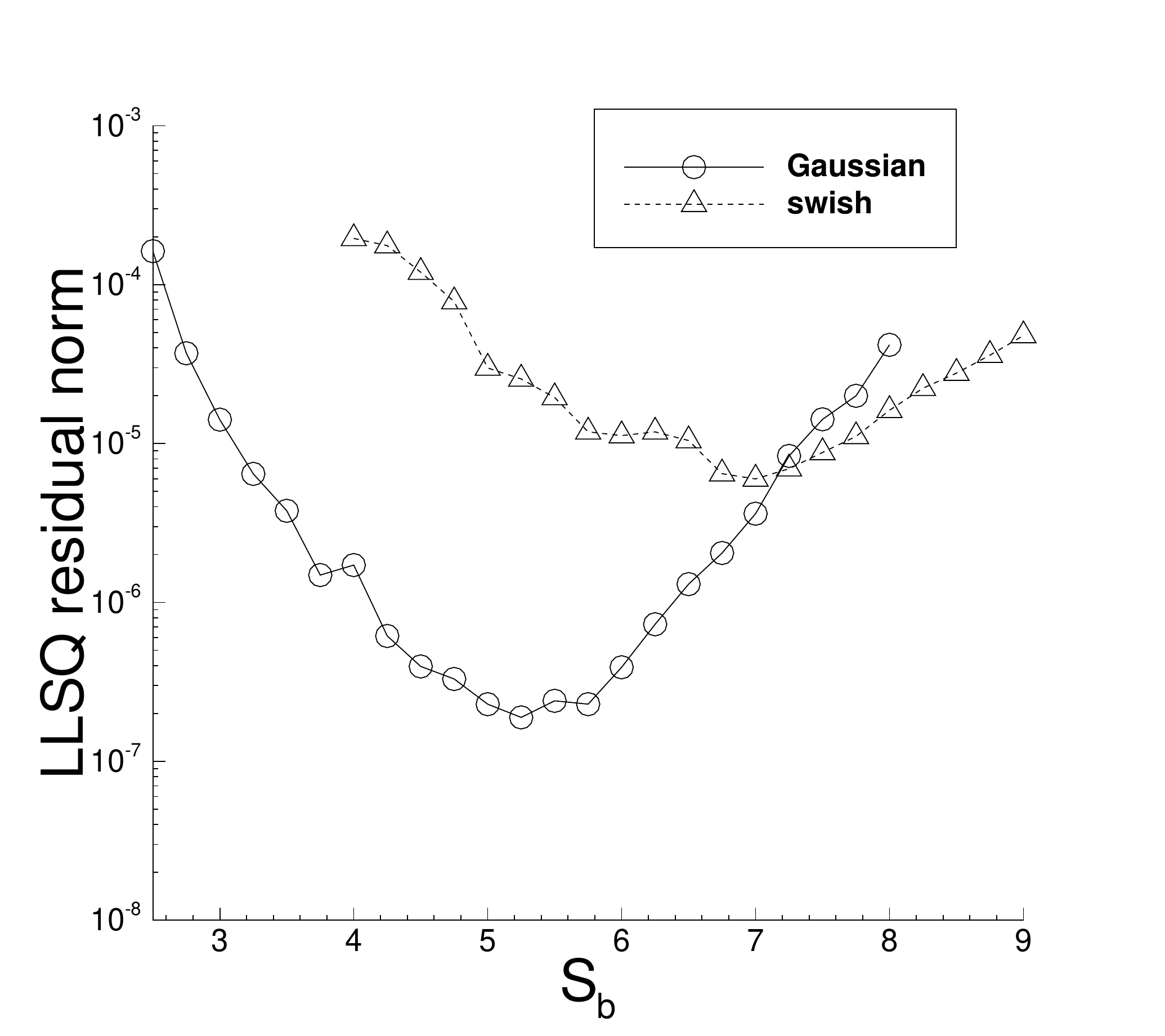}(a)
    \includegraphics[width=2in]{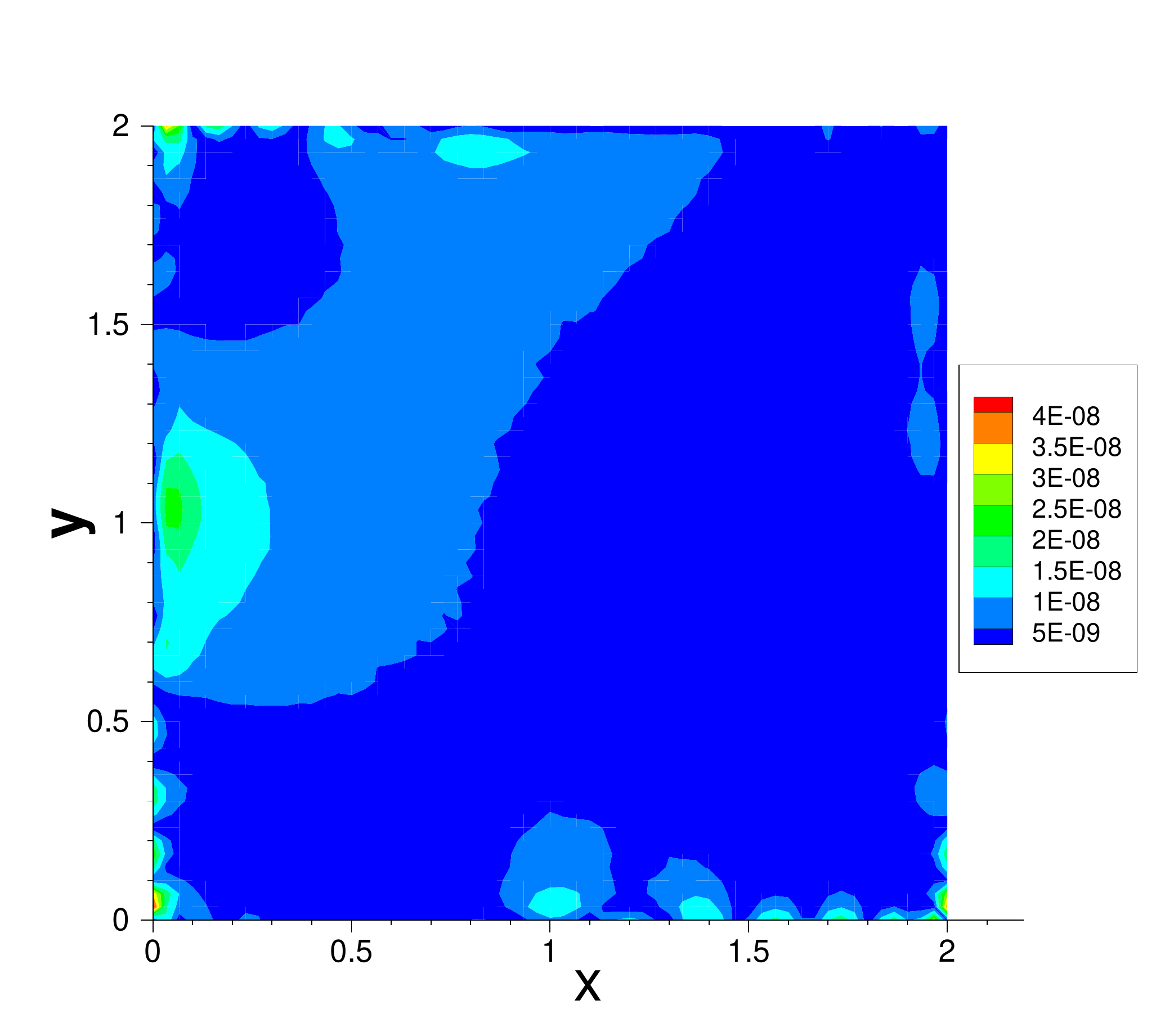}(b)
    \includegraphics[width=2in]{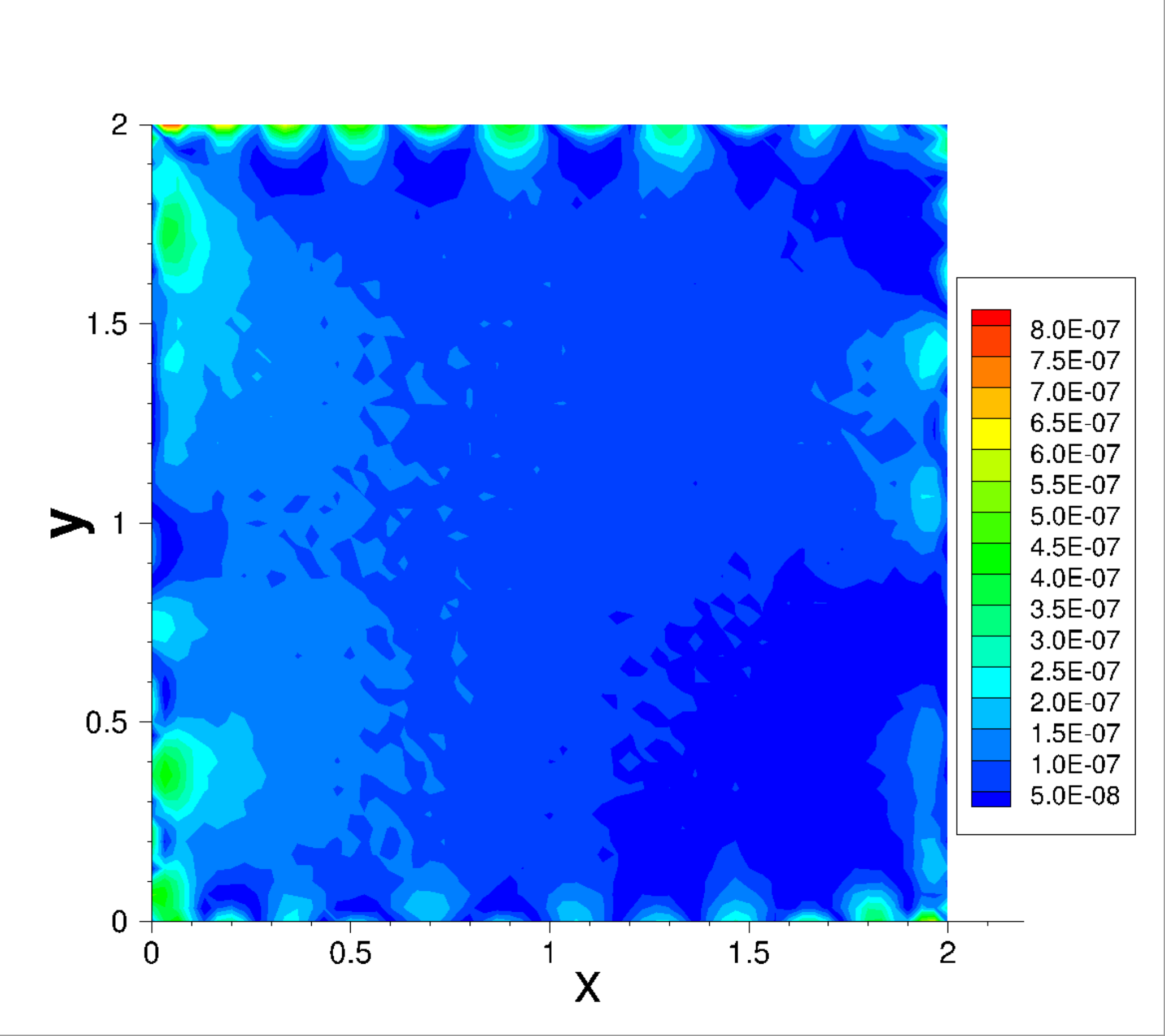}(c)
  }
  \centerline{
    \includegraphics[width=2in]{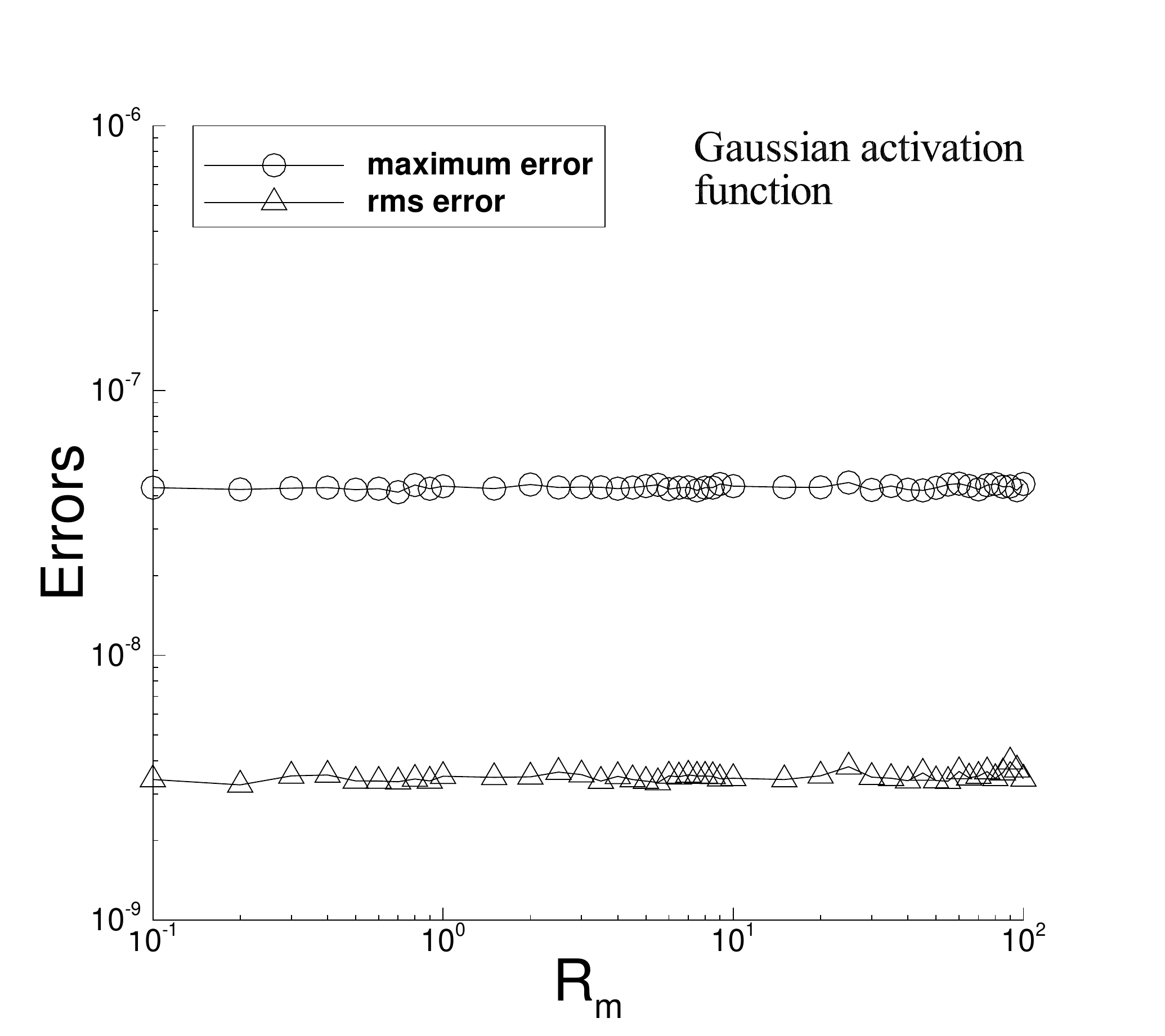}(d)
    \includegraphics[width=2in]{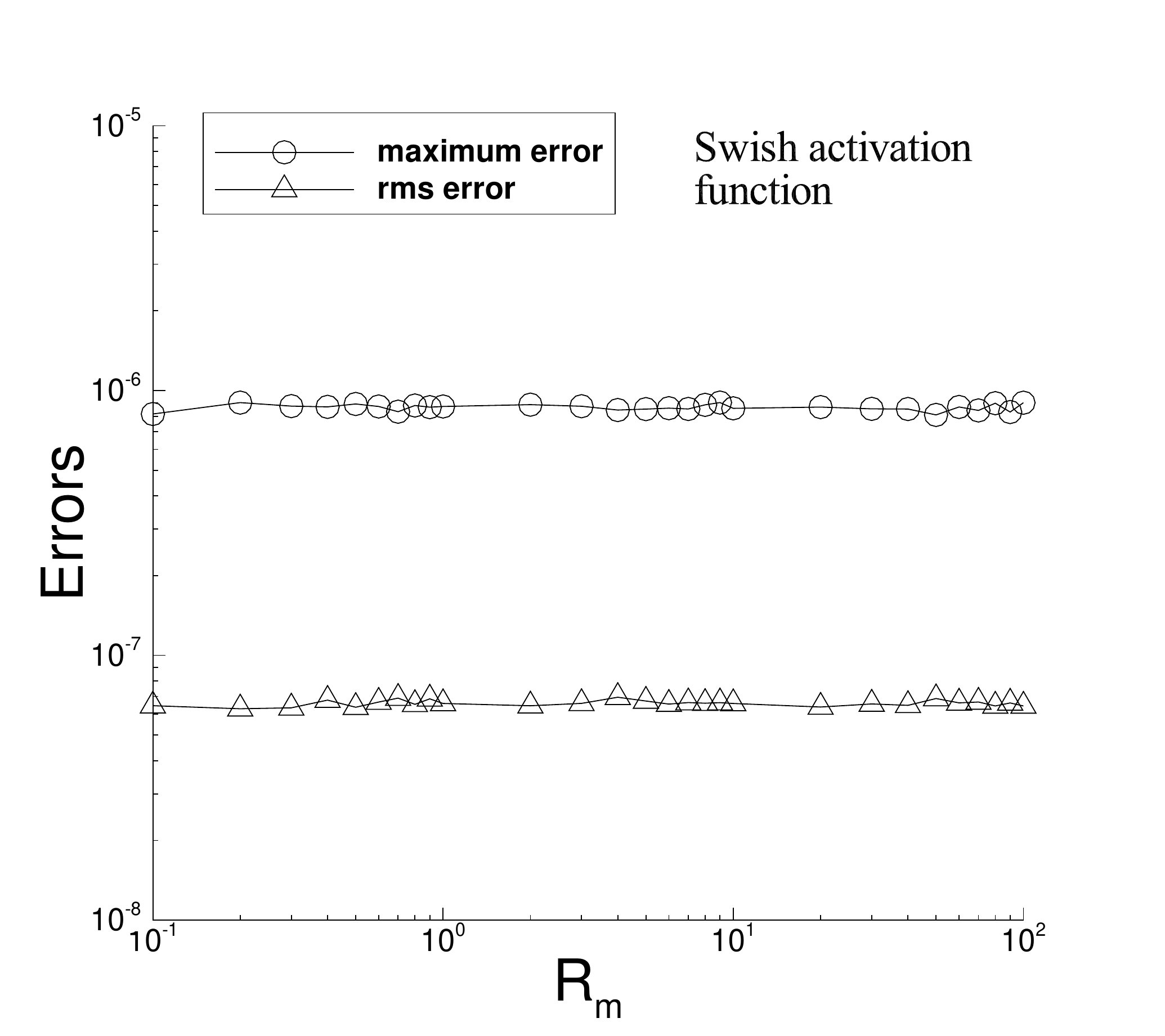}(e)
  }
  \caption{Poisson equation (non-monotonic activation functions):
    (a) LLSQ residual norms versus $S_b$, for estimating the best $S_b$ in modBIP
    with the Gaussian and swish activation functions.
    Error distributions of the ELM/modBIP solution obtained
    with (b) Gaussian
    and (c) swish activation functions.
    Maximum/rms errors of the ELM/modBIP solution versus $R_m$,
    obtained with
    (d) Gaussian
    and (e) swish activation functions.
    $R_m=10$ in (a,b,c) and is varied in (d,e).
    $S_b=5$ in (b,d) and $S_b=7$ in (c,e).
  }
  \label{fg_12}
\end{figure}

Figure \ref{fg_12} illustrates the ELM/modBIP results
attained with the Gaussian and the swish activation functions for
the hidden layers. It should be noted that the BIP algorithm~\cite{NeumannS2013}
breaks down with these activation functions because they do not have
an inverse.
In these tests, we have employed a network architecture $[2, 500, 1]$,
$Q=25\times 25$ uniform collocation points, either a fixed $R_m=10$
or a varying $R_m$ for generating the initial random coefficients,
and $S_c=S_b/2$ in modBIP.
Figure \ref{fg_12}(a) shows the LLSQ residual norms for estimating the $S_b$
parameter in modBIP, which suggest a value around $S_b\approx 5$ with
the Gaussian function and a value around $S_b\approx 7$ with the swish function.
Figures \ref{fg_12}(b) and (d) show the error distribution of the ELM/modBIP
solution with $R_m=10$, and its maximum/rms errors
corresponding to the initial random coefficients generated with different $R_m$,
computed using the Gaussian activation function with $S_b=5$ in modBIP.
Figures \ref{fg_12}(c) and (e) show the corresponding ELM/modBIP
results computed using the swish activation function with $S_b=7$ in modBIP.
These data indicate that the combined ELM/modBIP method produces highly accurate
results with these activation functions, and that its accuracy is insensitive
to the initial random coefficients.


\begin{figure}
  \centerline{
    \includegraphics[width=2.0in]{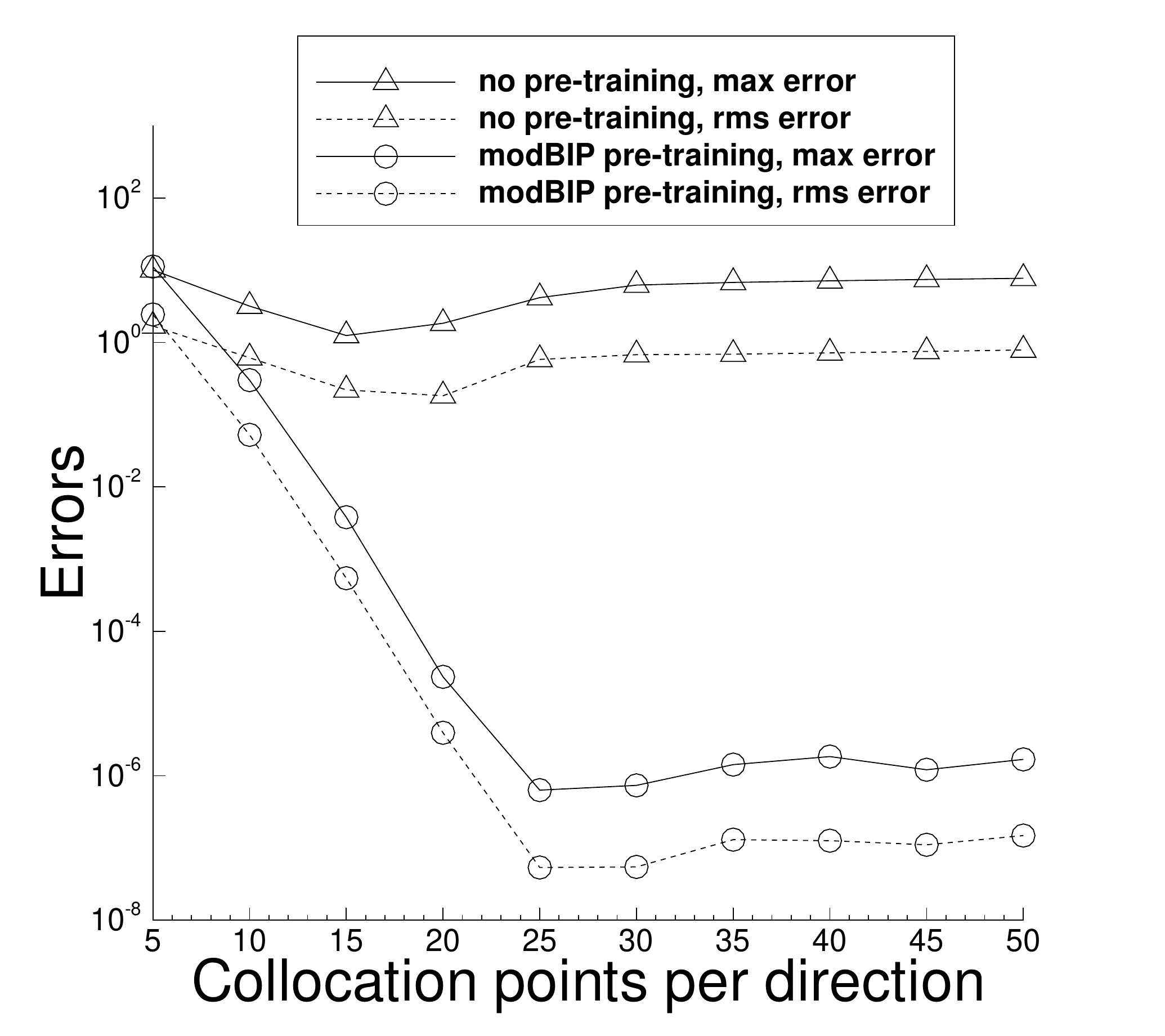}(a)
    \includegraphics[width=2.0in]{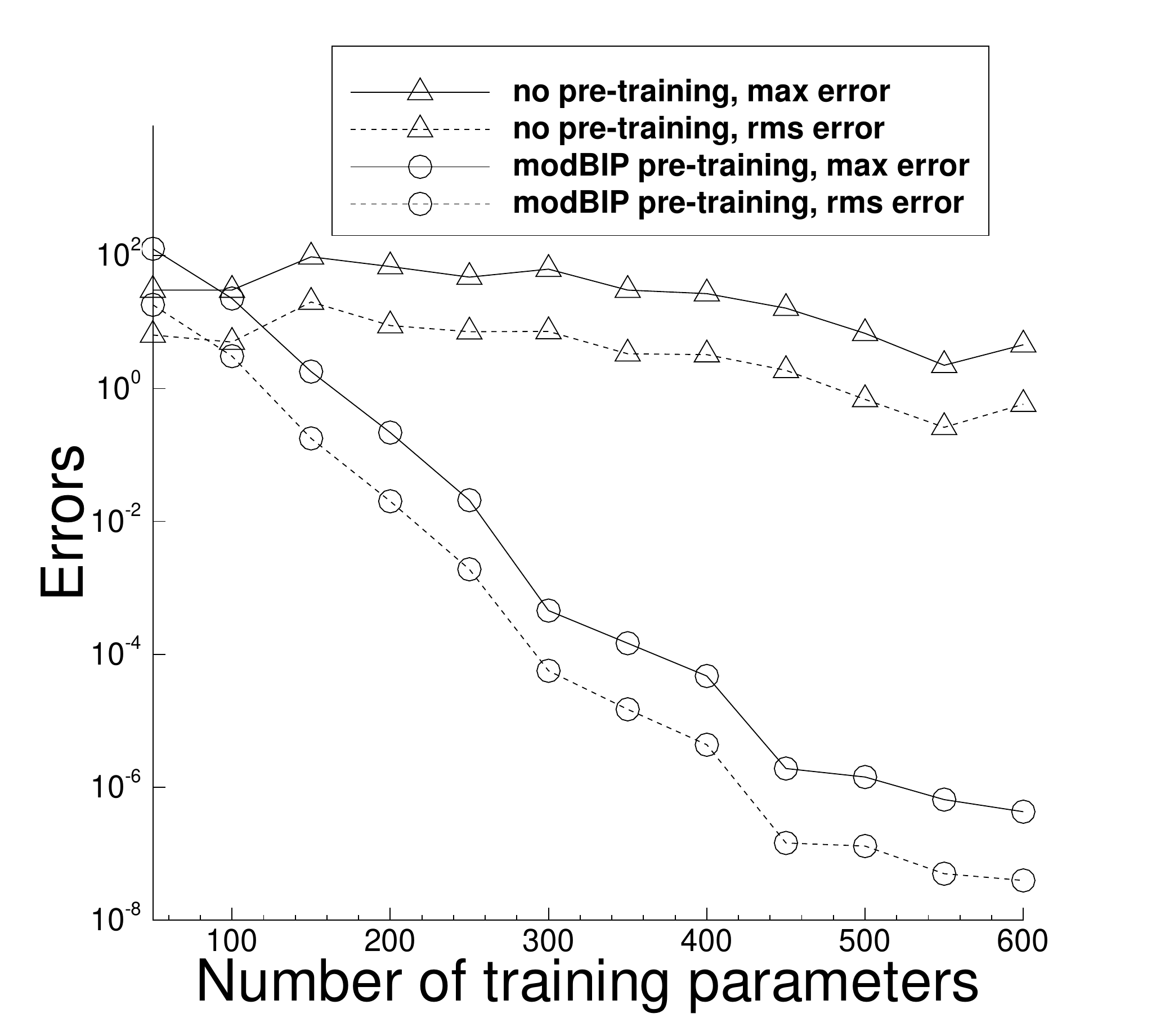}(b)
  }
  \centerline{
    \includegraphics[width=2.0in]{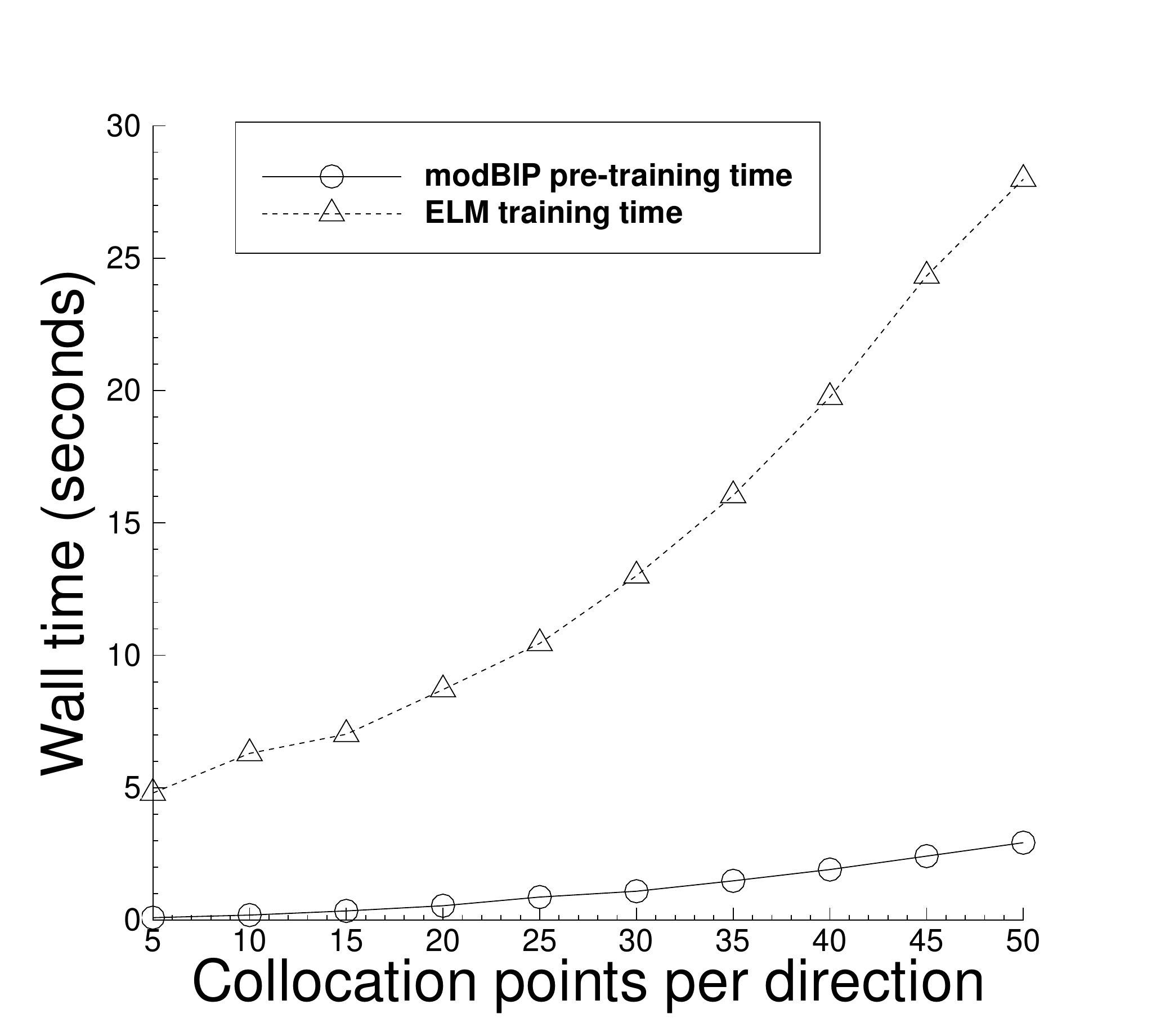}(c)
    \includegraphics[width=2.0in]{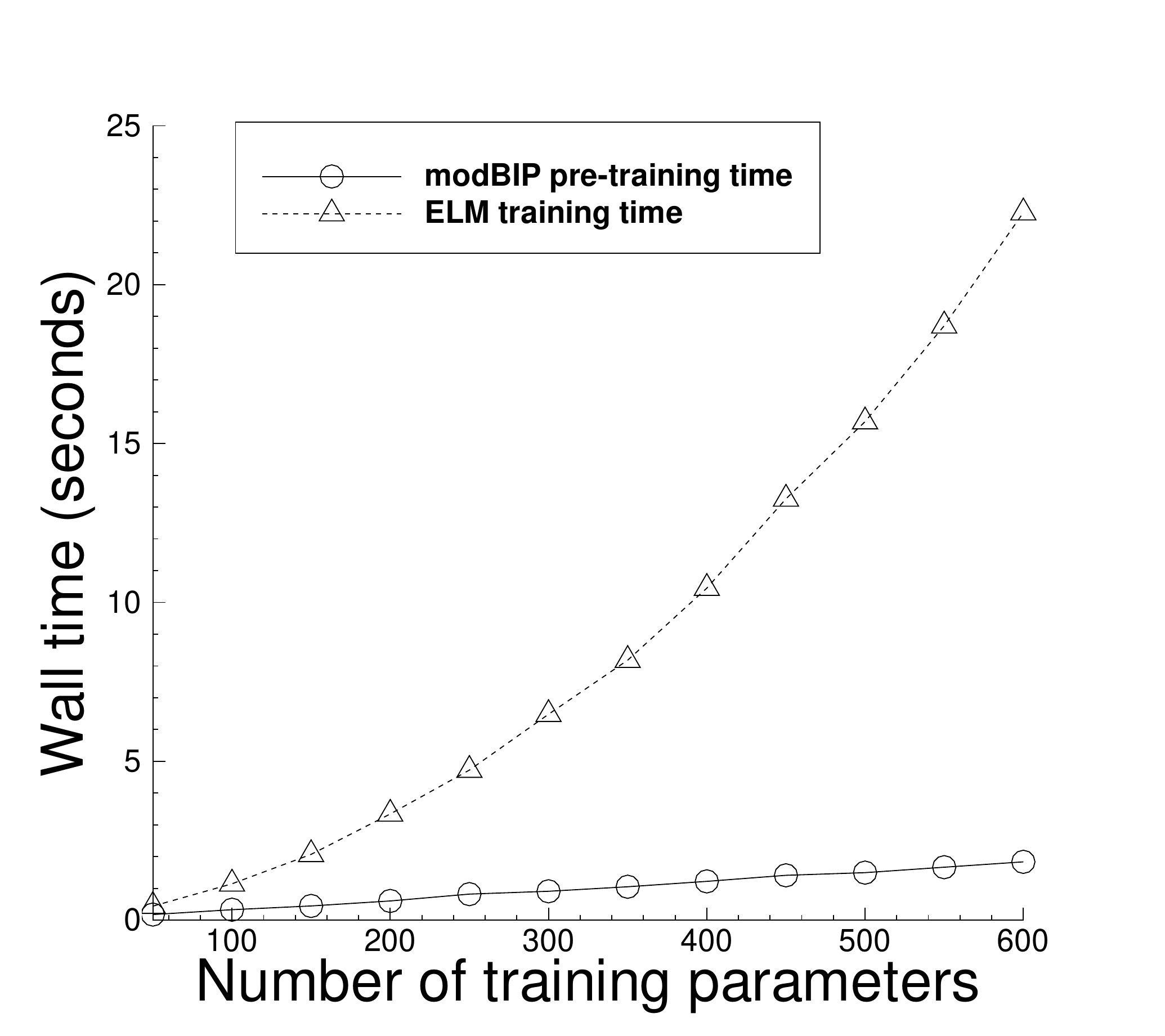}(d)
  }
  \caption{Poisson equation:  The maximum/rms errors of the ELM/modBIP solution
    as a function of (a) the number of collocation points in each direction
    and (b) the number of training parameters, attained with no pre-training and with modBIP
    pre-training of the random coefficients.
    The modBIP pre-training time of the random coefficients, and
    the ELM training time of the neural network, as a function of (c) the number of
    collocation points in each direction and (d) the number of training
    parameters.
  }
  \label{fg_13}
\end{figure}

Figure \ref{fg_13} compares the accuracy of the combined ELM/modBIP
method and the ELM method with no pre-training of the random coefficients,
and also examines the computational cost of the modBIP pre-training of
the random coefficients.
In this set of tests, we employ a neural network architecture $[2, M, 1]$,
where the number of training parameters $M$ is either fixed at $M=500$
or varied between $M=50$ and $M=600$. We employ
the $\tanh$ activation function for the
hidden layer, and $Q=Q_1\times Q_1$ uniform collocation points in the domain, where
$Q_1$ denotes the number of collocation points in $x$/$y$ directions
and is either fixed at $Q_1=35$ or varied between $Q_1=5$ and $Q_1=50$.
The initial random coefficients  are generated with $R_m=10$, and
we employ $S_b=3$ and $S_c=S_b/2$ in modBIP.
Figure \ref{fg_13}(a) shows the maximum and rms errors of the ELM solutions 
as a function of $Q_1$, obtained
without pre-training and with modBIP pre-training of the random coefficients
and with a fixed $M=500$.
Figure \ref{fg_13}(b) shows the maximum/rms ELM errors
as a function of $M$, obtained with no pre-training and with modBIP pre-training
and with a fixed $Q_1=35$ for the collocation points.
The ELM solution obtained without pre-training the random coefficients
generated with $R_m=10$ is not accurate, and increasing the number of collocation
points or the training parameters results in little or
no improvement in the accuracy.
In contrast, the errors of the combined ELM/modBIP method decrease exponentially
as the number of collocation points per direction $Q_1$ or the number of
training parameters $M$ increases. The errors are observed to saturate at a level
around $10^{-8}\sim 10^{-6}$ as $Q_1$ increase beyond $25$ for this case.

Figures \ref{fg_13}(c) and (d) show the modBIP pre-training time of the random
coefficients and the ELM training time of the neural network as
a function of the number of collocation points in each direction ($Q_1$)
and the number of training parameters ($M$), respectively.
Both the modBIP pre-training time and the ELM network training time
increase with increasing number of collocation
points in the domain and increasing number of training parameters in
the neural network. But the modBIP pre-training cost increases much more slowly than
the latter. The modBIP pre-training cost of the random coefficients
is only a fraction of the ELM network training cost.
For example, in the range of collocation points tested here, the modBIP
pre-training time is about $2\sim 10 \%$ of the ELM network training time.
These results suggest that the modBIP pre-training of the random coefficients
is not significant in terms of the overhead it induces. 
It should be noted that the modBIP pre-training cost can be further reduced,
because logically pre-training the random coefficients only needs to be performed
once (the first time) for a given a network architecture and the input collocation
points. The pre-trained hidden-layer
random coefficients can be saved and used directly for
subsequent ELM computations.


\begin{figure}
  \centerline{
    \includegraphics[width=1.5in]{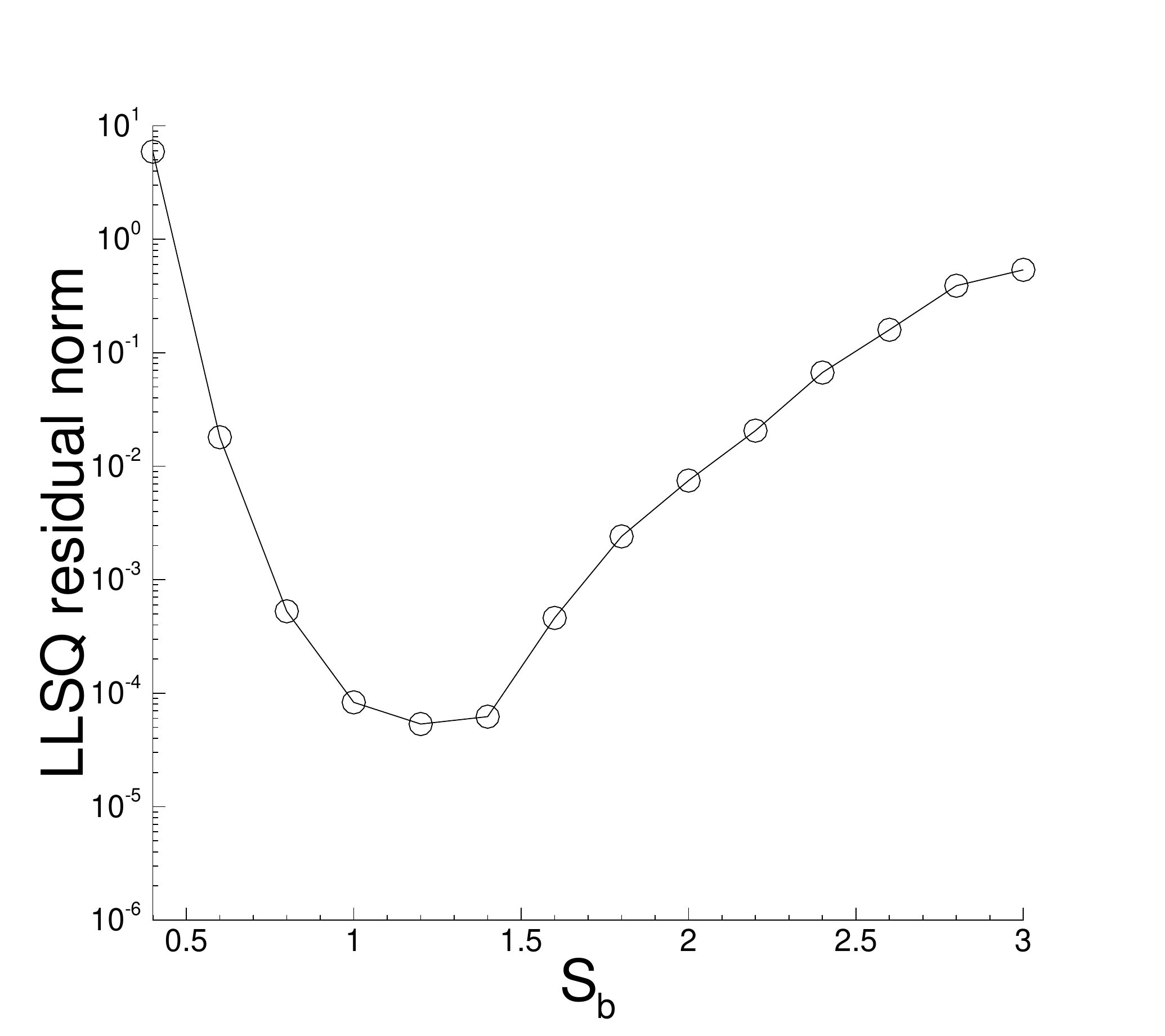}(a)
    \includegraphics[width=1.5in]{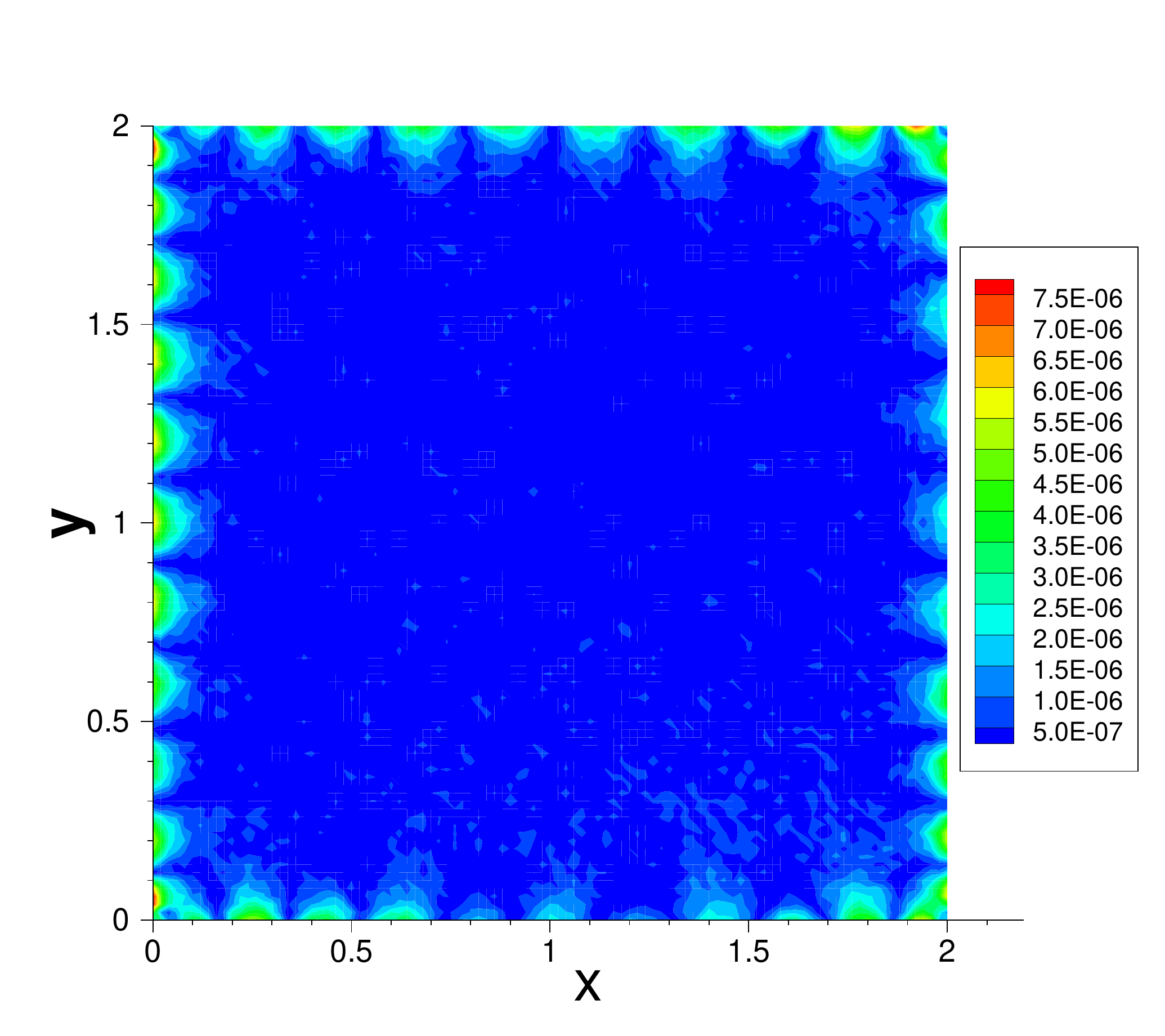}(b)
    \includegraphics[width=1.5in]{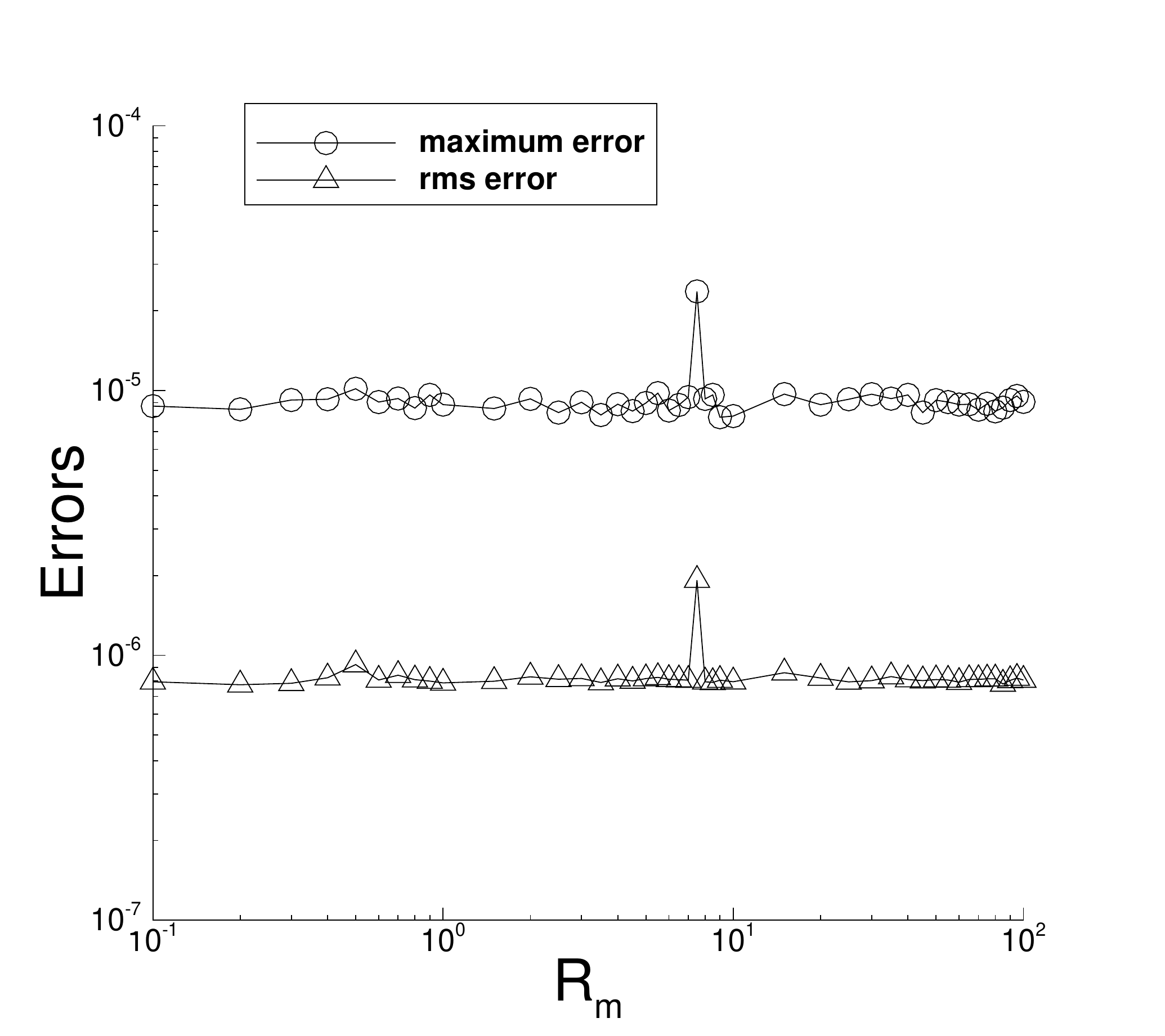}(c)
    \includegraphics[width=1.5in]{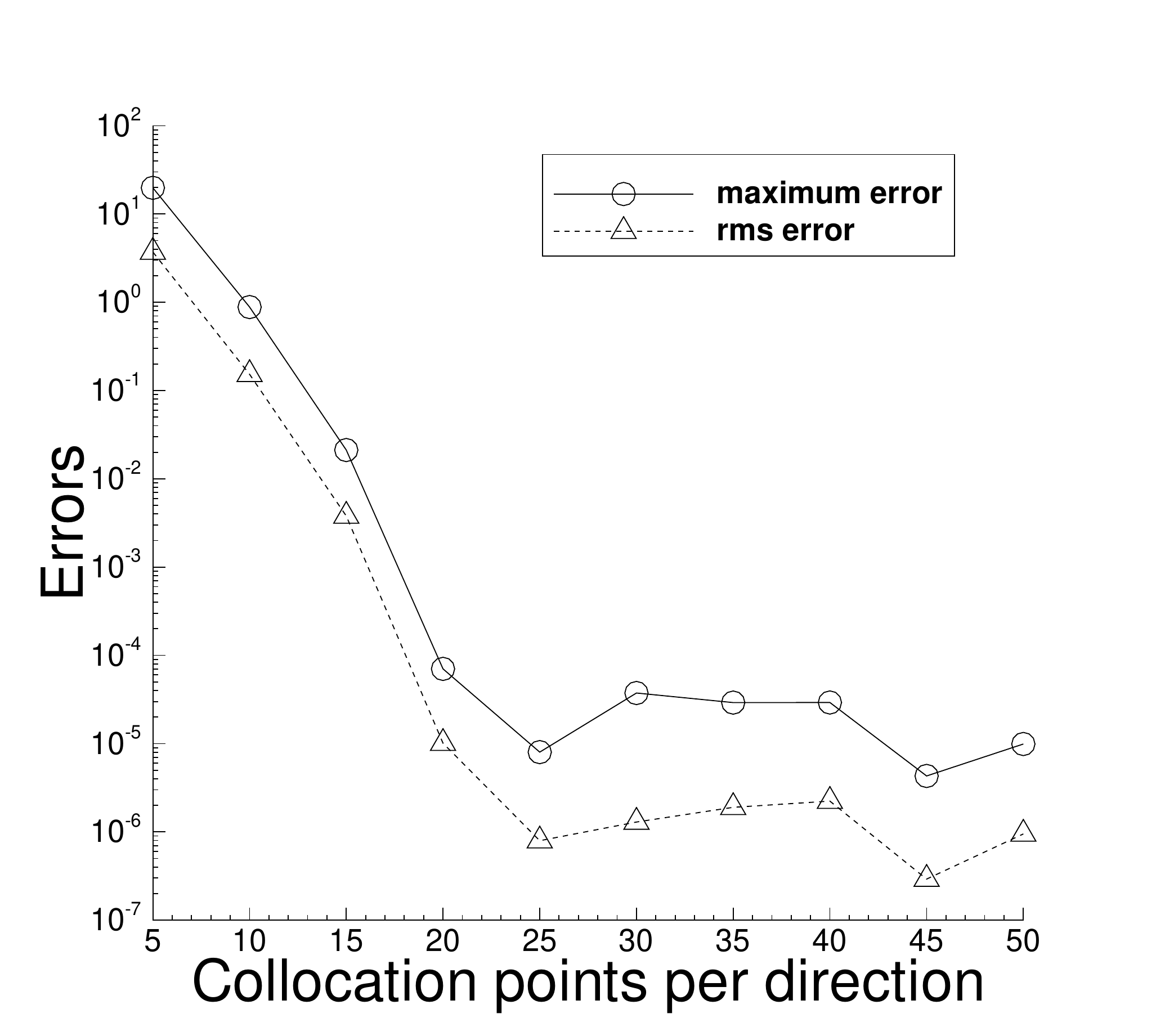}(d)
  }
  \caption{Poisson equation ($3$ hidden layers in neural network):
    (a) LLSQ residual norms versus $S_b$, for estimating
    the best $S_b$ in modBIP.
    (b) Error distribution of the ELM/modBIP solution.
    Maximum/rms errors of the ELM/modBIP solution as a function of (c) $R_m$,
    and (d) the number of collocation points in each direction.
    $S_b=1$ in (b,c,d) and is varied in (a).
    $R_m=10$ in (a,b,d) and is varied in (c).
    $Q=25\times 25$ in (a,b,c) and is varied in (d).
    $M=500$ in (a,b,c,d).
  }
  \label{fg_14}
\end{figure}


\begin{figure}
  \centerline{
    \includegraphics[width=1.5in]{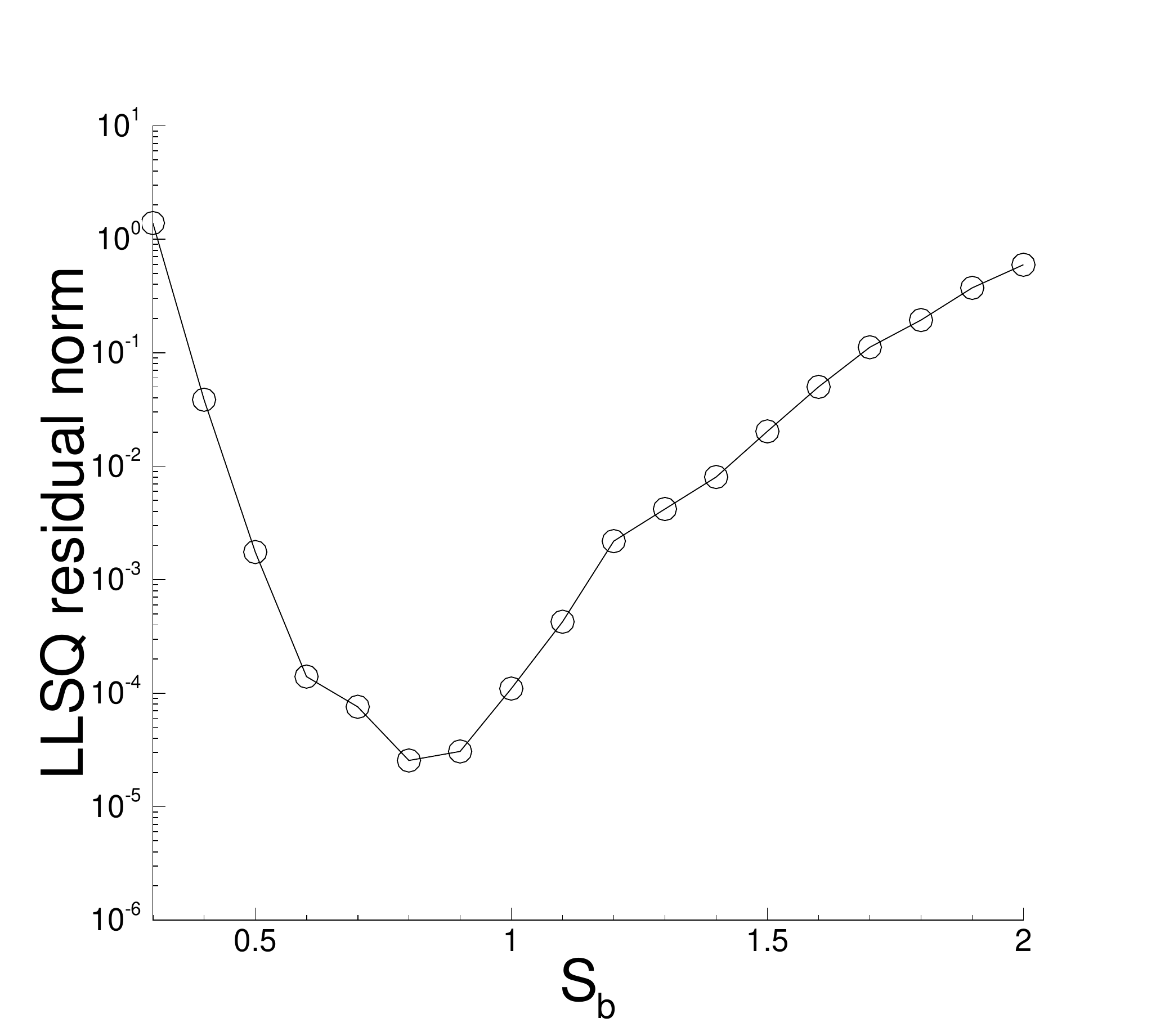}(a)
    \includegraphics[width=1.5in]{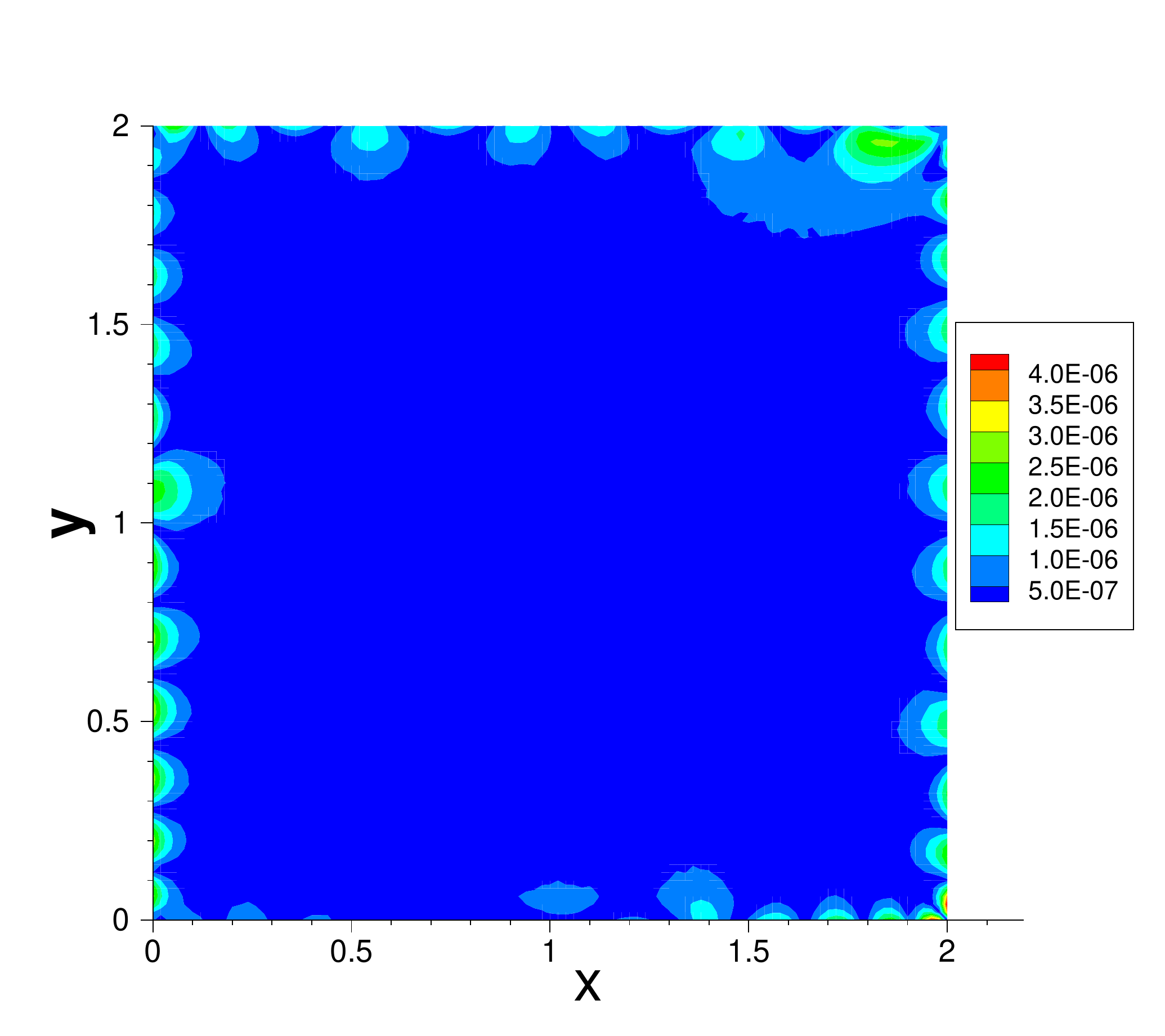}(b)
    \includegraphics[width=1.5in]{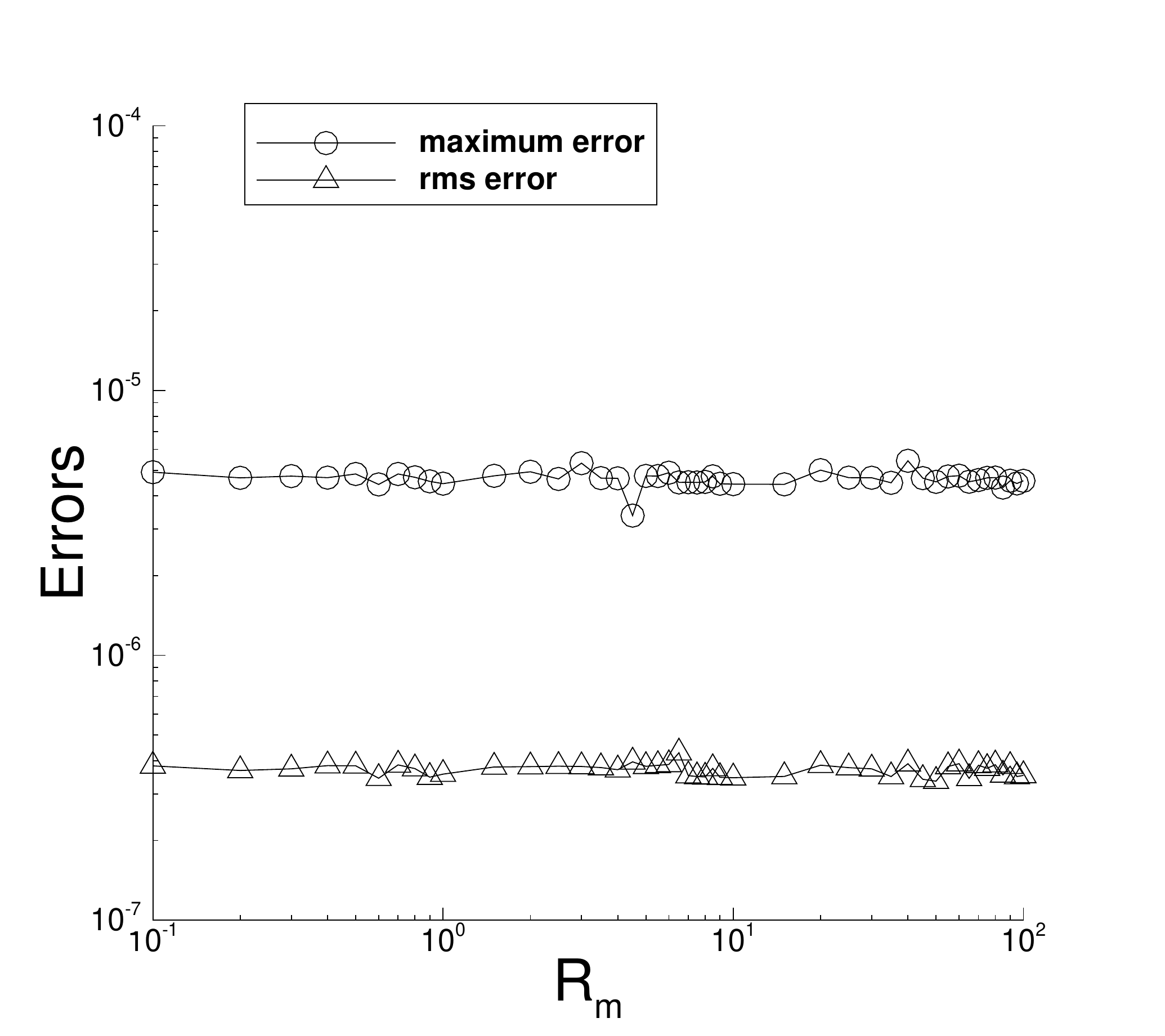}(c)
    \includegraphics[width=1.5in]{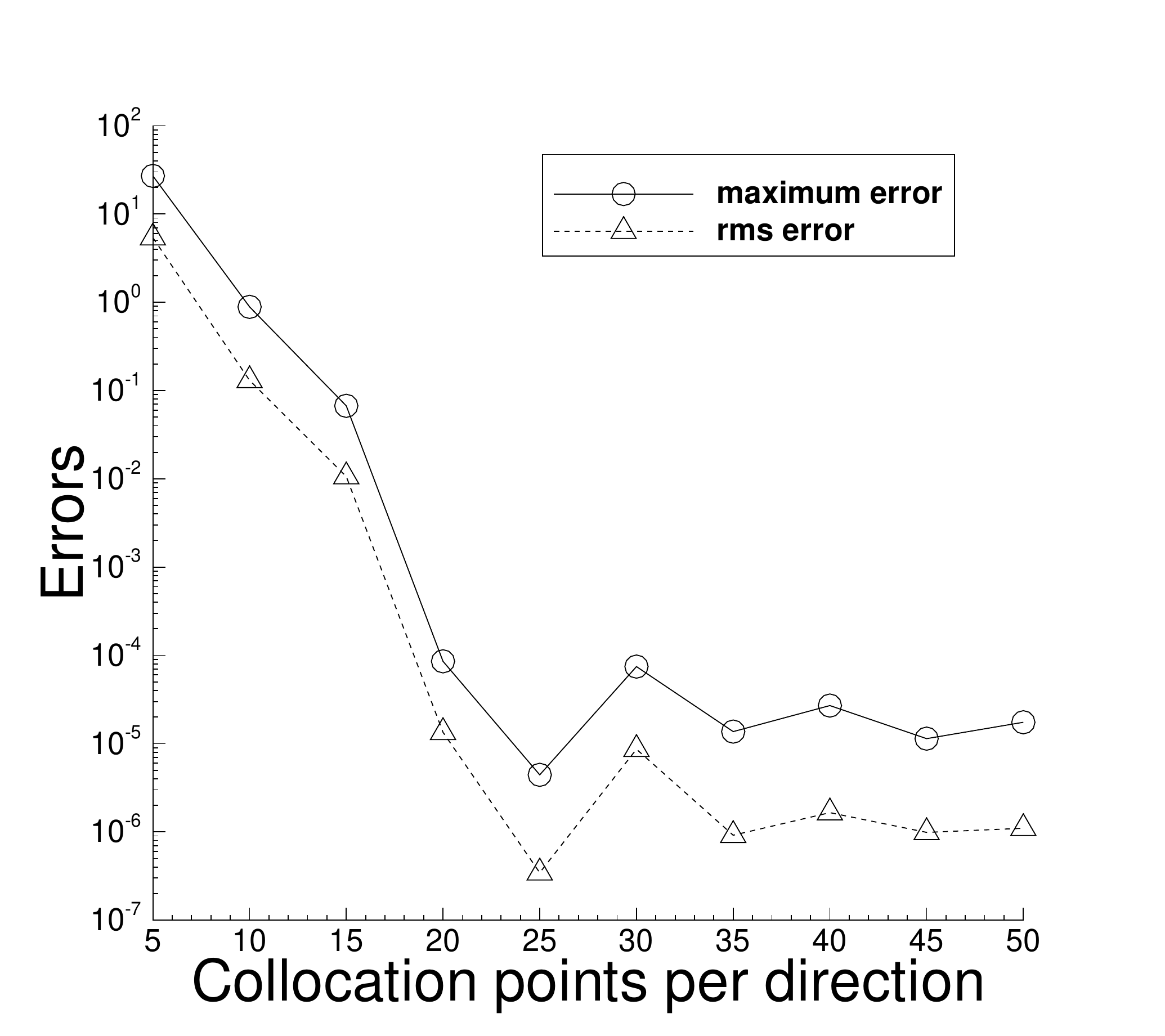}(d)
  }
  \caption{Poisson equation (5 hidden layers in neural network):
    (a) LLSQ residual norm for estimating the best $S_b$ in modBIP.
    (b) Error distribution of the ELM/modBIP solution.
    The maximum/rms errors of ELM/modBIP solution as a function of (c) $R_m$, and
    (d) the number of collocation points in each direction.
    $S_b=0.8$ in (b,c,d) and is varied in (a).
    $R_m=10$ in (a,b,d) and is varied in (c).
    $Q=25\times 25$ in (a,b,c) and is varied in (d).
    $M=500$ in (a,b,c,d).
  }
  \label{fg_15}
\end{figure}


We next test the combined ELM/modBIP method for
solving the Poisson equation with
multiple hidden layers in the neural network.
Figures \ref{fg_14} and \ref{fg_15} illustrate the ELM/modBIP simulation
results obtained using neural networks containing
$3$ hidden layers, with an architecture
$[2, 50, 50, 500, 1]$, and $5$ hidden layers, with an architecture
$[2, 50, 50, 50, 50, 500, 1]$, respectively.
The activation function is $\tanh$ in the hidden layers.
In these tests the initial random coefficients are generated on
$[-R_m,R_m]$ with $R_m$ either fixed at $R_m=10$ or varied between
$R_m=0.1$ and $R_m=100$.
We employ $Q=Q_1\times Q_1$ uniform collocation points, where $Q_1$
is fixed at $Q_1=25$ or varied between $Q_1=5$ and $Q_1=50$.
In modBIP we employ $S_c=S_b/2$ and determine $S_b$ based on
the procedure from Remark~\ref{rem_1}.

Figure \ref{fg_14}(a) shows the LLSQ residual norms for estimating
$S_b$, suggesting a value
around $S_b\approx 1$ for modBIP with three hidden layers in the neural network.
Figures \ref{fg_14}(b) to (d) show the error distribution, and the maximum/rms
errors in the domain of the ELM/modBIP solution as a function of $R_m$
and $Q_1$, obtained with $S_b=1$ in modBIP.
The specific parameter values for each plot
are provided in the caption of this figure.
Figure \ref{fg_15}(a) shows the LLSQ residual norms computed with
$5$ hidden layers in the neural network, suggesting a value
around $S_b\approx 0.8$ in this case.
Figures \ref{fg_15}(b,c,d) show the ELM/modBIP results obtained with $5$ hidden
layers in the neural network and  $S_b=0.8$ in modBIP,
which correspond to those of Figures \ref{fg_14}(b,c,d).
These results confirm that the combined ELM/modBIP method
produces accurate simulation results with multiple
hidden layers in the neural network.
The characteristics of exponential convergence and
insensitivity to $R_m$ are similar to what has been observed with
single-hidden-layer neural networks.

\begin{figure}
  \centerline{
    \includegraphics[width=2.0in]{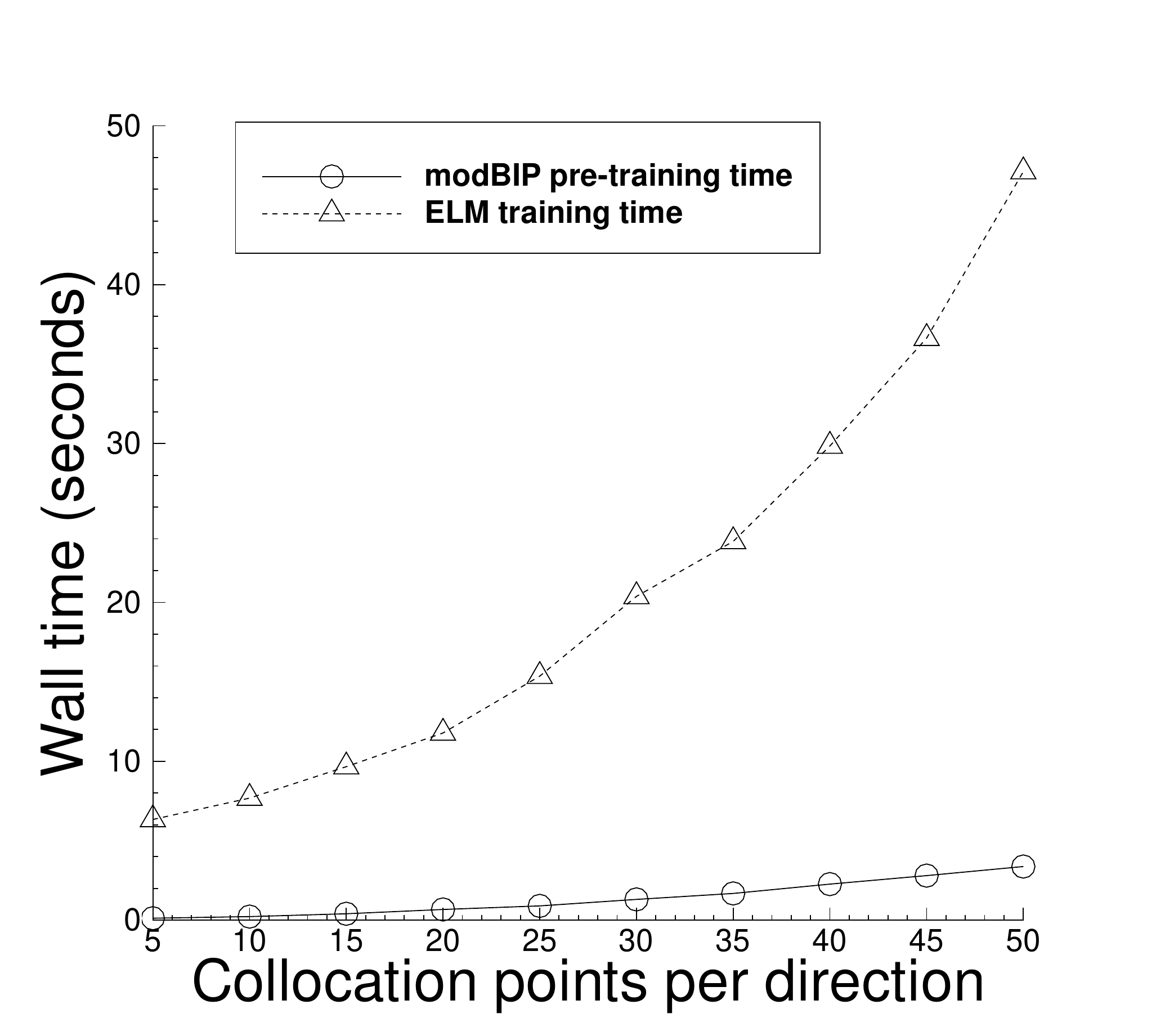}(a)
    \includegraphics[width=2.0in]{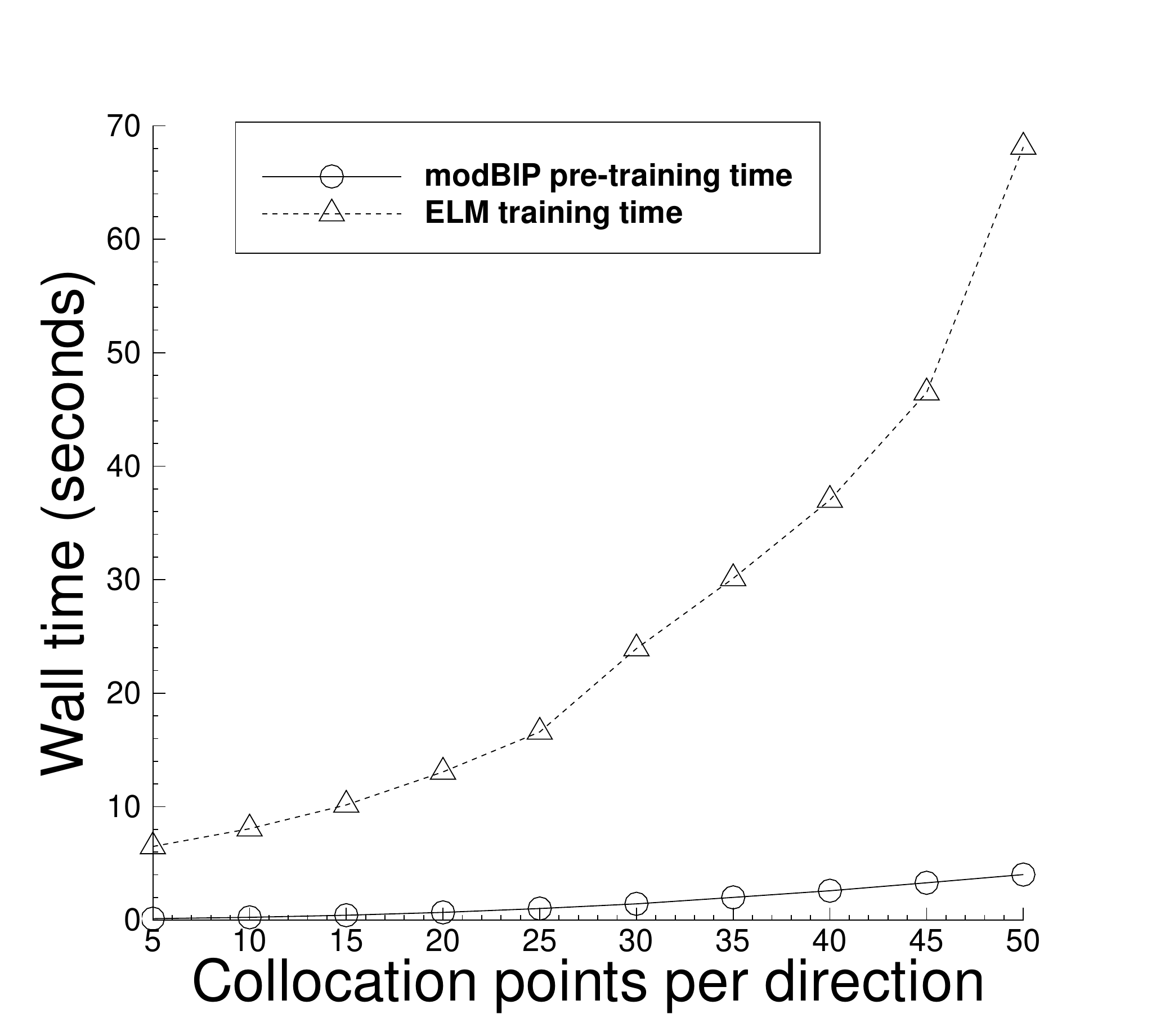}(b)
  }
  \caption{Poisson equation: The modBIP pre-training time of the random coefficients
    and the ELM training time of the neural network as a function of
    the number of collocation points in each direction,
    obtained with $3$ hidden layers (a) and $5$ hidden layers (b)
    in the neural network.
    The network architecture of (a) is
    $[2, 50, 50, 500, 1]$ and the parameter values correspond to those of
    Figure \ref{fg_14}(d).
    The network architecture of (b) is  $[2, 50, 50, 50, 50, 500, 1]$ and
    the parameter values correspond to those of Figure \ref{fg_15}(d).
  }
  \label{fg_15_a}
\end{figure}

Figure \ref{fg_15_a} is a study of the computational cost
of the modBIP pre-training of the random coefficients and the ELM training
of the neural network with multiple hidden layers in the neural network.
It shows the modBIP pre-training time and the ELM network training time
as a function of the number of collocation points in each direction,
corresponding to $3$ and $5$ hidden layers in the neural network.
The network architecture and the parameter values in Figures \ref{fg_15_a}(a)
and (b) are identical to those of Figure \ref{fg_14}(d) and
Figure \ref{fg_15}(d), respectively.
In the range of collocation points tested here,
the modBIP pre-training cost is approximately $2\sim 8\%$ of
the ELM network training cost with $3$ hidden layers,
and approximately $2\sim 7\%$ of the ELM network training cost
with $5$ hidden layers in the neural network.

\subsection{Wave Equation}
\label{sec:wave}

\begin{figure}
  \centerline{
    \includegraphics[height=1.9in]{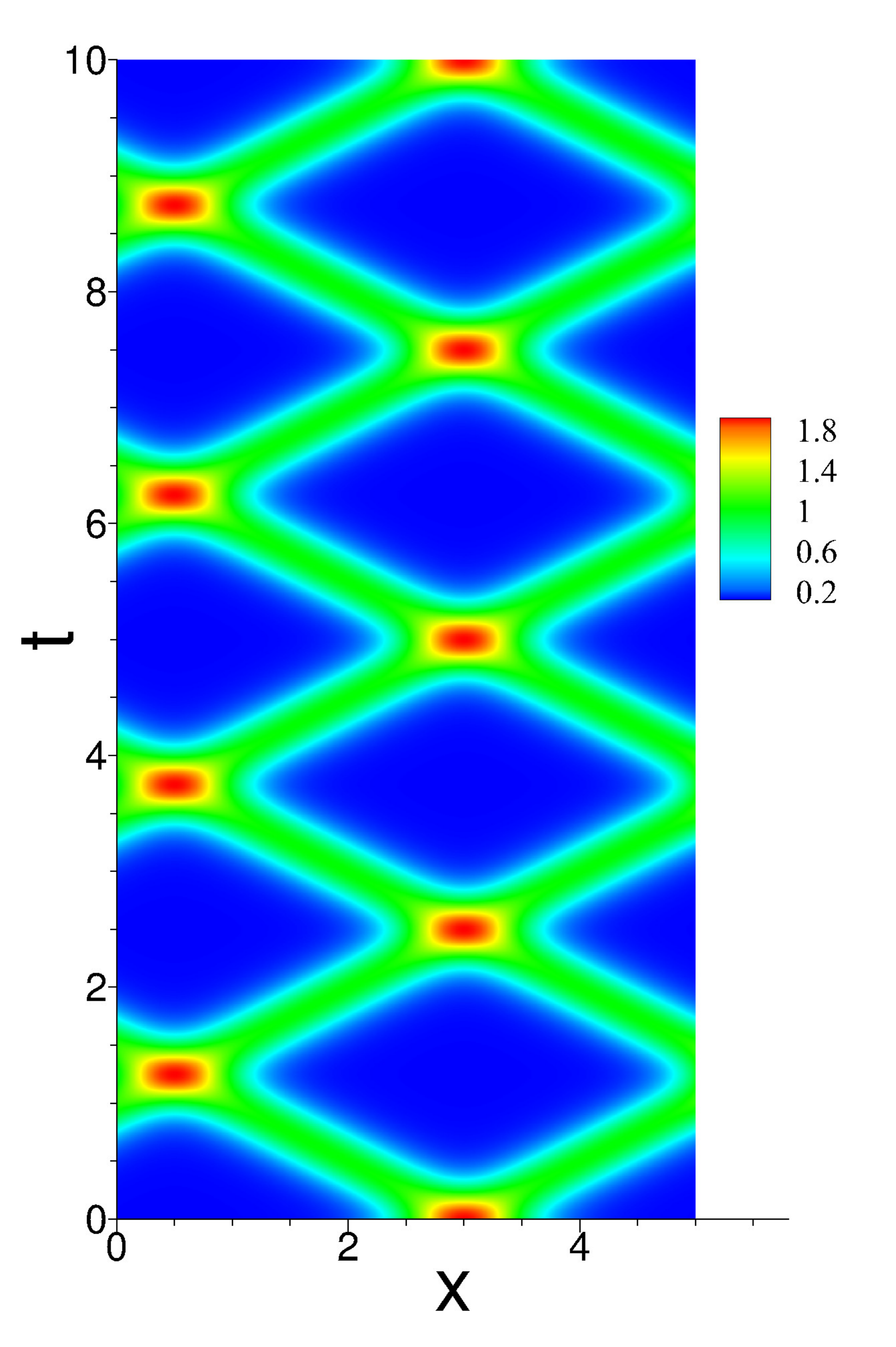}(a)
    \includegraphics[height=2.in]{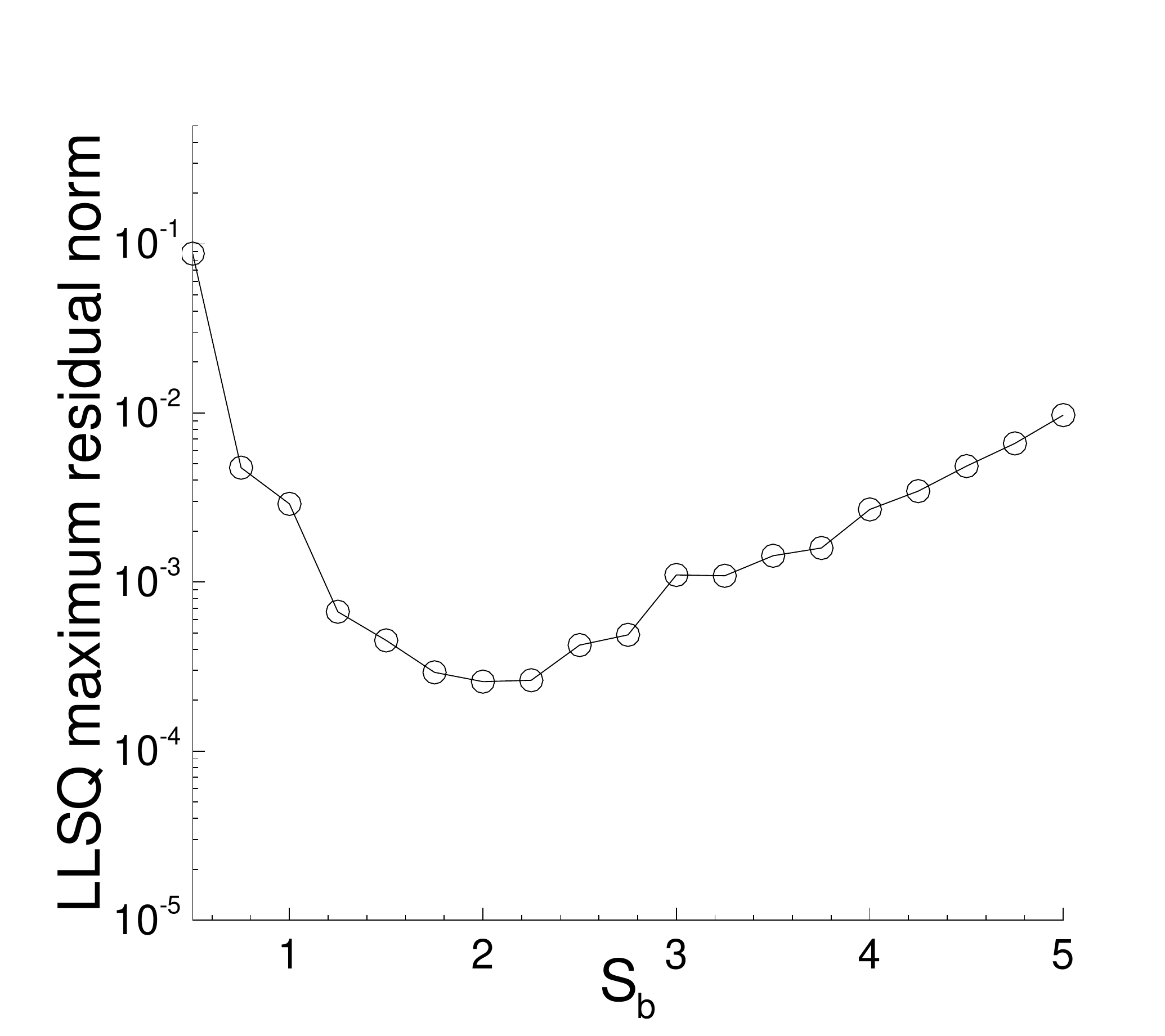}(b)
    \includegraphics[height=2.in]{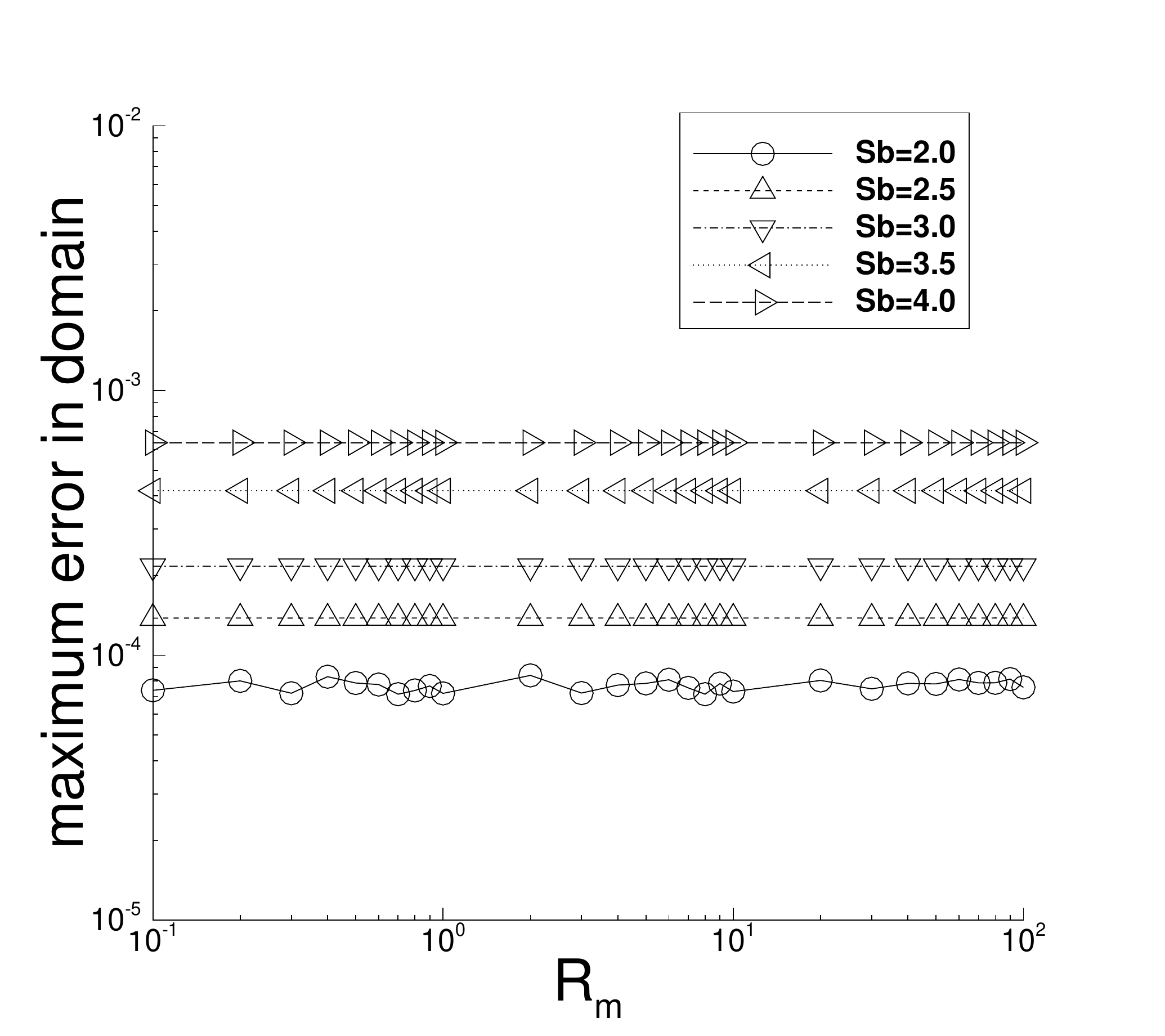}(c)
  }
  \caption{Wave equation:
    (a) Distribution of the exact solution in the spatial-temporal plane.
    (b) The maximum residual norm of the LLSQ problem versus 
    $S_b$, for estimating the best $S_b$ in modBIP.
    (c) The maximum error in the overall domain as a function of $R_m$,
    corresponding to several $S_b$ values in modBIP
    for pre-training the random coefficients.
  }
  \label{fg_16}
\end{figure}

We next test the ELM/modBIP method using the
one-dimensional second-order wave equation (plus time).
Consider the spatial-temporal domain,
$\Omega=\{(x,t)\ |\ x\in[0,5],\ t\in[0,10]  \}$, and
the initial/boundary-value problem with the wave equation on this domain,
\begin{subequations}\label{eq_16}
  \begin{align}
    &
    \frac{\partial^2 u}{\partial t^2} - c^2\frac{\partial^2 u}{\partial x^2} = 0, \\
    &
    u(0,t) = u(5,t), \\
    &
    \frac{\partial}{\partial x}u(0,t) = \frac{\partial}{\partial x}u(5,t), \\
    &
    u(x,0) = 2\sech^3\left[\frac{3}{\delta_0}(x-x_0) \right], \\
    &
    \left.\frac{\partial u}{\partial t}\right|_{(x,0)} = 0,
  \end{align}
\end{subequations}
where $u(x,t)$ is the field solution to be solved for, periodic boundary conditions are
imposed on $x=0$ and $5$,
$c$ is the wave speed,
$x_0$ is the initial peak location of the wave, and the constant $\delta_0$ controls
the width of the wave profile.
The constant parameters assume the following values for this problem:
\begin{equation*}
  c = 2, \quad
  \delta_0 = 2, \quad
  x_0 = 3.
\end{equation*}
This problem has the following solution,
\begin{equation}
  \left\{
  \begin{split}
    &
    u(x,t) = \sech^3\left[\frac{3}{\delta_0}\left(-\frac52+\xi \right) \right]
    + \sech^3\left[\frac{3}{\delta_0}\left(-\frac52+\eta \right) \right], \\
    &
    \xi = \text{mod}\left(x-x_0+ct+\frac52,5  \right), \quad
    \eta = \text{mod}\left(x-x_0-ct+\frac52,5 \right),
  \end{split}
  \right.
\end{equation}
where mod refers to the modulo operation. The two terms in this solution represent
the leftward- and rightward-traveling waves, respectively.
Figure \ref{fg_16}(a) shows the distribution of this solution in
the spatial-temporal plane.


To simulate this problem, we employ the block time-marching scheme and the local
extreme learning machines (locELM) developed in~\cite{DongL2020}. 
We first divide the spatial-temporal domain $\Omega$ along the temporal
direction into $20$ uniform time blocks, and the system \eqref{eq_16}
is computed on each time block individually and successively (see Remark~\ref{rem_6}).
We partition the spatial-temporal domain of each time block
into $4$ uniform sub-domains along the $x$ direction, and represent
the field solution $u(x,t)$ on each sub-domain by a local
feed-forward neural network~\cite{DongL2020}.
$C^1$ continuity conditions are imposed on
the sub-domain boundaries. The configuration of the local neural
networks follows that of the ELM. The weight/bias coefficients in
the hidden layers of the local neural networks are initialized as uniform
random values generated on $[-R_m,R_m]$, which are pre-trained by
modBIP and then fixed afterwards. The output layers of the local neural networks
are linear (with zero bias), and the weight coefficients therein are the training
parameters and can be determined by
a least squares computation. The local neural networks are coupled with
one another due to the $C^1$ continuity conditions~\cite{DongL2020},
and need to be trained together as a whole system.
By enforcing the system of equations, boundary/initial conditions,
and the $C^1$ continuity conditions on a set of collocation points inside
each sub-domain and on the domain and sub-domain boundaries, we arrive
at a system of linear algebraic equations about the training parameters,
which can be solved by the linear least squares method. We refer to~\cite{DongL2020}
for more detailed discussions of the locELM method and the block
time marching scheme.

After the random coefficients in the local neural networks are initialized,
we use modBIP to pre-train the random coefficients in each local neural network
(see Algorithm \ref{alg_1}), and then employ the pre-trained
random coefficients in the locELM computation for the training parameters
based on the linear least squares method. We refer to the overall method
as the combined locELM/modBIP method.

We first consider a single hidden layer in the local neural networks,
whose architecture each is characterized by $[2, 250, 1]$ and with $\tanh$
as the activation function for the hidden layers. The two nodes in the input
layer represent the spatial-temporal coordinates $x$ and $t$, and the single
node in the output layer represents the field function $u(x,t)$ on the
corresponding sub-domain.
We employ $Q=25\times 25$
uniform collocation points on each sub-domain, and $S_c=S_b/2$ in
the modBIP algorithm. $S_b$ in modBIP is determined by the procedure
from Remark~\ref{rem_1}.

Figure \ref{fg_16}(b) shows the maximum residual norms, among the $20$
time blocks, of the linear least squares (LLSQ) problem as a function
of $S_b$, for estimating the best $S_b$ in modBIP.
In this set of tests, the initial random coefficients are generated
with $R_m=50$. The data suggest a value around $S_b\approx 2$ for modBIP.
Figure \ref{fg_16}(c) shows the maximum error in the entire spatial-temporal
domain of the locELM/modBIP solution as a function of $R_m$,
obtained with several $S_b$ values in modBIP.
We observe the familiar insensitivity of the locELM/modBIP error with
respect to $R_m$ (or the initial random coefficients) in the neural network.

\begin{figure}
  \centerline{
    \includegraphics[width=1.5in]{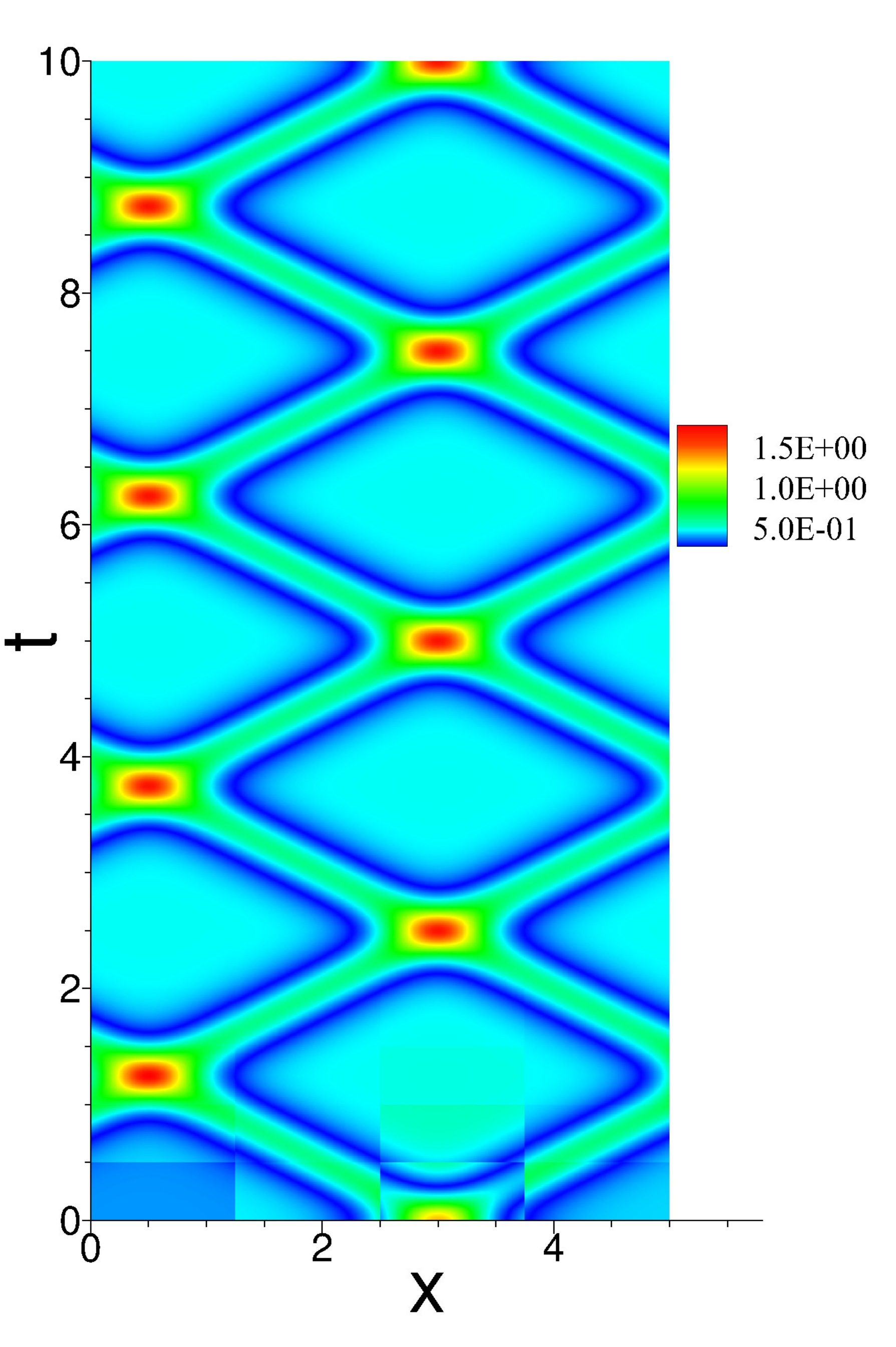}(a)
    \includegraphics[width=1.5in]{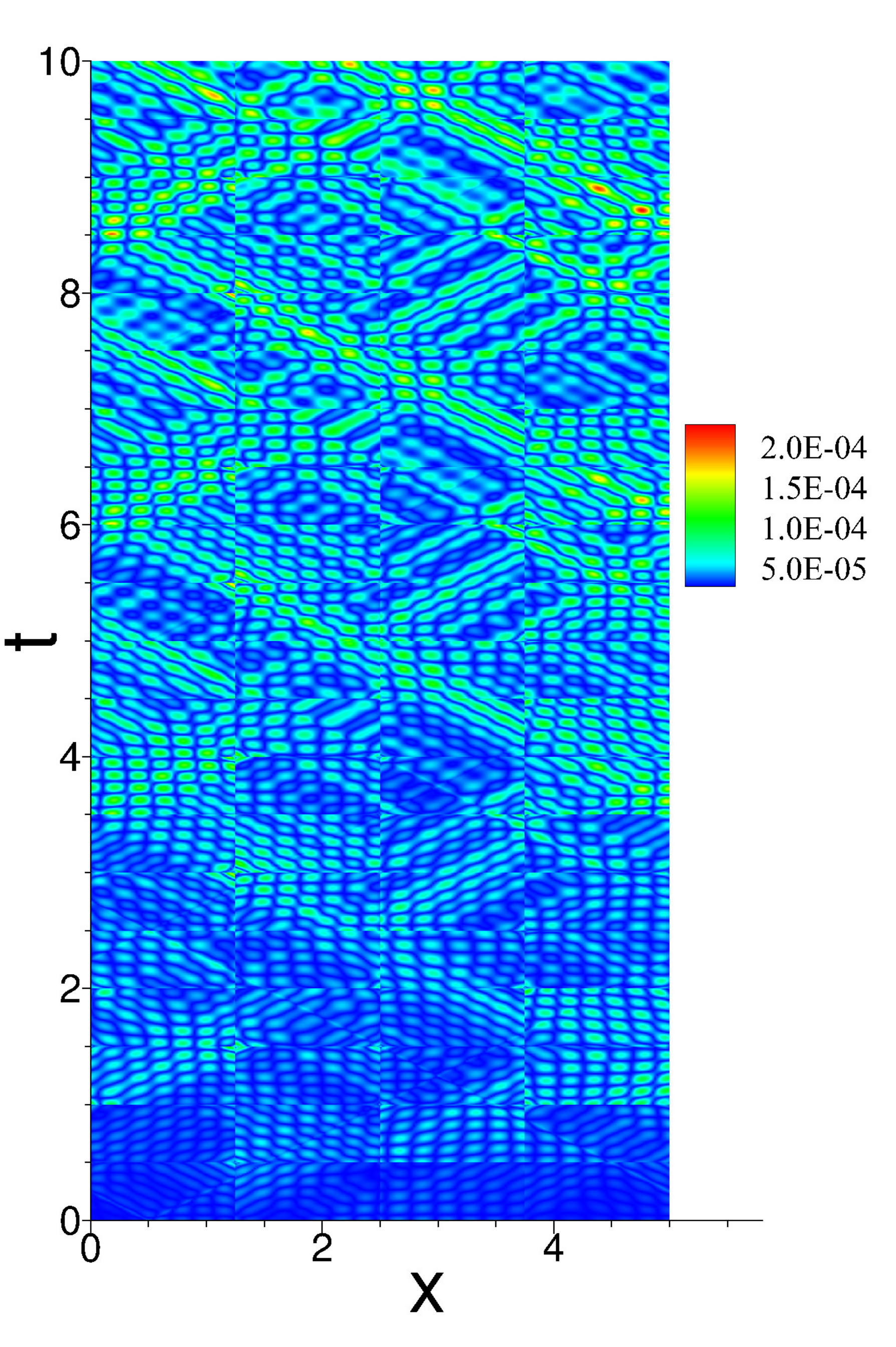}(b)
    \includegraphics[width=1.5in]{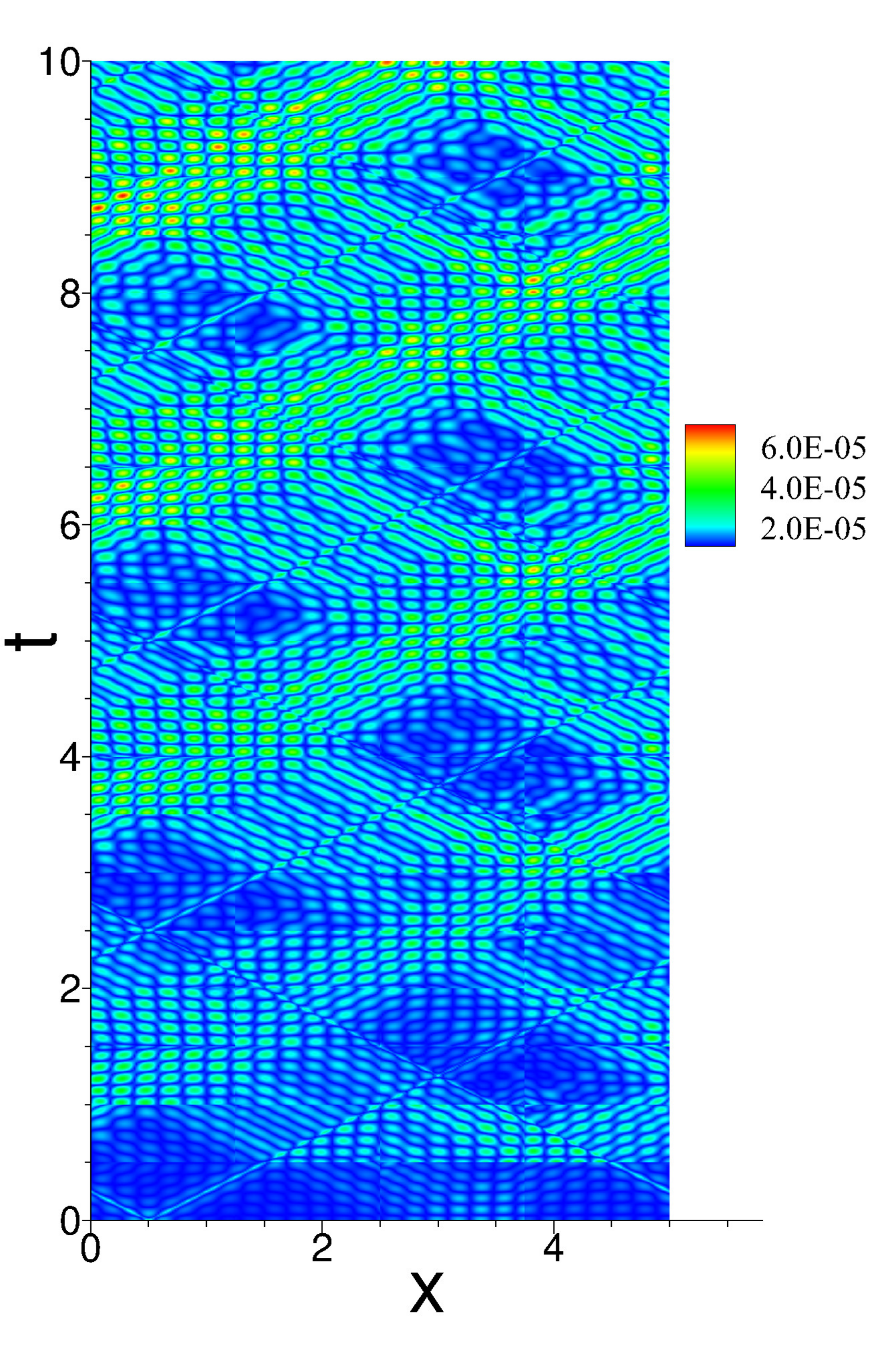}(c)
  }
  \caption{Wave equation:
    Distributions of the absolute error of the locELM solution obtained with
    (a) no pre-training, (b) BIP pre-training, and (c) modBIP pre-training
    of the random coefficients.
  }
  \label{fg_17}
\end{figure}

Figure \ref{fg_17} compares distributions in the spatial-temporal plane
of the absolute error of the locELM
solution obtained with no pre-training, and with BIP and modBIP pre-training
of the random coefficients in the local neural networks.
In this set of tests, we employ local neural networks with an architecture
$[2, 250, 1]$, the $\tanh$ activation function for the hidden layers,
and $Q=31\times 31$ uniform collocation points on each sub-domain.
The initial random coefficients in the local neural networks are generated
with $R_m=50$. With BIP, we generate target samples
by a normal distribution with a random mean from $[-1,1]$ and a standard deviation
$0.5$~\cite{NeumannS2013}. With modBIP we employ $S_b=2$ and $S_c=S_b/2$
in Algorithm~\ref{alg_1}. The locELM solution obtained without
pre-training of the random coefficients generated by $R_m=50$
exhibits no accuracy for this problem (Figure \ref{fg_17}(a)).
On the other hand, the locELM solutions obtained with BIP and modBIP
pre-training are quite accurate, and the combined locELM/modBIP
solution is observed to be
notably more accurate than the locELM/BIP one (with error levels
$10^{-5}$ versus $10^{-4}$).

\begin{figure}
  \centerline{
    \includegraphics[width=2in]{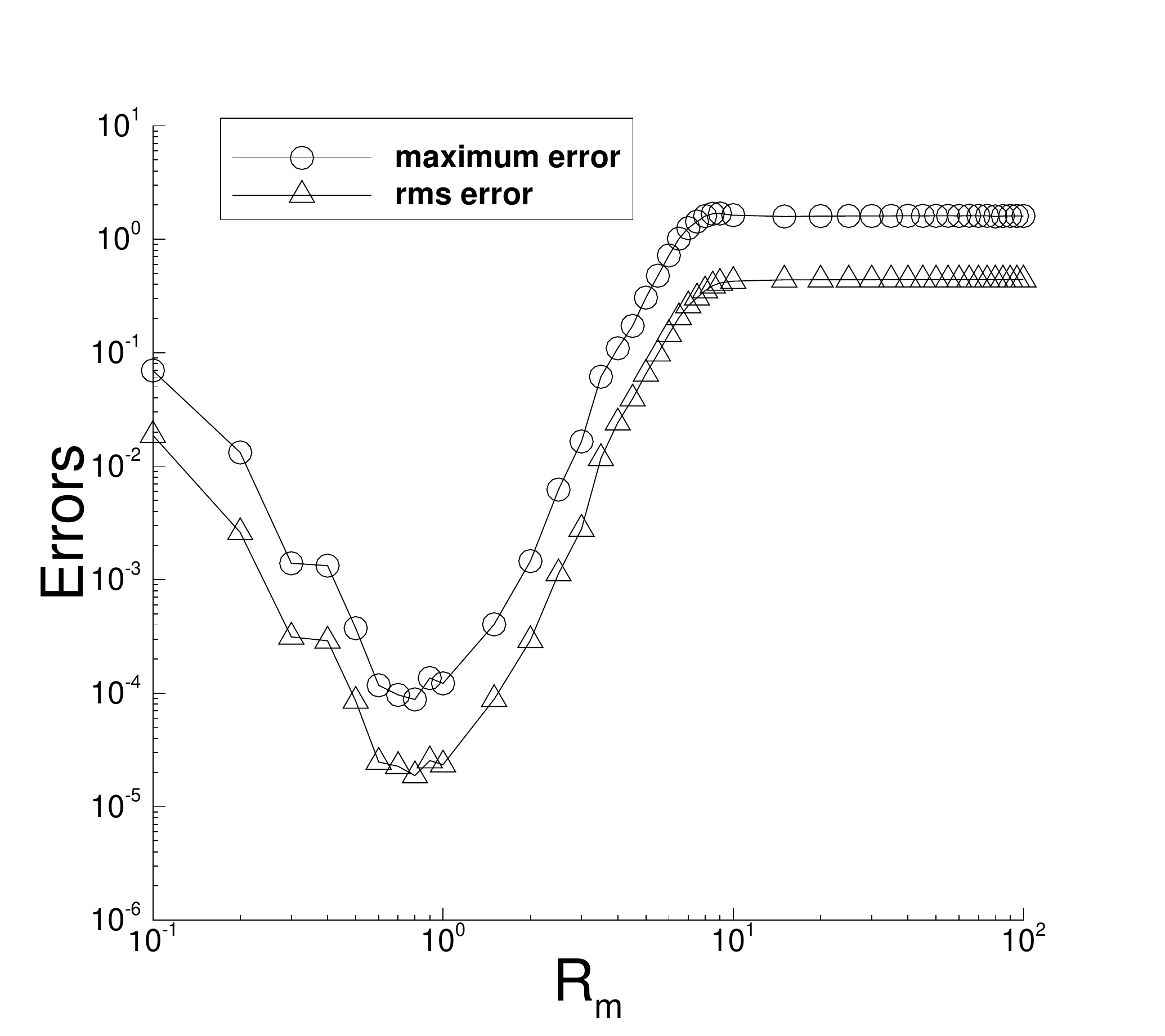}(a)
    \includegraphics[width=2in]{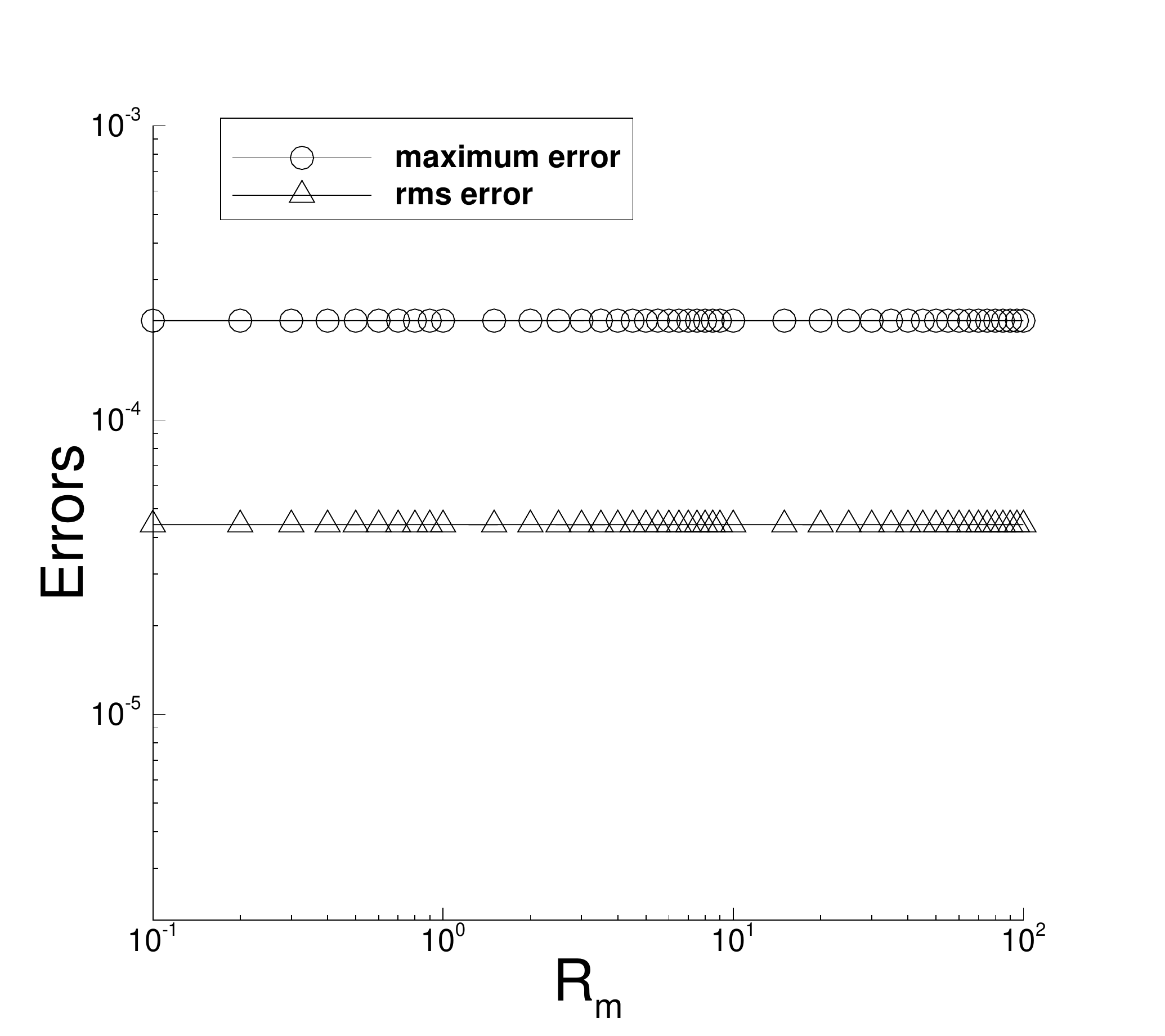}(b)
    \includegraphics[width=2in]{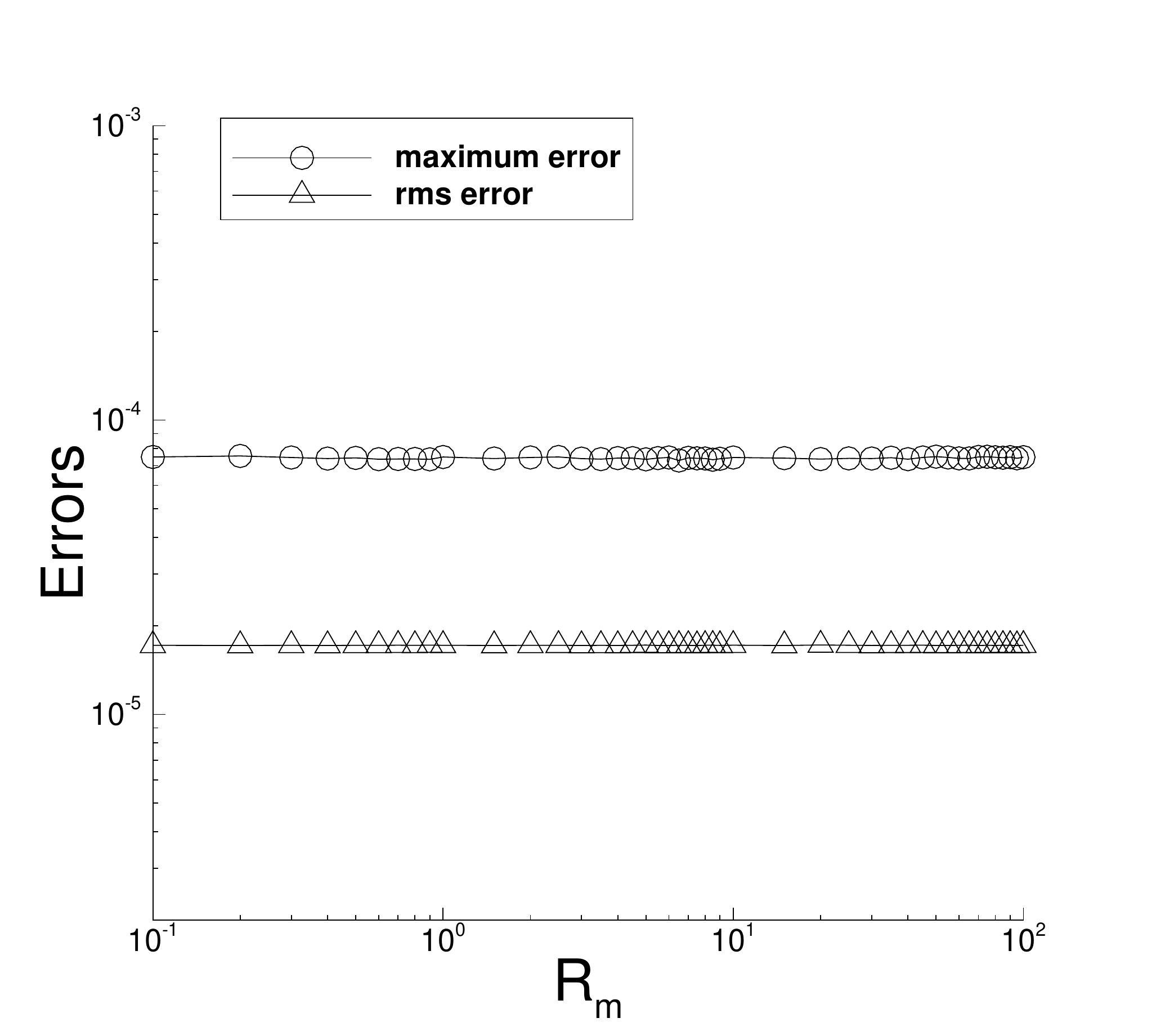}(c)
  }
  \caption{Wave equation: Maximum and rms errors of the locELM solution in the domain
    as a function of $R_m$, obtained with (a) no pre-training,
    (b) BIP pre-training, and (c) modBIP pre-training of the random coefficients.
  }
  \label{fg_18}
\end{figure}

Figure \ref{fg_18} is a further comparison of the locELM methods
with no pre-training and with BIP and modBIP pre-training of the random
coefficients.
Plotted here are the maximum and rms errors in the overall domain
of the locELM solution obtained with these three cases
as a function of $R_m$ for generating the initial random coefficients.
The local neural-network architecture and related parameters,
the collocation points, and the parameters for the BIP and modBIP algorithms
are the same as those for Figure \ref{fg_17},
except that here the $R_m$ is varied systematically between $R_m=0.1$
and $R_m=100$ in this set of computations.
Without pre-training of the initial random coefficients, one
can observe a strong influence of $R_m$ on the locELM solution accuracy
(Figure \ref{fg_18}(a)).
On the other hand, the pre-training of the random coefficients by either
modBIP or BIP essentially eliminates the dependence of the
solution error on $R_m$ (i.e.~the initial random coefficients).
The combined locELM/modBIP method is again observed to be
more accurate than locELM/BIP.


\begin{figure}
  \centerline{
    \includegraphics[height=2.0in]{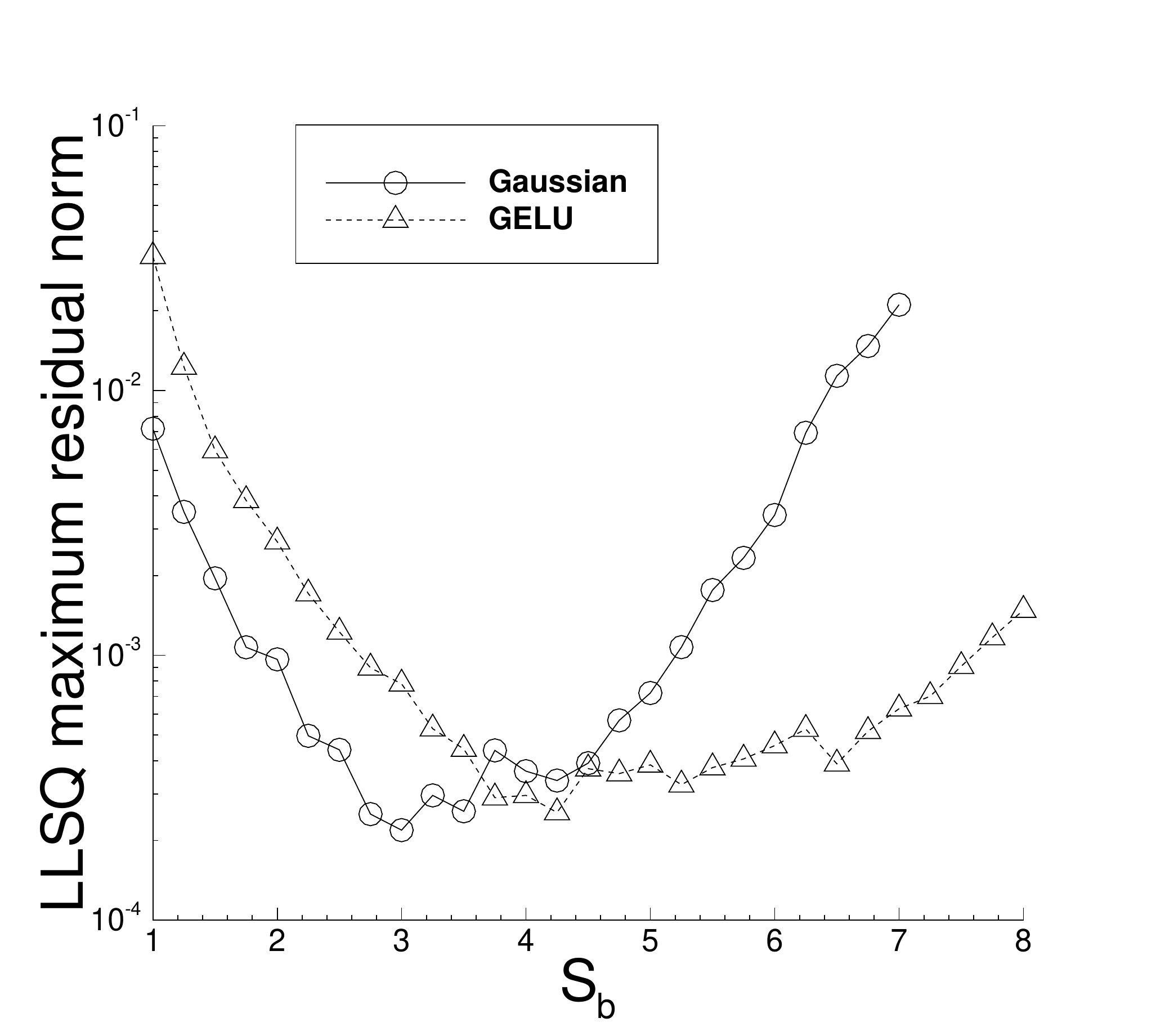}(a)
    \includegraphics[height=1.9in]{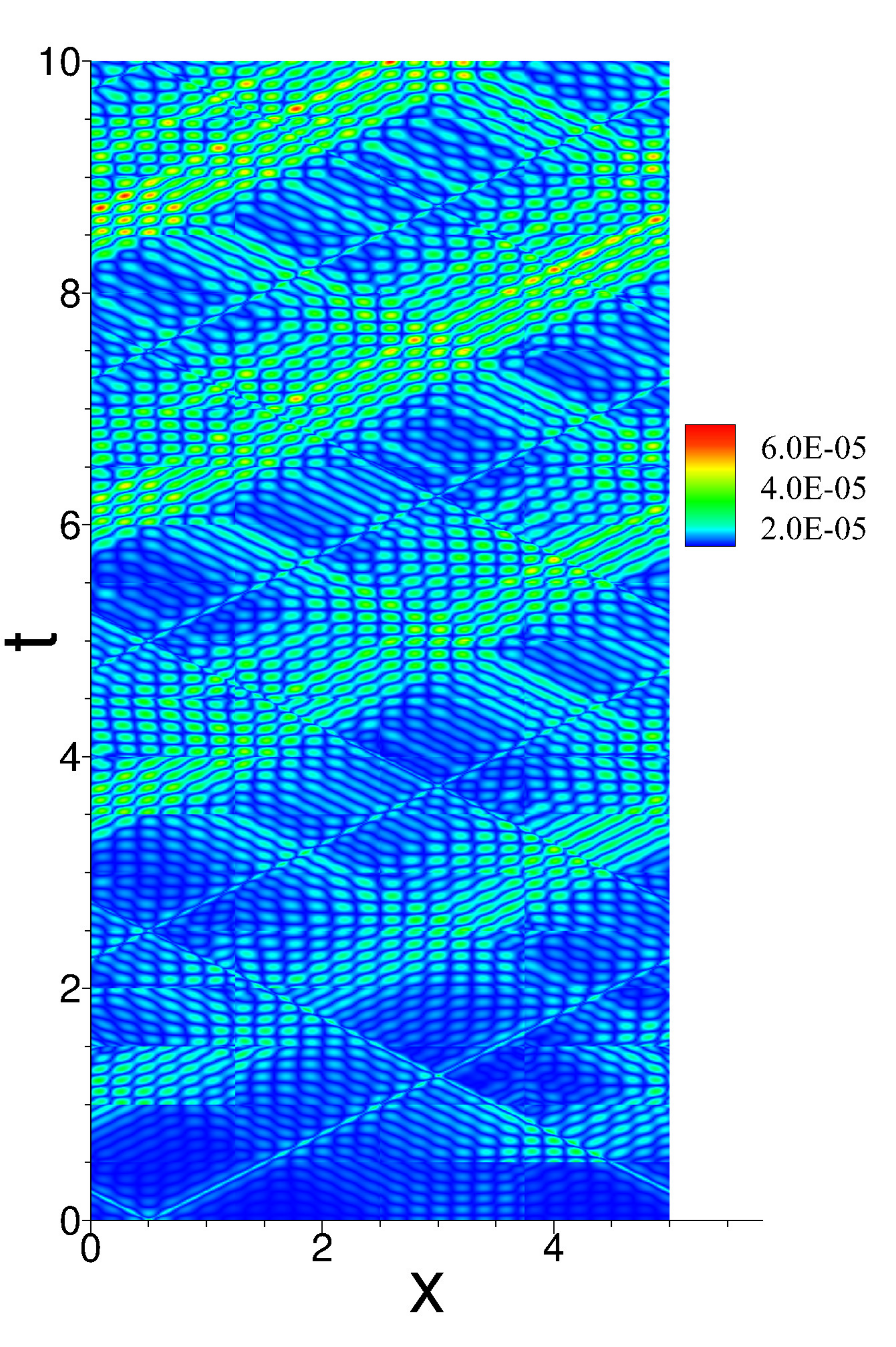}(b)
    \includegraphics[height=1.9in]{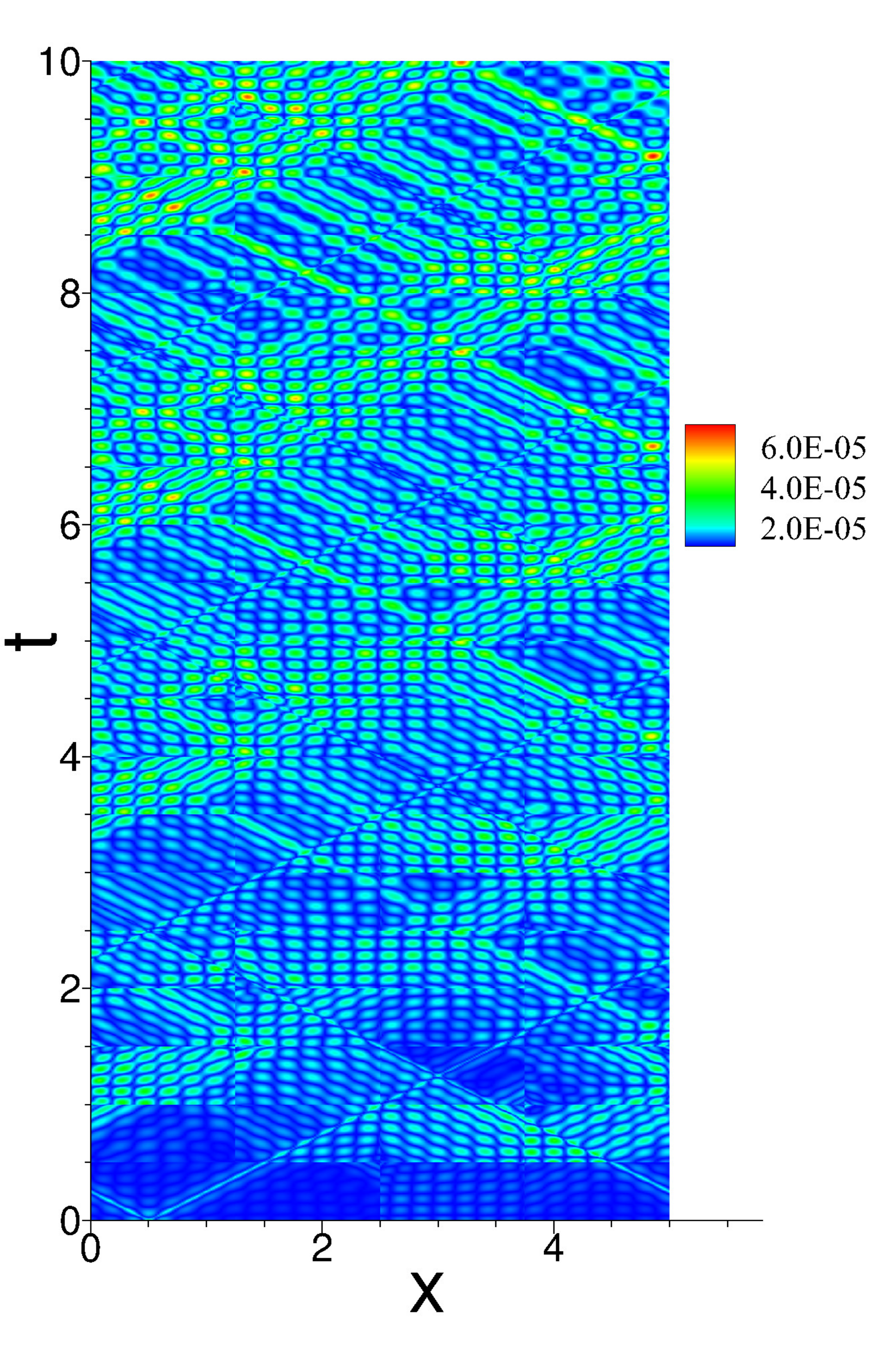}(c)
  }
  \centerline{
    \includegraphics[width=1.5in]{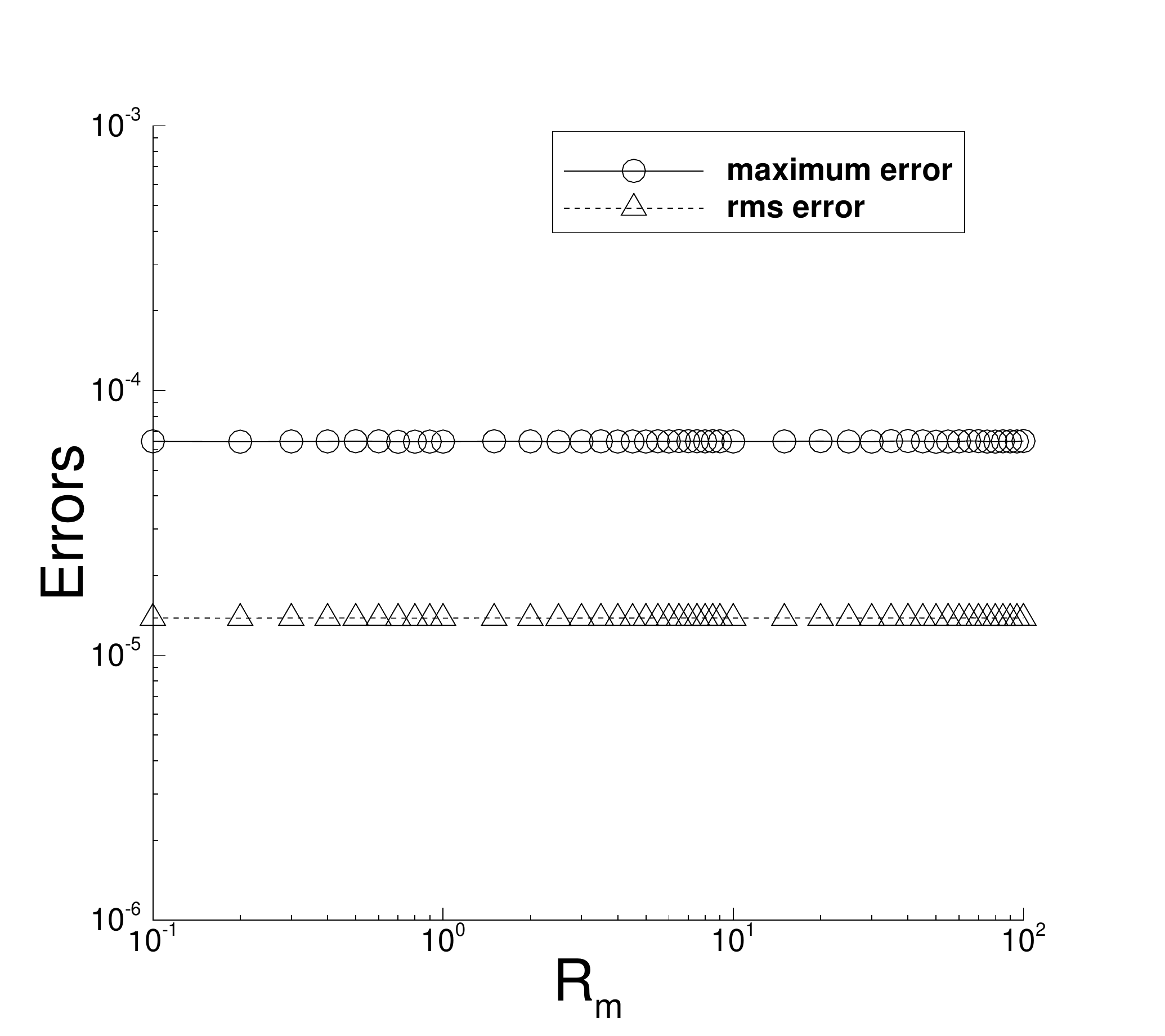}(d)
    \includegraphics[width=1.5in]{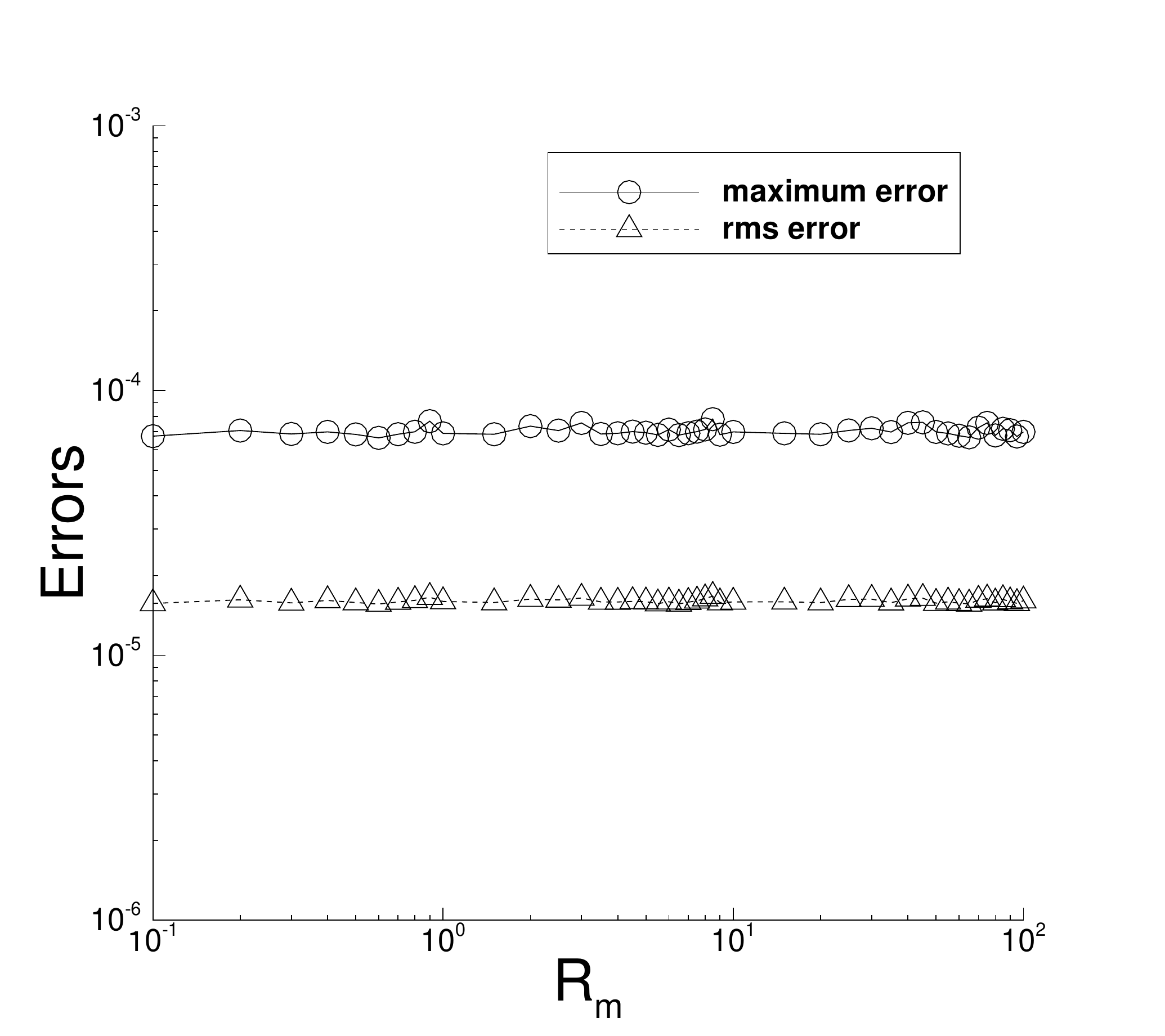}(e)
  }
  \caption{Wave equation (non-monotonic activation functions):
    (a) The LLSQ maximum residual norm  for estimating $S_b$ in modBIP,
    with the Gaussian and GELU~\cite{HendrycksG2020} activation functions.
    Error distributions of the locELM/modBIP solution
    obtained with (b) the Gaussian, and
    (c) the GELU activation functions.
    The maximum/rms errors of the locELM/modBIP solution as a function of
    $R_m$ obtained using (d) the Gaussian, 
    and (e) the GELU activation functions.
    $S_b=3$ in (b,d) with the Gaussian function, and $S_b=4$ in (c,e)
    with GELU, and $S_b$ is varied in (a).
    $R_m=50$ in (a,b,c) and is varied in (d,e).
  }
  \label{fg_19}
\end{figure}

Figure \ref{fg_19} demonstrates the ability of the current modBIP algorithm
to work with non-invertible activation functions. By contrast, the
BIP algorithm breaks down if such activation functions are present
in the neural network. Specifically, this figure examines the locELM/modBIP
simulation results obtained using the Gaussian function
and the Gaussian error linear unit (GELU)~\cite{HendrycksG2020}
as the activation functions in the hidden layers of the local neural networks.
Here each local neural network has an architecture $[2,250,1]$,
with Gaussian or GELU as the activation function for the hidden layer,
and we employ $Q=31\times 31$ uniform collocation points on each sub-domain.
The initial random coefficients in the local neural networks are
generated with either a fixed $R_m=50$ or with $R_m$ varied between
$R_m=0.1$ and $R_m=100$, and are pre-trained using modBIP with $S_c=S_b/2$.
Figure \ref{fg_19}(a) shows the LLSQ maximum residual norms (among the $20$
time blocks) for estimating the $S_b$ in modBIP,
which suggest a value around $S_b\approx 3$ with the Gaussian activation function
and a value around $S_b\approx 4$ with the GELU activation function.
Figures \ref{fg_19}(b) and (c) show the error distributions
the locELM/modBIP solution obtained using the Gaussian and GELU activation
functions, respectively, with a fixed $R_m=50$ for generating the initial
random coefficients.
Figures \ref{fg_19}(d) and (e) are the maximum and rms errors of the locELM/modBIP
solution in the overall domain as a function of $R_m$,
obtained with the Gaussian and GELU activation functions, respectively.
In the plots (b)-(d), we employ $S_b=3$ with the Gaussian function and
$S_b=4$ with GELU in the modBIP algorithm.
It is evident that the combined locELM/modBIP produces
accurate simulation results with the Gaussian and GELU activation functions,
and that its errors are insensitive to the initial random coefficients.


\begin{figure}
  \centerline{
    \includegraphics[width=2in]{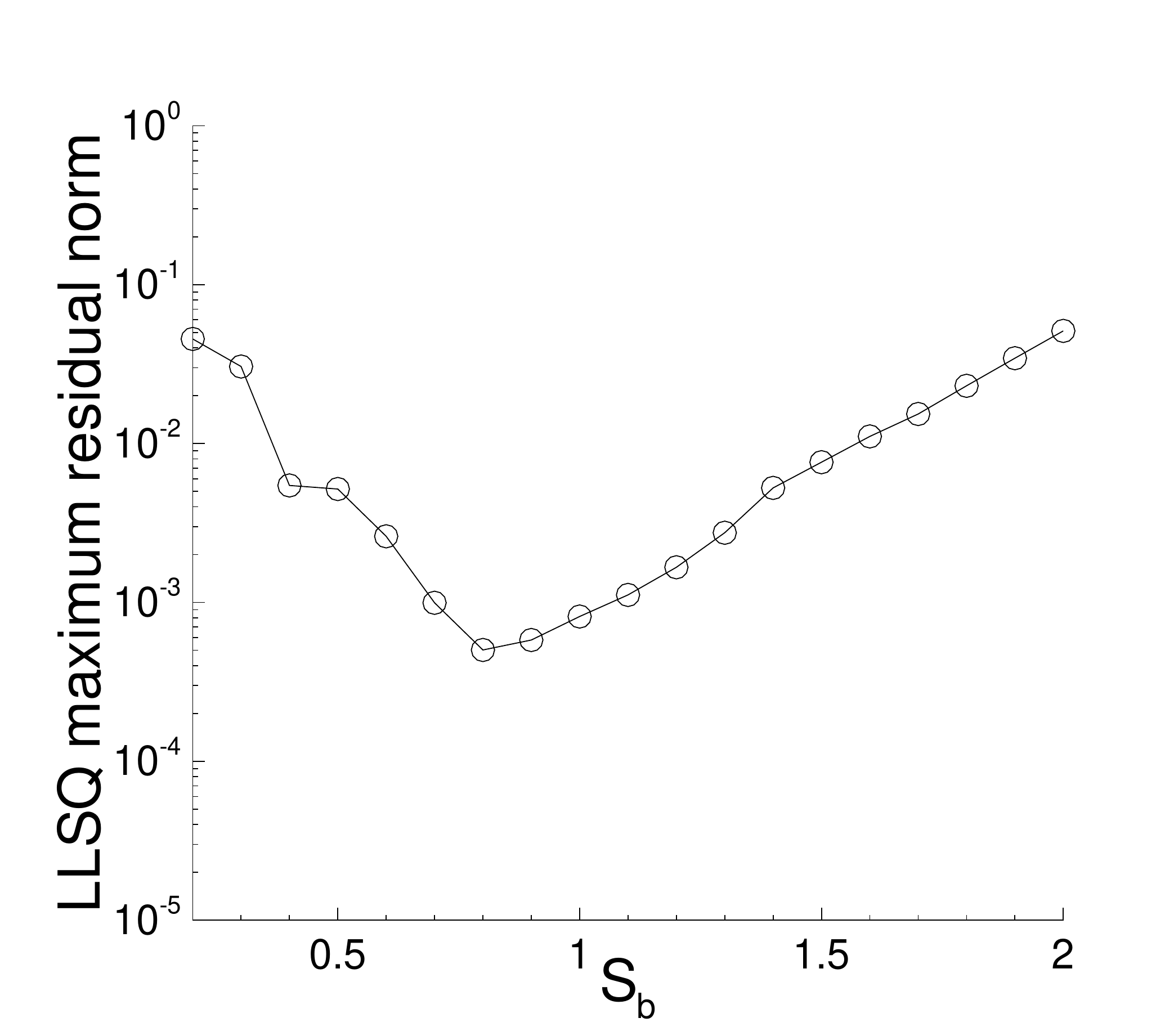}(a)
    \includegraphics[width=2in]{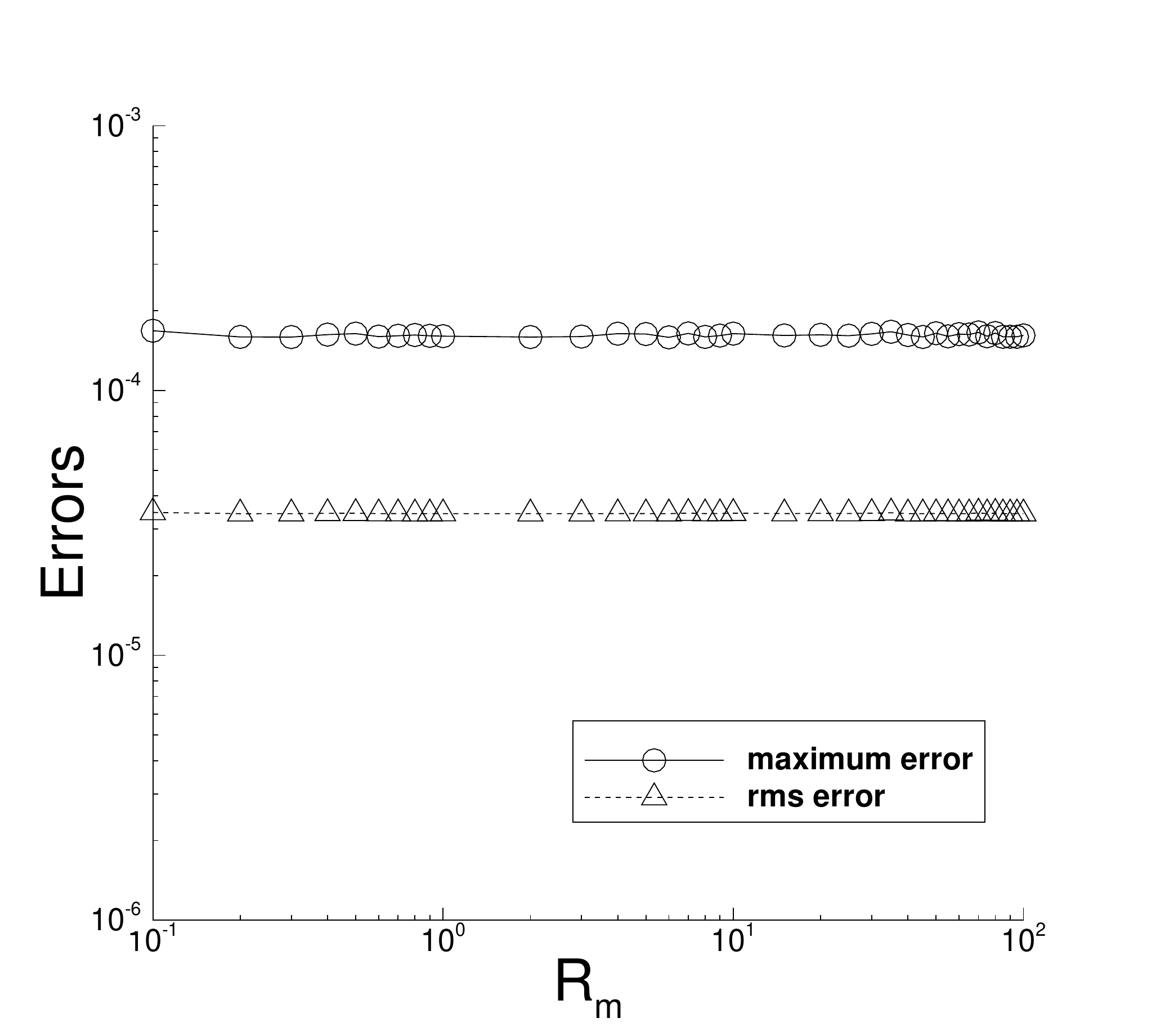}(b)
  }
  \centerline{
    \includegraphics[width=2in]{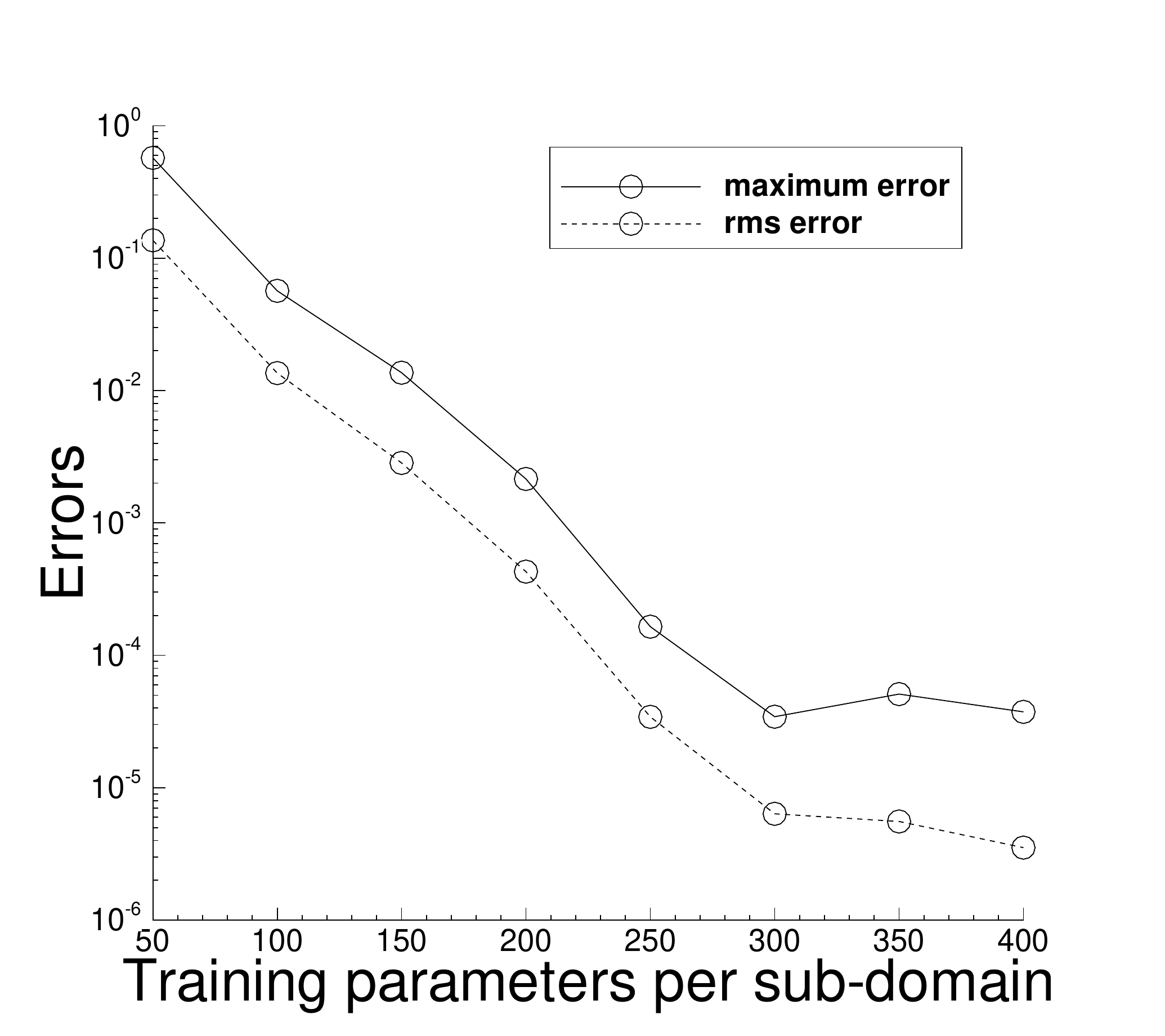}(c)
    \includegraphics[width=2in]{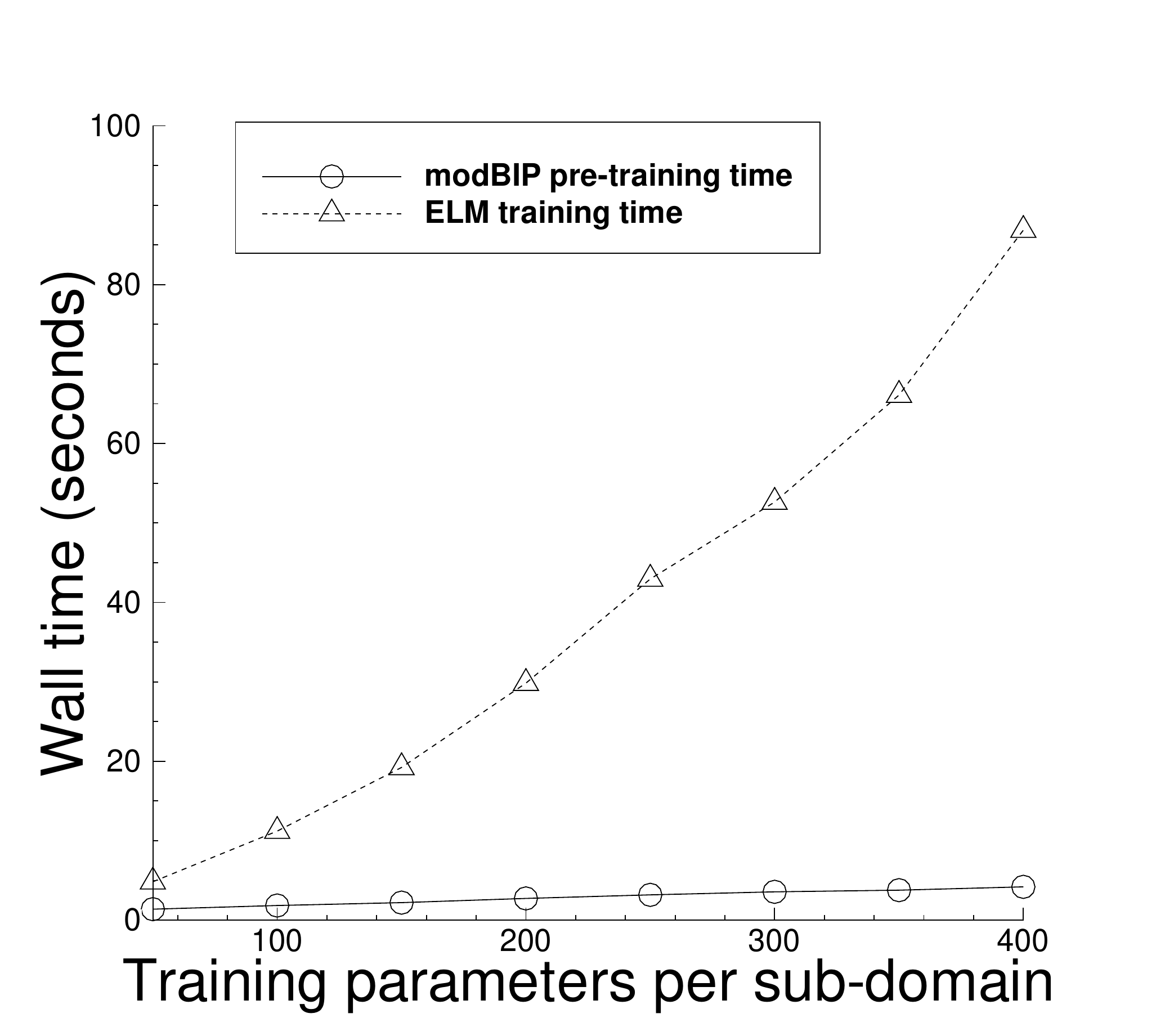}(d)
  }
  \caption{Wave equation (3 hidden layers in neural network):
    (a) The LLSQ maximum residual norm for estimating the best $S_b$ in modBIP.
    The maximum and
    rms errors of the locELM/modBIP solution as a function of (b) $R_m$, 
    and (c) the number of training parameters
    per sub-domain.
    (d) The modBIP pre-training time and the ELM network training time
    as a function of the number of training parameters per sub-domain.
    $S_b=0.8$ in (b,c,d), and is varied in (a).
    $R_m=50$ in (a,c,d), and is varied in (b).
    Local neural network architecture
    is $[2, 50, 50, M, 1]$, with $M=250$ in (a,b) and varied in (c,d).
    $Q=31\times 31$ in (a,b,c,d).
  }
  \label{fg_20}
\end{figure}

Finally, Figure \ref{fg_20} is an illustration of the locELM/modBIP results
with $3$ hidden layers in the local neural networks.
In this group of tests, we employ an architecture $[2, 50, 50, M, 1]$ in
all the local neural networks with the $\tanh$ activation function
for the hidden layers, where $M$ is either fixed at $M=250$
or varied between $M=50$ and $M=400$.
We again employ $20$ uniform time blocks, $4$ uniform sub-domains within
each time block, and a set of $Q=31\times 31$ uniform collocation
points on each sub-domain.
The initial random coefficients are generated with $R_m$ either fixed
at $R_m=50$ or varied between $R_m=0.1$ and $R_m=100$.
Figure \ref{fg_20}(a) shows the LLSQ maximum residual norm for estimating
$S_b$, suggesting a value around $S_b\approx 0.8$ in modBIP.
Figures \ref{fg_20}(b) and (c) show the maximum and rms
errors in the overall domain of the locELM/modBIP solution as
a function of $R_m$ and the number of training parameters per sub-domain $M$,
respectively. The results signify the insensitivity of the locELM/modBIP
errors with respect to the initial random coefficients,
and the exponential decrease in the errors with increasing training parameters
in the neural network.
Figure \ref{fg_20}(d) shows the modBIP pre-training time of the random coefficients and
the ELM training time of the neural network as a function of
the number of training parameters per sub-domain.
While both the modBIP pre-training time
and the ELM training time grows with increasing number of training parameters,
the growth rate of the modBIP pre-training time is much slower
than the ELM training time.
For example, as the number of training parameters per sub-domain
increases from $50$ to $400$,
the modBIP pre-training time increases from around $1.4$ seconds
to about $4.2$ seconds, while the ELM training time increases
from about $4.8$ seconds to about $87$ seconds.

\subsection{Diffusion Equation}
\label{sec:diffu}

\begin{figure}
  \centerline{
    \includegraphics[height=2.1in]{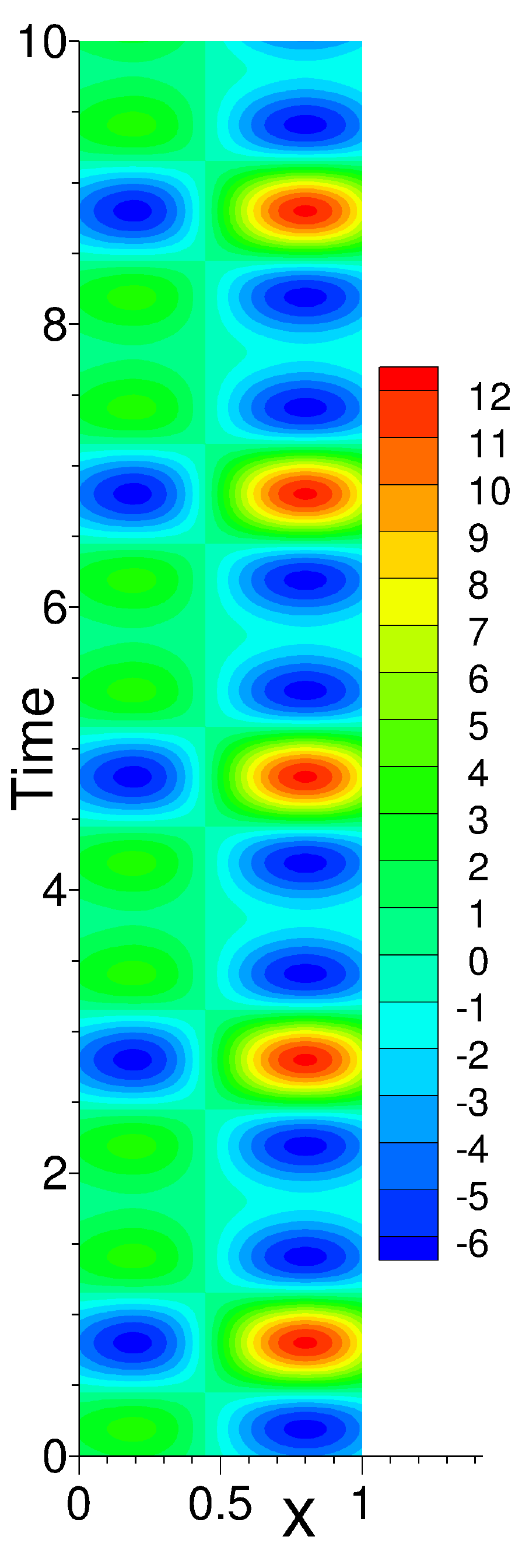}(a)
    \includegraphics[height=2.2in]{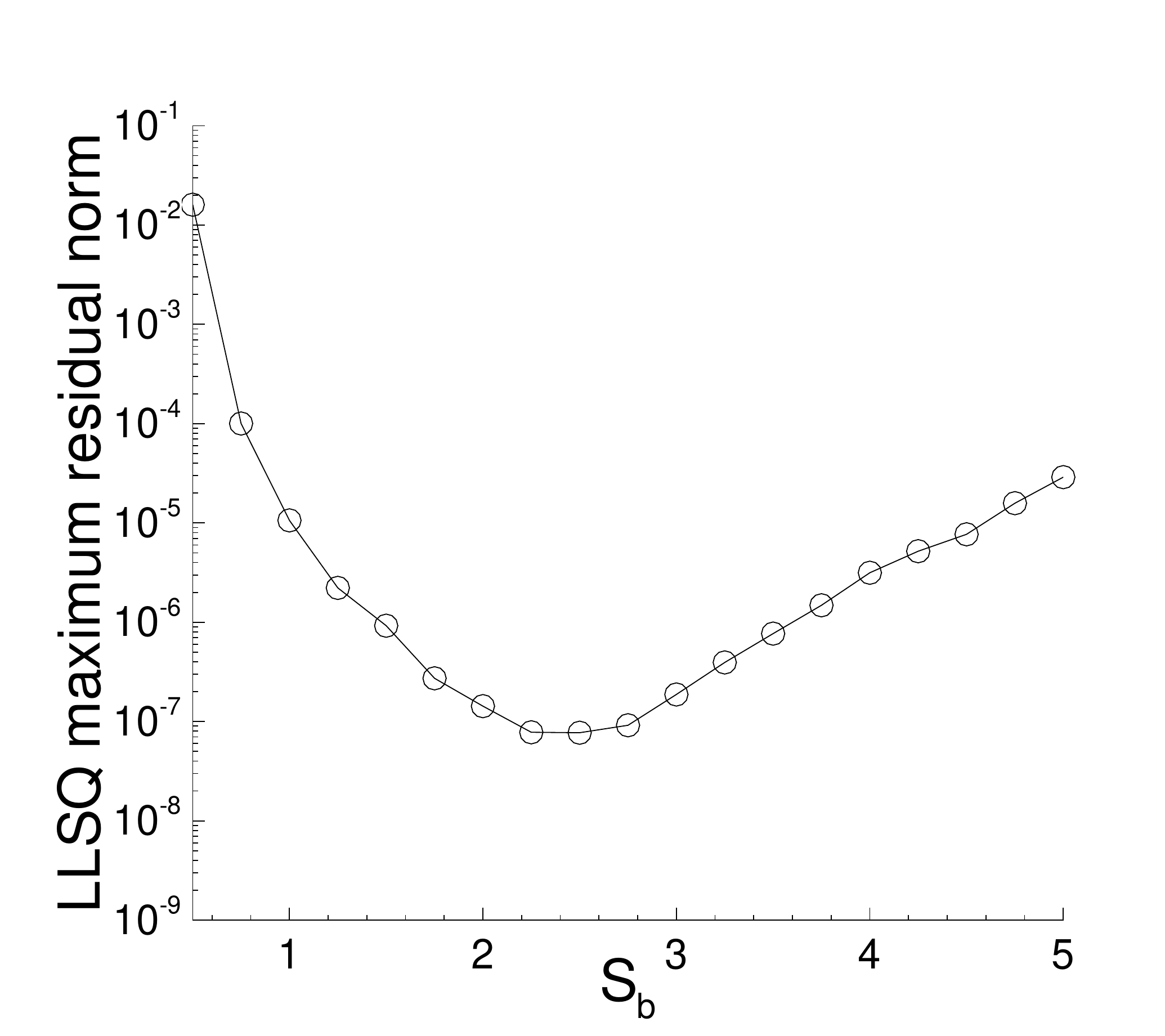}(b)
    \includegraphics[height=2.2in]{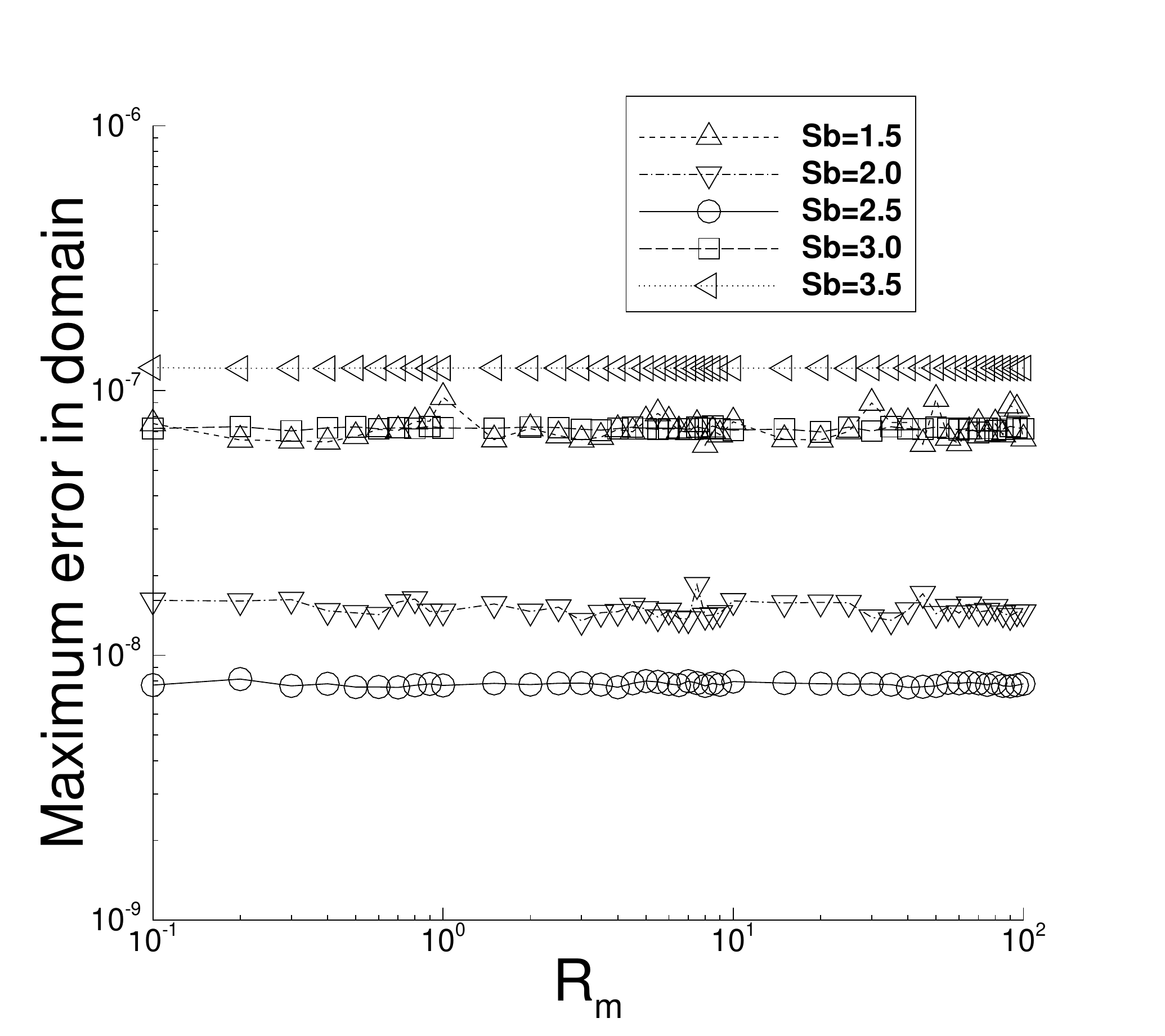}(c)
  }
  \caption{Diffusion equation: (a) Distribution of the exact solution
    in the spatial-temporal plane.
    (b) The maximum residual norm of
    the linear least squares (LLSQ) problem as a function of $S_b$,
    for estimating the best $S_b$ in modBIP.
    (c) The maximum error in the domain of the ELM/BIP solution
    as a function of $R_m$,
    corresponding to several $S_b$ values in modBIP.
    $R_m=100$ in (b) and is varied in (c).
  }
  \label{fg_21}
\end{figure}

As another example we test the combined ELM/modBIP method using
the unsteady diffusion equation. Consider the spatial-temporal
domain, $\Omega=\{\ (x,t)\ |\ x\in[0,1],\ t\in[0,10]  \}$,
and the following initial/boundary value problem,
\begin{subequations}\label{eq_18}
  \begin{align}
    &
    \frac{\partial u}{\partial t} - \nu\frac{\partial^2u}{\partial x^2} = f(x,t),
    \\
    &
    u(0,t) = g_1(t), \\
    &
    u(1,t) = g_2(t), \\
    &
    u(x,0) = h(x),
  \end{align}
\end{subequations}
where $u(x,t)$ is the field solution to be solved for, the constant $\nu=0.01$
denotes the diffusion coefficient, $f(x,t)$ is a prescribed
source term, $g_1$ and $g_2$ are
the Dirichlet boundary distributions at $x=0$ and $x=1$, and $h(x)$ is
the initial distribution.
We employ the following manufactured solution to this problem,
\begin{equation} \label{eq_19}
  u(x,t) = \left[2\cos\left(\pi x+\frac{\pi}{5}\right)
    +\frac32\cos\left(2\pi x-\frac{3\pi}{5}  \right)\right]
  \left[2\cos\left(\pi t+\frac{\pi}{5}\right)
    +\frac32\cos\left(2\pi t-\frac{3\pi}{5}  \right)\right].
\end{equation}
Accordingly, the source term $f(x,t)$, the boundary/initial distributions $g_1$, $g_2$
and $h$ are chosen such that the expression \eqref{eq_19}
satisfies the system \eqref{eq_18}.
Figure \ref{fg_21}(a) shows the distribution of the analytic
solution \eqref{eq_19} in the spatial-temporal $(x,t)$ plane.


We employ the block time-marching scheme and the combined ELM/modBIP
method to solve the system \eqref{eq_18}.
We divide the spatial-temporal domain $\Omega$ along the temporal
direction into $10$ uniform time blocks, and solve the system~\eqref{eq_18}
on each time block individually and successively (see Remark~\ref{rem_6})
using the combined ELM/modBIP method (see Section~\ref{sec:pde}).

Let us first consider the neural network containing a single hidden layer,
with the architecture characterized by $[2, 300, 1]$,
the $\tanh$ activation function for the hidden layer and a linear output layer.
The input layer ($2$ nodes) denotes the spatial/temporal coordinates ($x$ and $t$),
and the output layer ($1$ node) represents the field solution $u(x,t)$.
We employ a set of $Q=25\times 25$ uniform grid (collocation) points
on each time block ($25$ points in both $x$ and $t$ directions)
as the training data points, which constitute the input data into
the neural network. The hidden-layer coefficients in the neural network are
initialized to uniform random values generated on $[-R_m,R_m]$, with
$R_m$ either fixed at $R_m=100$ or varied between $R_m=0.1$ and $R_m=100$
in the subsequent tests. The initial random coefficients are pre-trained by
modBIP with $S_c=S_b/2$ and $S_b$ determined by the procedure discussed
in Remark~\ref{rem_1}.

Figure \ref{fg_21}(b) shows the LLSQ maximum residual norm among the
$10$ time blocks for estimating the best $S_b$ in modBIP, which suggests
a value around $S_b\approx 2.5$.
This set of tests are performed with a fixed $R_m=100$ when generating
the random coefficients in the neural network.
Figure \ref{fg_21}(c) plots the maximum error in the overall domain
of the ELM/modBIP solution as a function of $R_m$, with several $S_b$
values around $S_b=2.5$ and $S_c=S_b/2$ in the modBIP algorithm.
The ELM/modBIP solution errors in this and the subsequent
figures are computed as follows.
After the neural network has been trained, we evaluate the neural network
on a set of $101\times 101$ uniform grid points on each time block
to obtain the numerical solution. The exact solution \eqref{eq_19}
is evaluated on the same set of grid points in the overall domain.
Then the maximum and the rms errors of the numerical solution against
the analytic solution can be computed based these data.
The error of the ELM/modBIP solution can be observed to be
insensitive to the initial random coefficients in the neural network,
irrespective of the $S_b$ parameter in modBIP.

\begin{figure}
  \centerline{
    \includegraphics[height=2.6in]{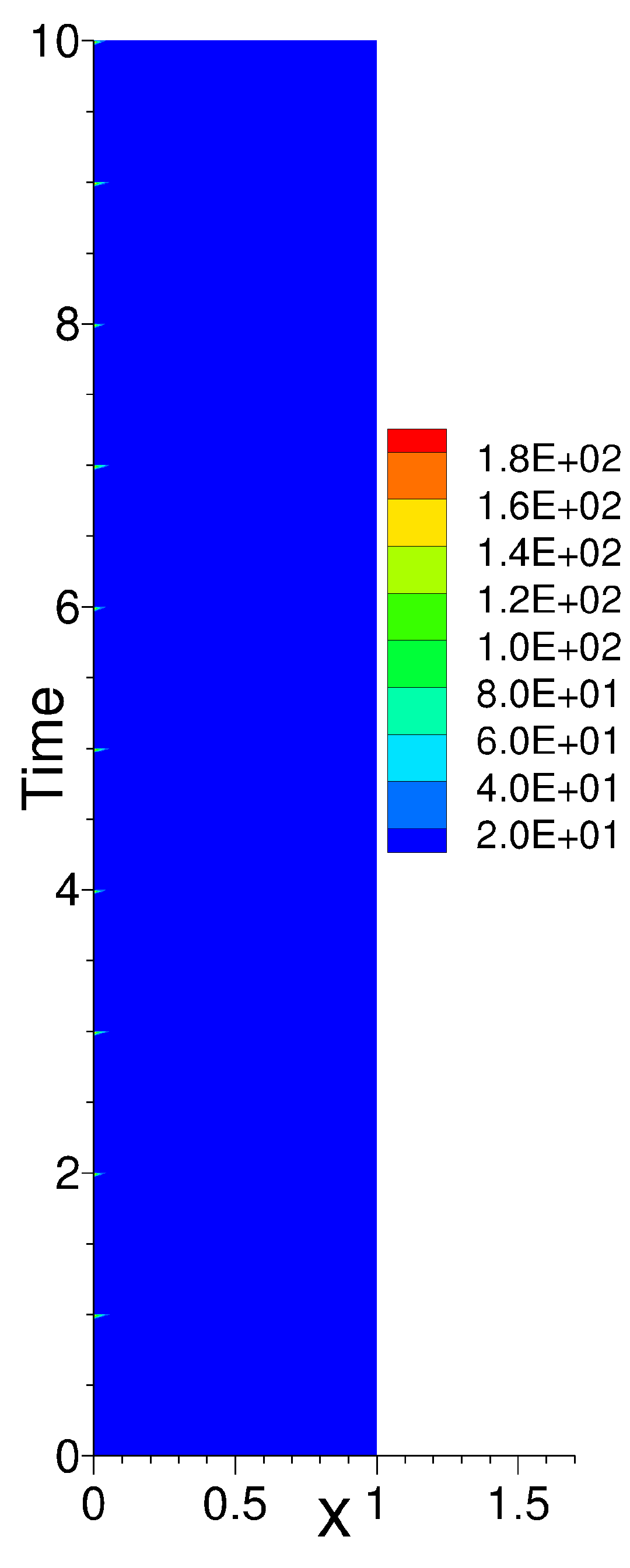}(a)
    \includegraphics[height=2.6in]{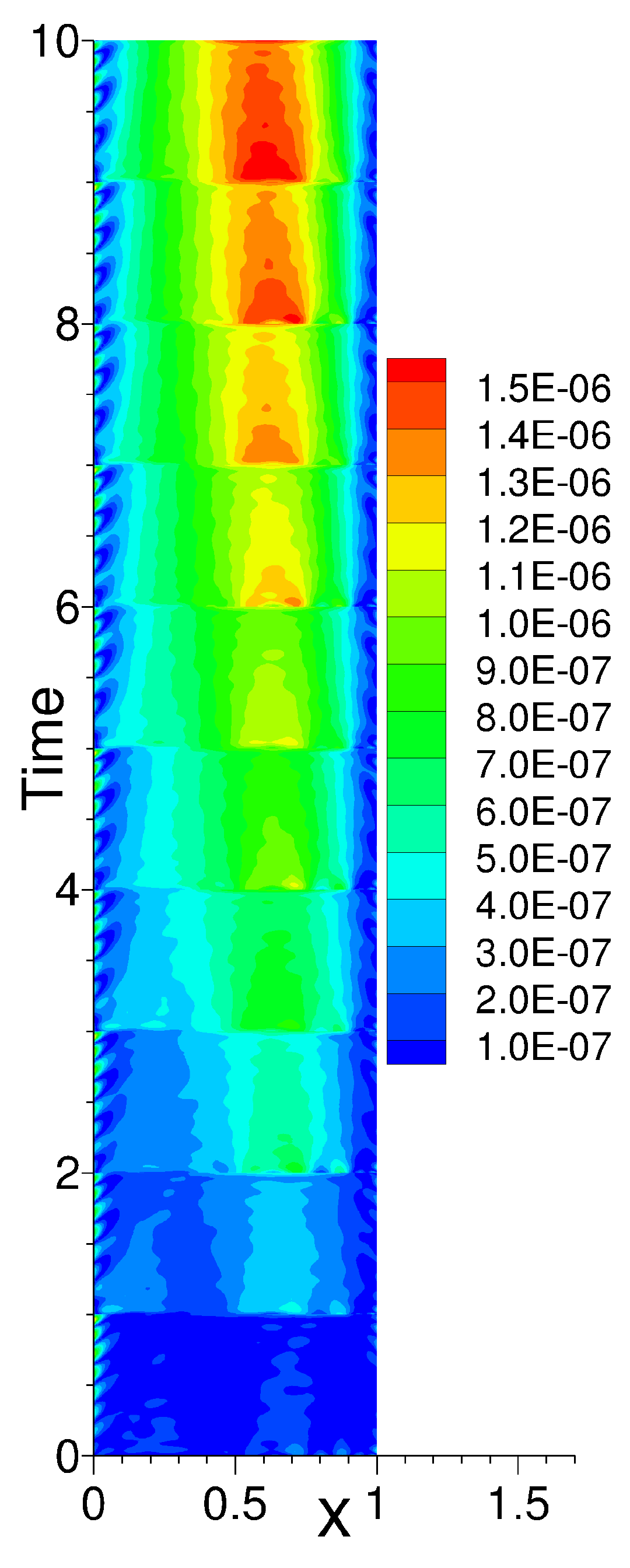}(b)
    \includegraphics[height=2.6in]{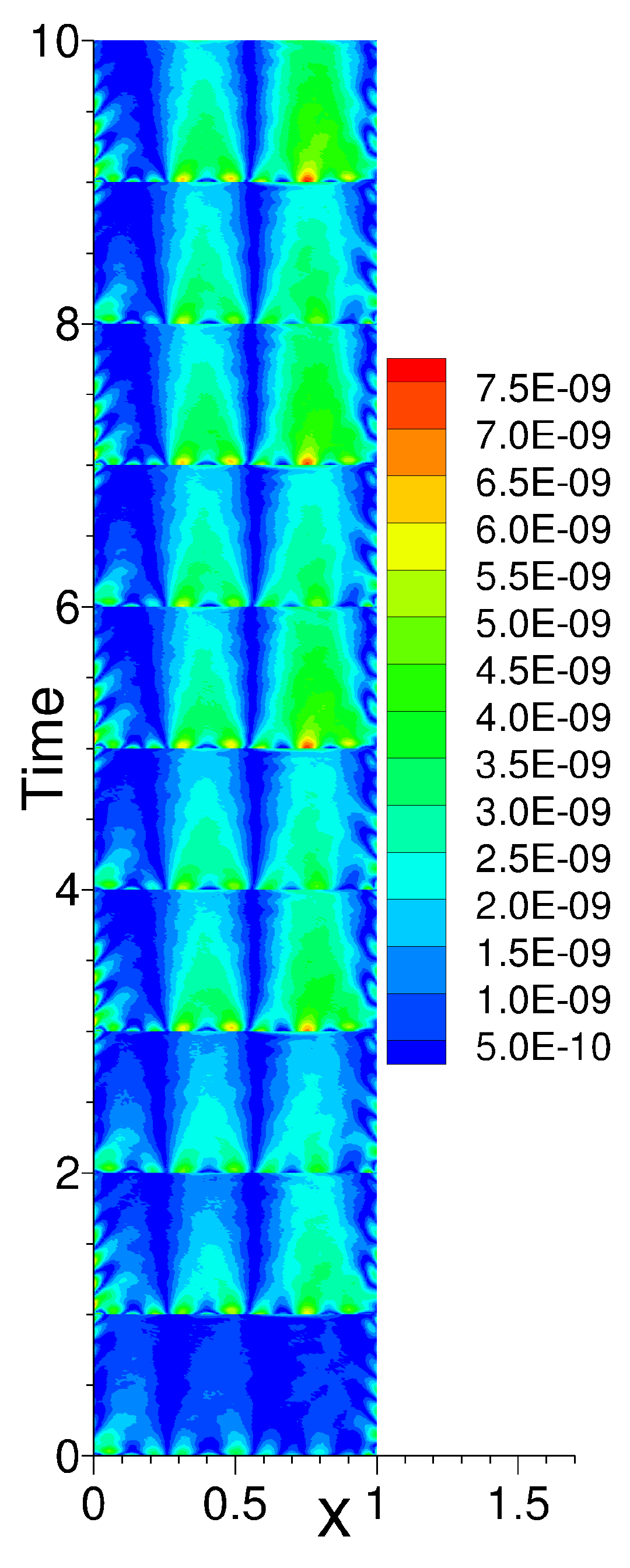}(c)
  }
  \caption{Diffusion equation: 
    Distributions of the absolute error of the ELM solution obtained
    with (a) no pre-training, (b) BIP pre-training, and (c) modBIP pre-training
    of the random coefficients.
  }
  \label{fg_22}
\end{figure}

\begin{figure}
  \centerline{
    \includegraphics[width=2in]{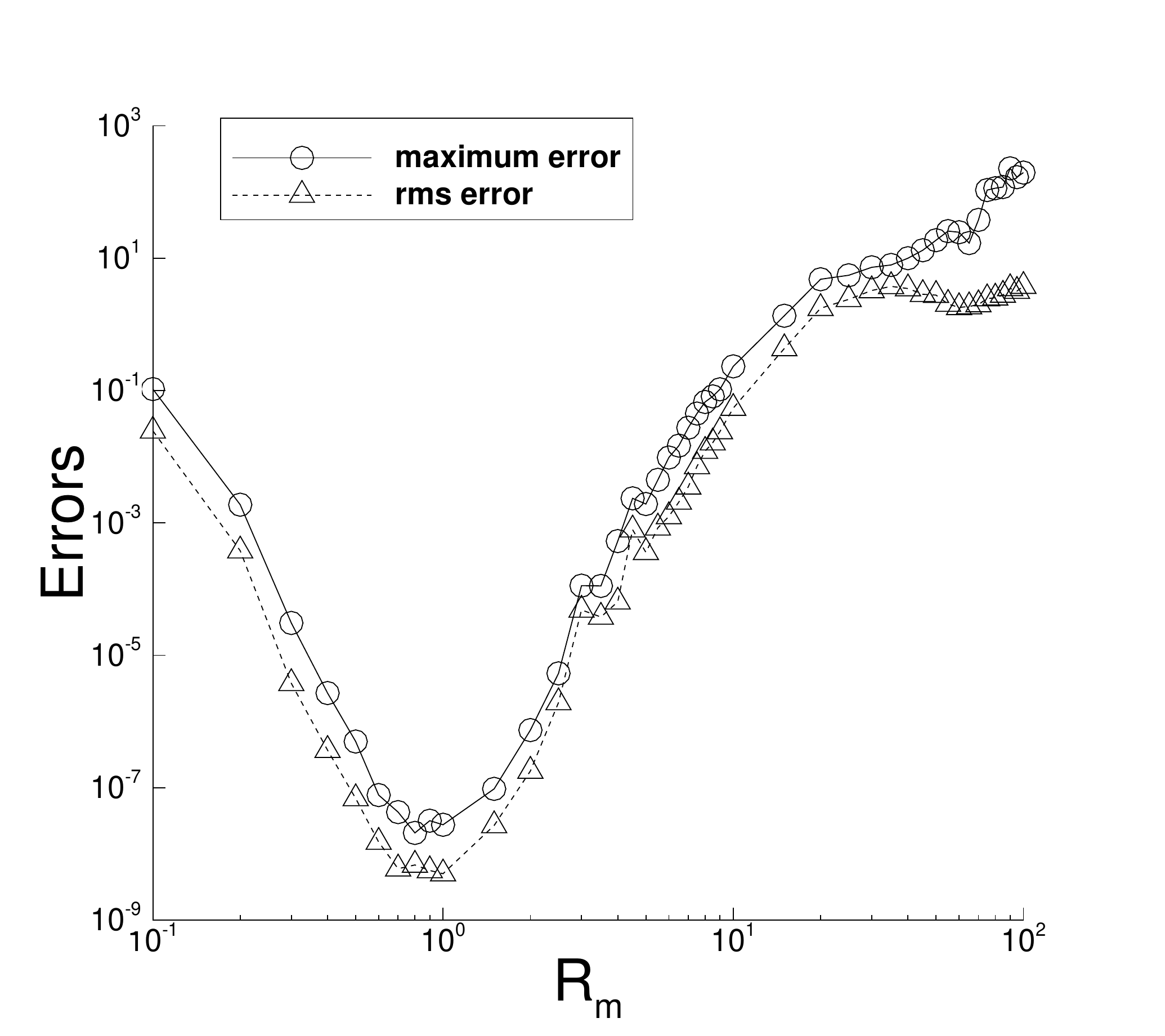}(a)
    \includegraphics[width=2in]{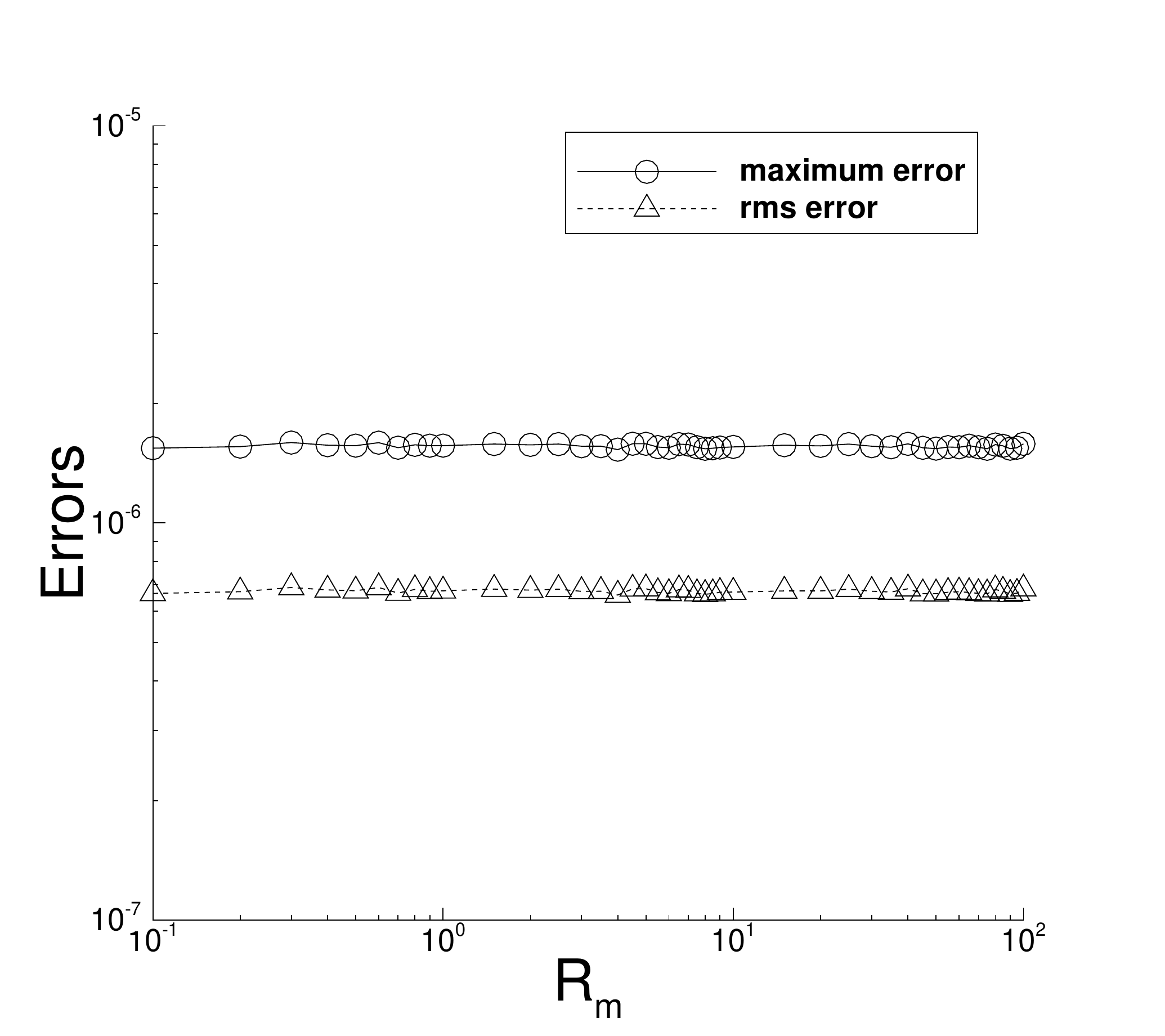}(b)
    \includegraphics[width=2in]{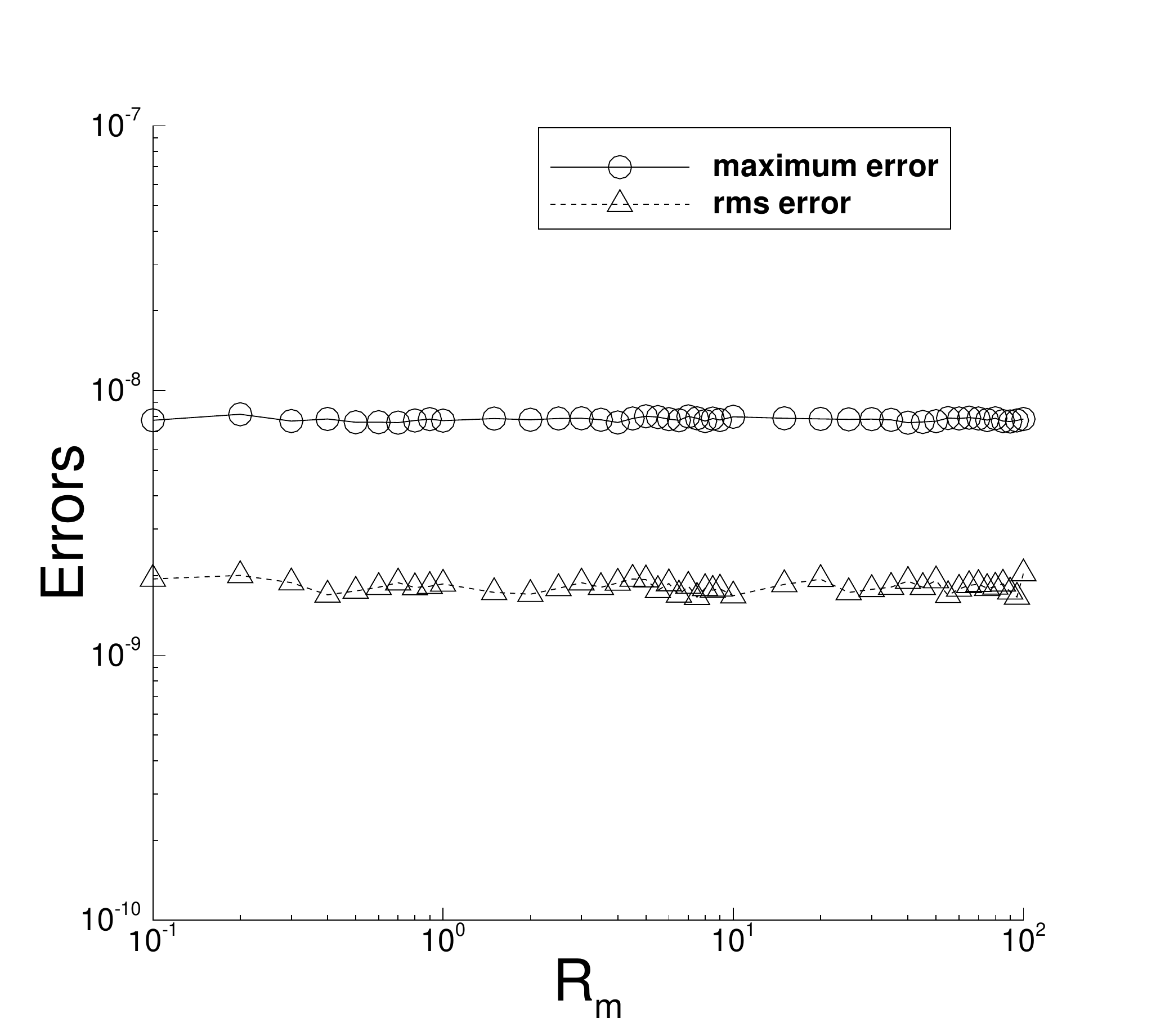}(c)
  }
  \caption{Diffusion equation: The maximum and rms errors of the ELM solution
    as a function of $R_m$ obtained with (a) no pre-training,
    (b) BIP pre-training, and (c) modBIP pre-training of the
    random coefficients.
  }
  \label{fg_23}
\end{figure}

In Figures \ref{fg_22} and \ref{fg_23} we compare the ELM method
with no pre-training and with the BIP and modBIP pre-training of
the random coefficients in the neural network.
In the case with no pre-training, the initial random coefficients
generated on $[-R_m,R_m]$ are directly used in the ELM computation.
In the case with BIP or modBIP pre-training, the initial random
coefficients are pre-trained first, and the updated random coefficients
are then used in the ELM computation.
With BIP, a normal distribution for
the target samples has been employed with a random mean on
$[-1, 1]$ and a standard deviation $0.5$~\cite{NeumannS2013}.
With modBIP we employ $S_c=S_b/2$ and $S_b=2.5$
in the Algorithm~\ref{alg_1}.
Figure \ref{fg_22} shows distributions in the spatial-temporal
plane of the absolute error of
the ELM solution obtained with no pre-training and with BIP/modBIP
pre-training of the random coefficients.
The initial random coefficients are generated with $R_m=100$
in this set of tests. The ELM solution obtained without pre-training
of the random coefficients
is totally off and inaccurate.
On the other hand, the ELM solutions with the initial random coefficients
pre-trained by BIP and modBIP are observed to be very accurate.
The ELM/modBIP solution is observed to be considerably more
accurate (by two or three orders of magnitude)
than the ELM/BIP solution.

Figure \ref{fg_23} shows the maximum and rms errors of the ELM solutions,
obtained with no pre-training and with BIP or modBIP pre-training of
the random coefficients, as a function of $R_m$ for generating
the initial random coefficients.
The accuracy of
the ELM solution without pre-training of the random coefficients
strongly depends on $R_m$, as is evident from Figure \ref{fg_23}(a).
In contrast, with the random coefficients pre-trained by
BIP or modBIP, the ELM solution accuracy becomes essentially
independent of $R_m$ (Figures \ref{fg_23}(b,c)).
The data again signify that the ELM/modBIP solution is much more
accurate than the ELM/BIP one.

\begin{figure}
  \centerline{
    \includegraphics[height=2.0in]{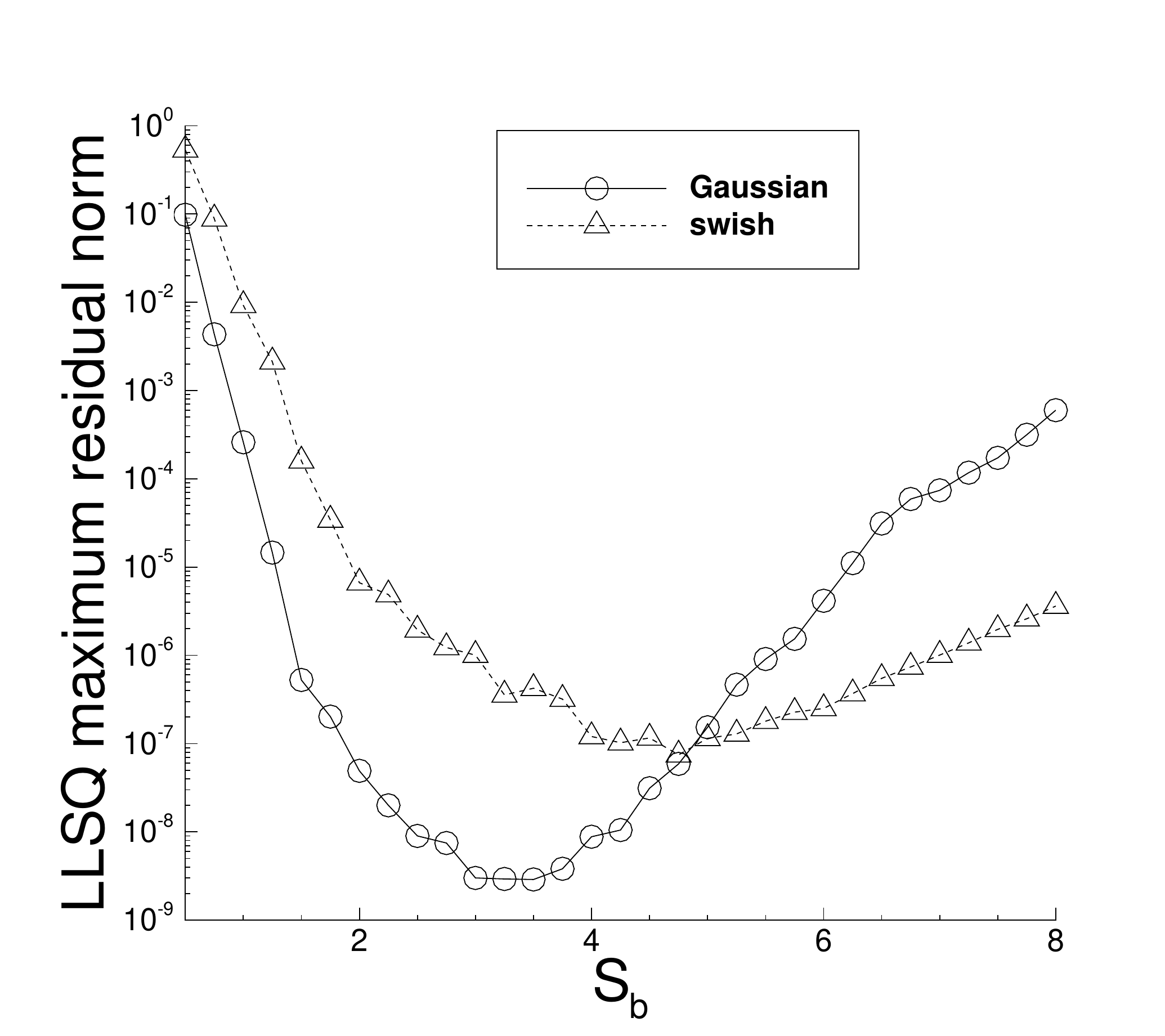}(a)\qquad
    \includegraphics[height=1.9in]{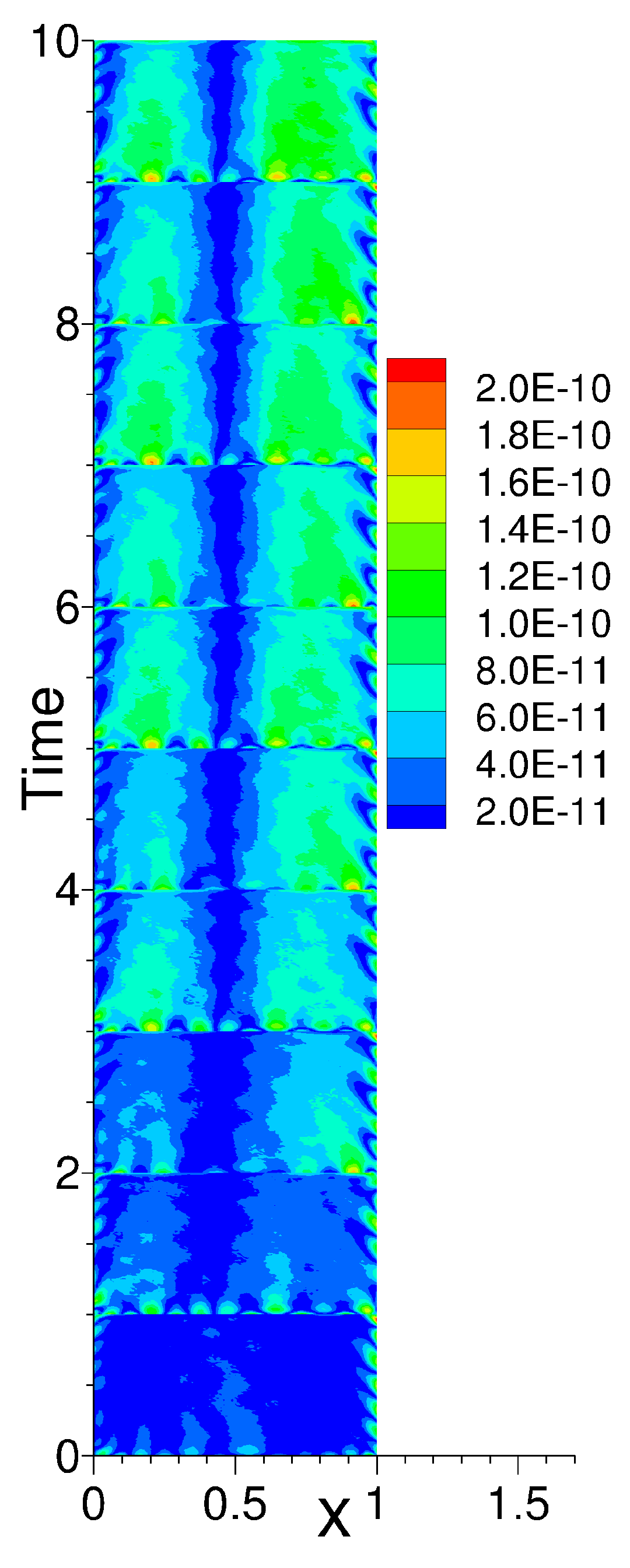}(b)\qquad
    \includegraphics[height=1.9in]{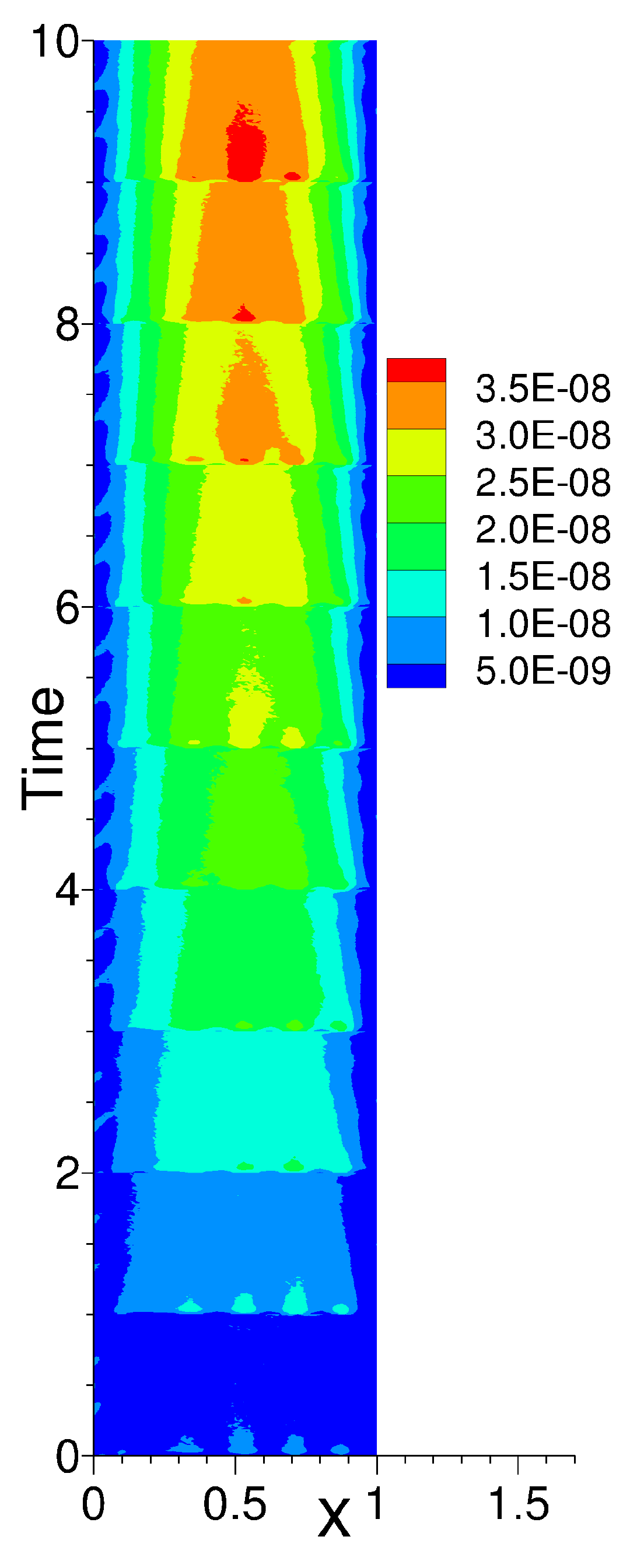}(c)
  }
  \centerline{
    \includegraphics[width=2.0in]{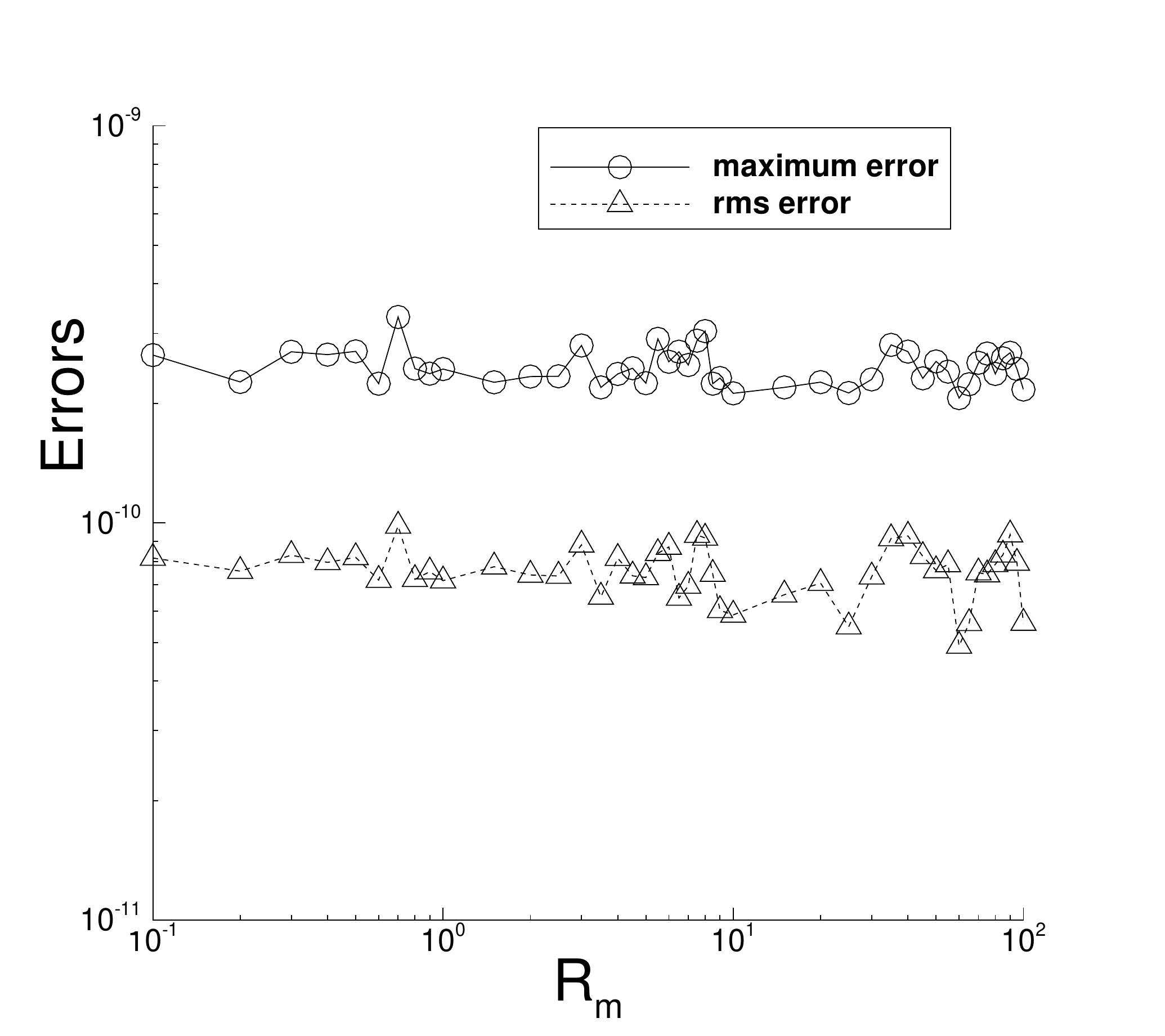}(d)
    \includegraphics[width=2.0in]{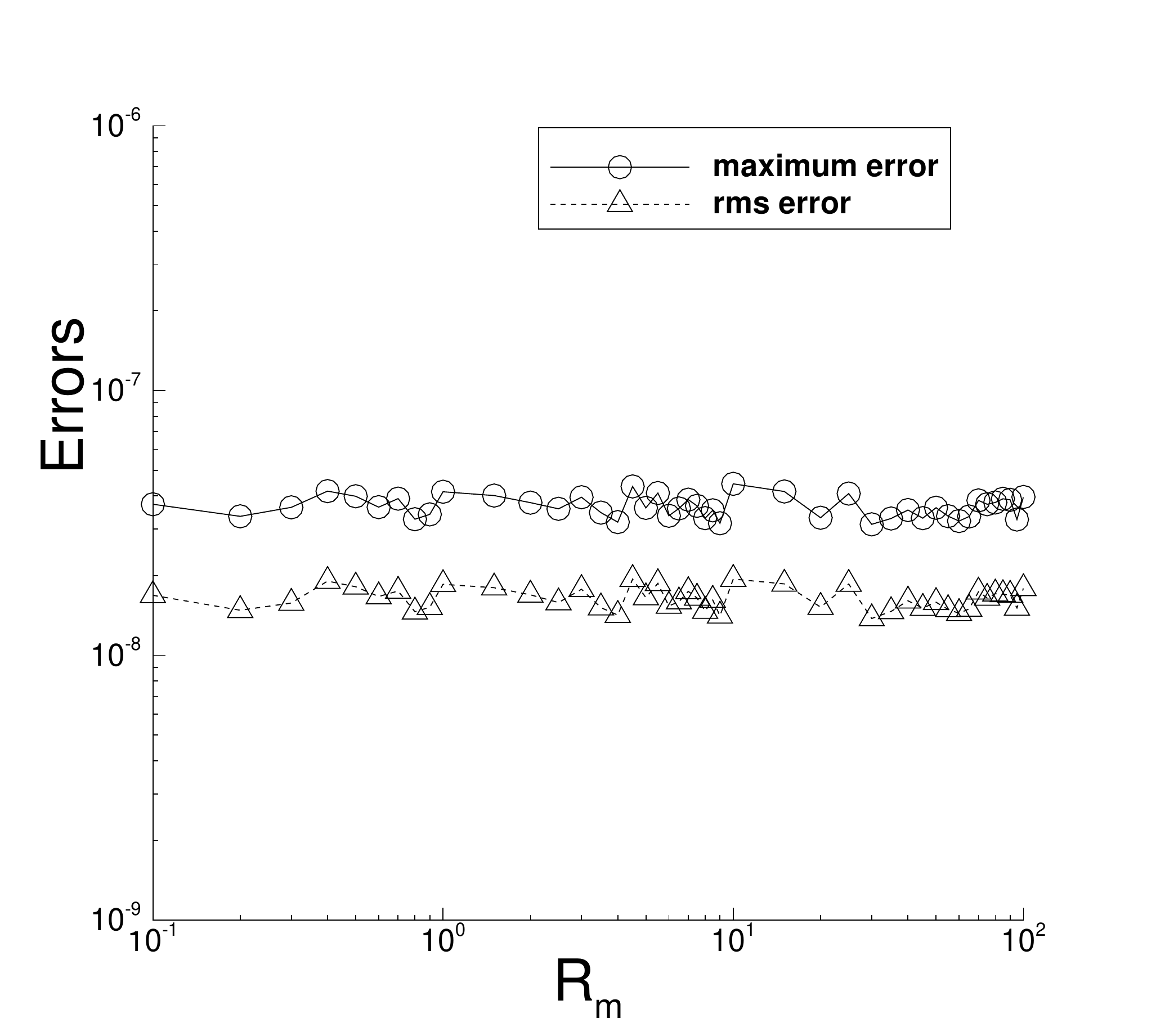}(e)
  }
  \caption{Diffusion equation (non-monotonic activation functions):
    (a) The LLSQ maximum residual norm versus $S_b$, for estimating
    the best $S_b$ in modBIP, with the Gaussian and swish activation functions.
    Distributions of the absolute error of
    the ELM/modBIP solution obtained with (b) the Gaussian, 
    and (c) swish
    activation functions.
    The maximum/rms errors in the domain as a function of $R_m$ corresponding
    to (d) the Gaussian, 
    and (e) swish activation functions.
    $R_m=100$ in (a,b,c) and is varied in (d,e).
    $S_b=3$ in (b,d) for the Gaussian function,
    and $S_b=5$ in (c,e) for the swish function,
    and $S_b$ is varied in (a).
  }
  \label{fg_24}
\end{figure}

Figure \ref{fg_24} demonstrates the capability of the combined ELM/modBIP
method to work with activation functions that do not have an inverse.
Here we have considered the Gaussian and swish activation functions in
the neural network. We again employ a
network architecture $[2,300,1]$, with $Q=25\times 25$ uniform collocation
points on each time block. The initial random coefficients are generated
with $R_m$ fixed at $R_m=100$ or varied between $R_m=0.1$ and $R_m=100$.
Figure \ref{fg_24}(a) shows the LLSQ maximum residual norms for estimating
the $S_b$ in modBIP, which suggest a value around $S_b\approx 3$ with
the Gaussian activation function and $S_b\approx 5$ with the swish activation
function. Figures \ref{fg_24}(b) and (c) depict the error distributions
in the spatial-temporal plane of the ELM/modBIP solution
corresponding to the Gaussian activation function (with $S_b=3$ in modBIP)
and the swish activation function (with $S_b=5$ in modBIP).
One can observe that the combined ELM/modBIP method, especially
with the Gaussian activation function, has produced very
accurate results.
The initial random coefficients are generated with $R_m=100$ in
the plots (a,b,c).
Figures \ref{fg_24}(d) and (e) show the maximum/rms errors in
the overall domain of the ELM/modBIP solution as a function of $R_m$,
corresponding to the Gaussian and swish activation functions,
respectively. The solution errors are not sensitive to the
initial random coefficients in the neural network.
It should be noted that the BIP algorithm breaks down
with the type of activation functions tested here.


\begin{figure}
  \centerline{
    \includegraphics[height=2.0in]{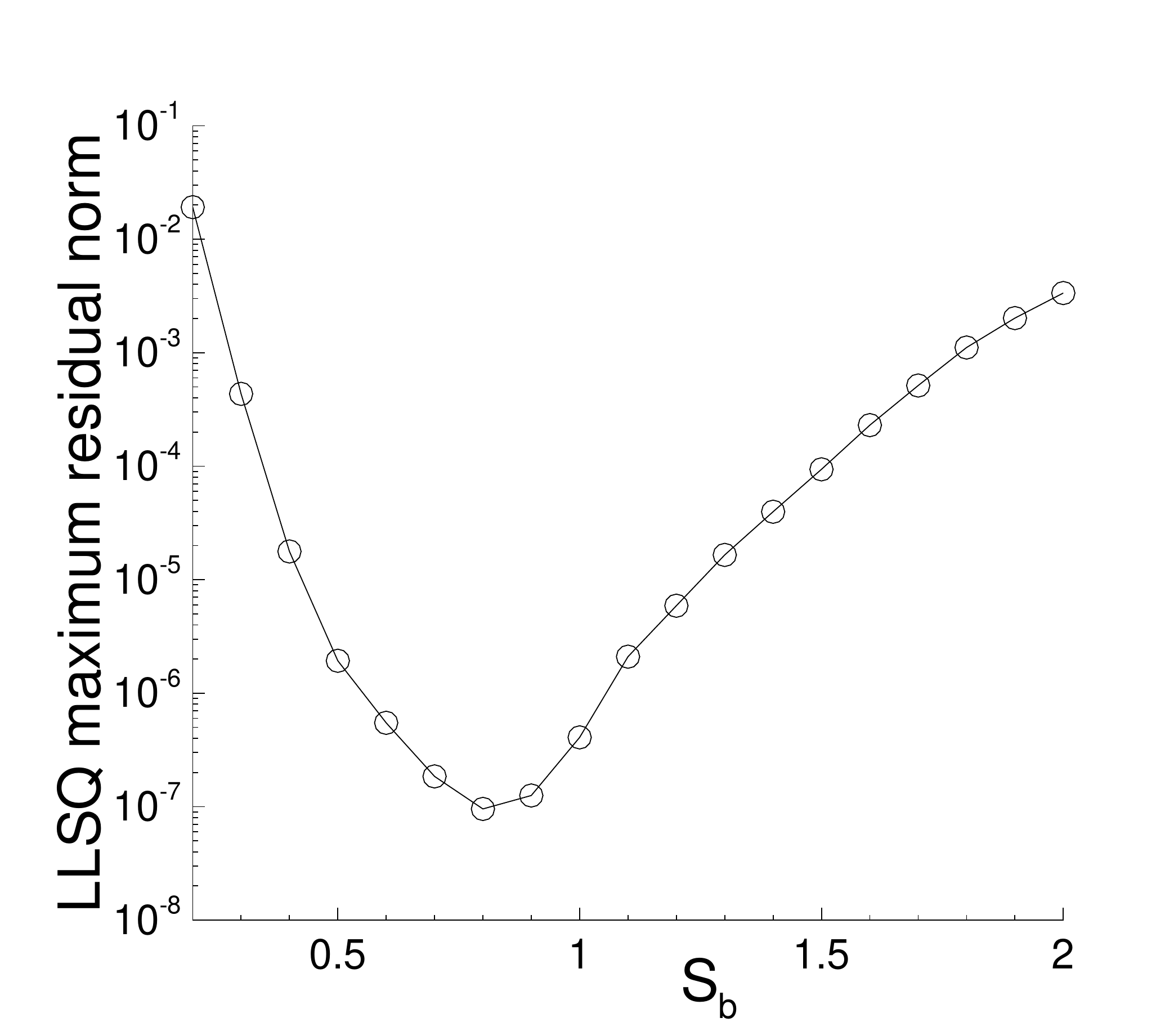}(a)
    \includegraphics[height=1.9in]{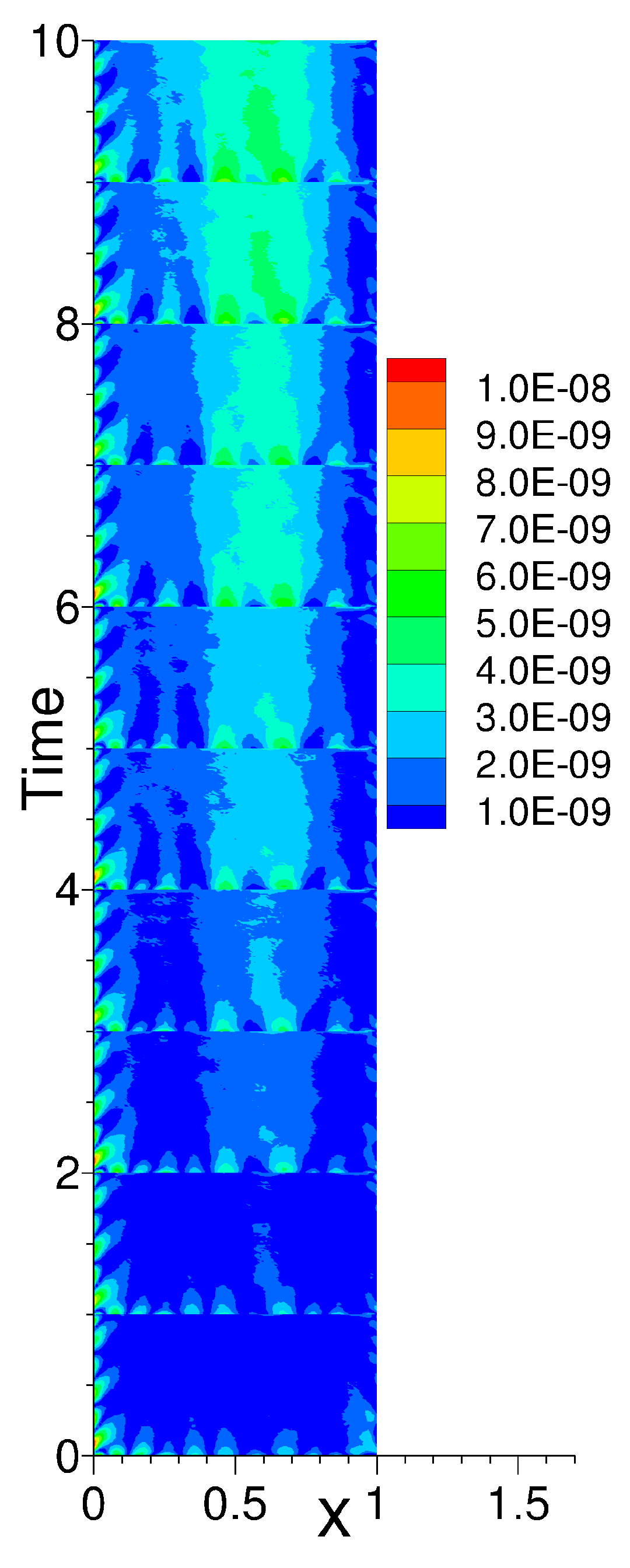}(b)
    \includegraphics[height=2.in]{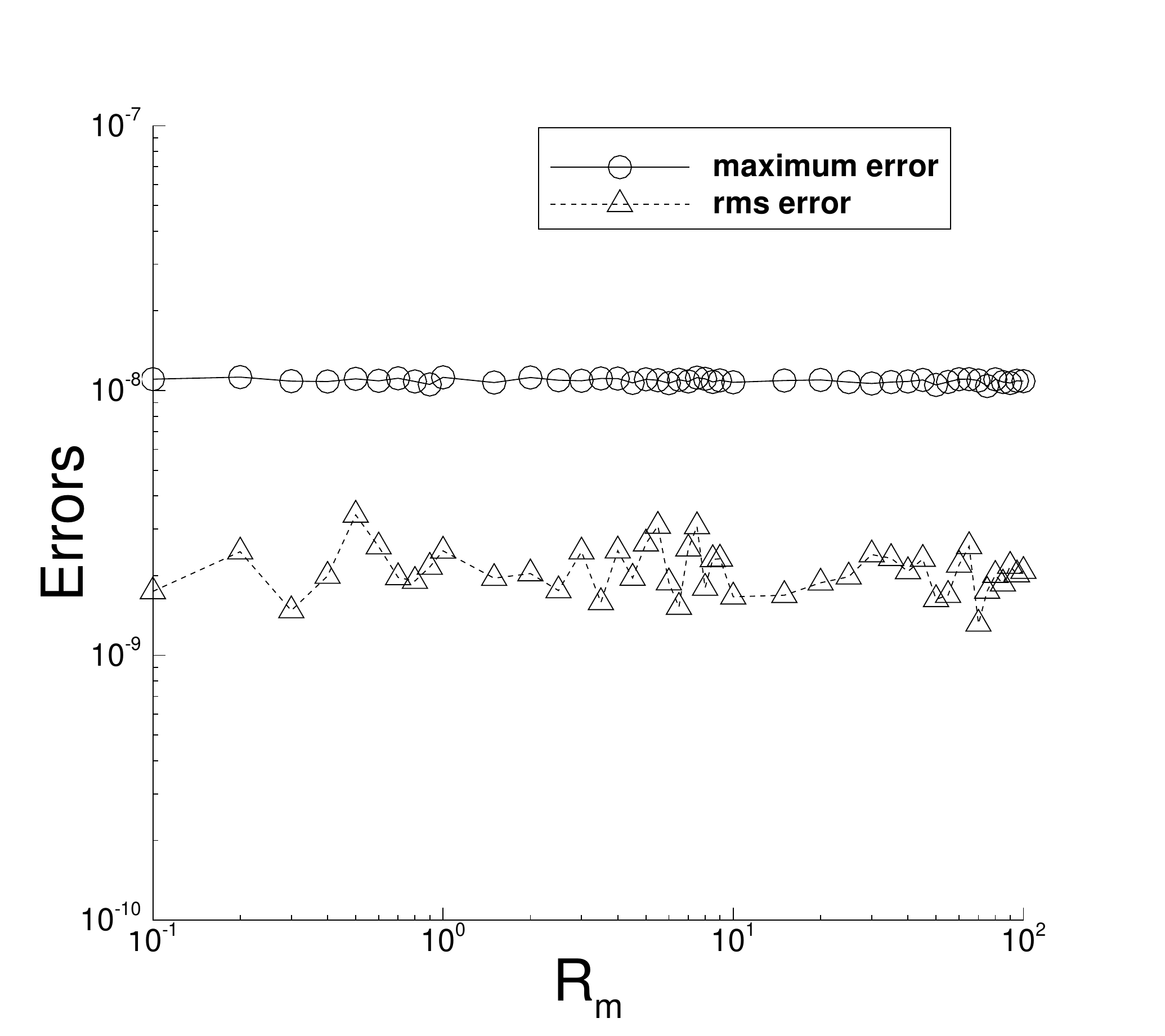}(c)
  }
  \centerline{    
    \includegraphics[width=2in]{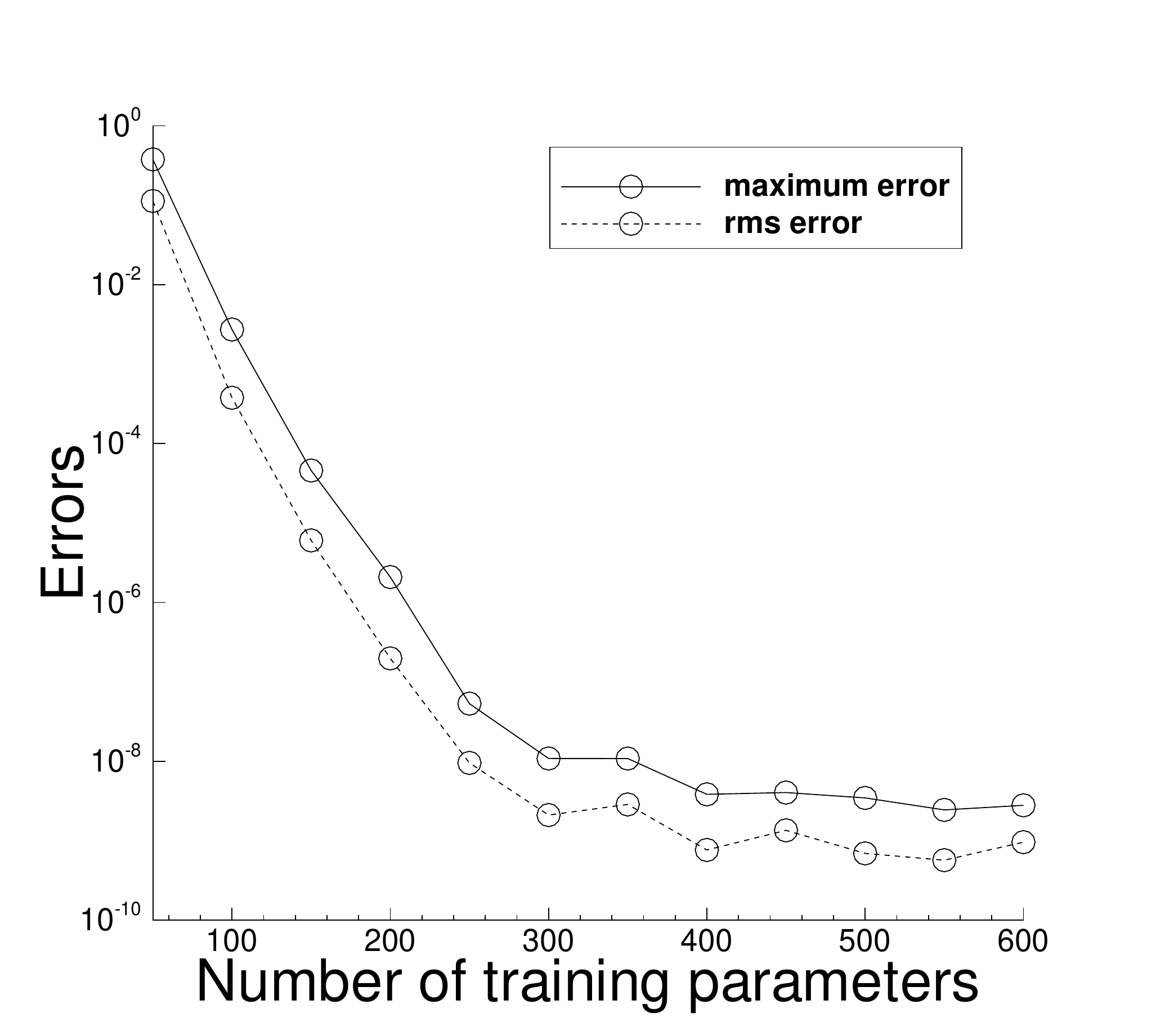}(d)
    \includegraphics[width=2in]{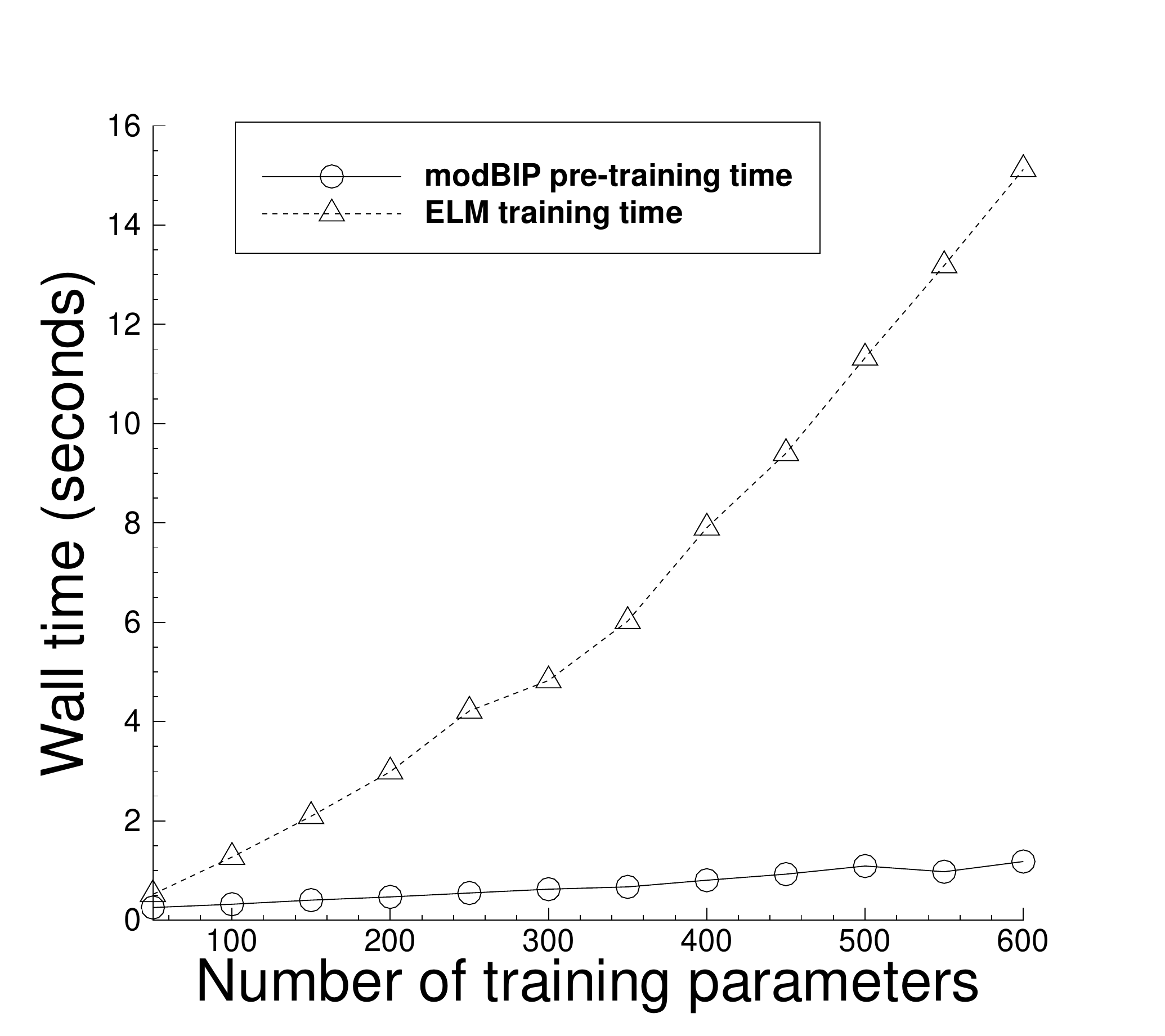}(e)
  }
  \caption{Diffusion equation (3 hidden layers in neural network):
    (a) The maximum residual norm of the LLSQ problem for estimating the 
    $S_b$ in modBIP.
    (b) Distribution of the absolute error of the ELM/modBIP solution.
    The maximum and rms errors in the domain as a function of (c) $R_m$,
    and (d)
    the number of training parameters $M$.
    (e) The modBIP pre-training time and the ELM network training time
    as a function of the number of training parameters $M$.
    $S_b=0.8$ in (b,c,d,e) and is varied in (a).
    $R_m=100$ in (a,b,d,e) and is varied in (c).
    $M=300$ in (a,b,c) and is varied in (d,e).
    $Q=25\times 25$ in (a,b,c,d,e).
  }
  \label{fg_25}
\end{figure}

Finally, Figure \ref{fg_25} illustrates the ELM/modBIP simulation results
using multiple hidden layers in the neural network.
Here $10$ uniform time blocks are used in the domain for block time marching.
The neural network architecture is characterized by $[2, 50, 50, M, 1]$,
where the number of training parameters $M$ is fixed at $M=300$
or varied between $M=50$ and $M=600$. The $\tanh$ activation function is
used for all the hidden layers. We employ $Q=Q_1\times Q_1$ uniform
collocation points within each time block, where $Q_1$ is either fixed at $Q_1=25$
or varied between $Q_1=5$ and $Q_1=50$. The initial random coefficients are
generated with a fixed $R_m=100$ or with $R_m$ varied between
$R_m=0.1$ and $R_m=100$. These random coefficients are pre-trained by modBIP
with $S_c=S_b/2$ and $S_b$ estimated by the procedure in Remark~\ref{rem_1}.
The specific parameter values for each plot are provided in the caption
of the figure.
Figure \ref{fg_25}(a) shows the LLSQ residual norms versus $S_b$
for estimating the best $S_b$ in modBIP, which suggests a value
around $S_b\approx 0.8$ in the algorithm.
Figure \ref{fg_25}(b) illustrates the error distribution of
the ELM/modBIP solution in the spatial-temporal plane, demonstrating that
the method produces accurate simulation results.
Figure \ref{fg_25}(c) depicts the maximum and rms errors in the overall domain
of the ELM/modBIP solution as a function of $R_m$ for generating
the initial random coefficients, indicating that the accuracy of the ELM/modBIP
method is not sensitive to the initial random coefficients
with multiple hidden layers in the neural network.
Figure \ref{fg_25}(d) shows the maximum/rms errors of the
ELM/modBIP solution as a function of 
the number of training parameters ($M$),
indicating an exponential decrease in the errors (before saturation)
with this method.
Figure \ref{fg_25}(e) shows the modBIP pre-training time and
the ELM network training time as a function of the number of
training parameters in the neural network.
The modBIP pre-training time grows very slowly
as the number of training parameters increases.
Its growth rate is much smaller compared with that of the ELM training time.

%% file: Summary.tex
\section{Concluding Remarks}
\label{sec:summary}


We have presented an effective algorithm (termed modBIP) for pre-training
the random hidden-layer coefficients of extreme learning machines (ELM),
and applied the combined ELM/modBIP method to function approximations
and solving partial differential equations.
The initial random coefficients 
in the neural network are first pre-trained by modBIP,
and the updated hidden-layer coefficients are then fixed and
employed in the least squares computation with the ELM method.
The modBIP algorithm is devised based on the
same principle as the batch intrinsic
plasticity (BIP) method, namely, by enhancing the information
transmission in every node of the neural network.

%

Suppose the neural network architecture is chosen, which may be shallow or deep,
the hidden-layer coefficients 
are initialized to random values, and that the input
data to the neural network are given.
The modBIP algorithm pre-trains the
random hidden-layer coefficients as follows.
For each node in each hidden layer of the neural network,
this algorithm first computes
the total (synaptic) input to this node for all
input data samples to the neural network.
The synaptic input samples are then mapped, by an affine mapping,
to a set of random
target samples on a random sub-interval of $[-S_b,S_b]$
with a minimum size,
where $S_b$ is a user-provided hyper-parameter.
The coefficients in the affine mapping are determined by solving
a linear least squares problem, and are used to update 
the random weight and bias coefficients associated with this node.
These operations are performed on each hidden layer, individually and
successively, and within each hidden layer node by node.
For a given particular problem,
the $S_b$ parameter in modBIP can be estimated with preliminary
simulations by computing the residual norm of
the least squares problem associated with the least squares solution
in ELM; see the procedure outlined in Remark~\ref{rem_1}.


The modBIP method differs from BIP~\cite{NeumannS2013} in one prominent
aspect: modBIP does not involve the activation function in its
algorithm. In contrast, the BIP method employs the inverse of
the activation function in its construction, and requires
the activation function to be invertible (or monotonic).
This limits the applicability of BIP to neural networks
with only monotonic activation functions. On the other hand,
the modBIP method can be applied with essentially any activation function,
including those often-used non-monotonic activation functions
such as the Gaussian function, swish function, Gaussian error linear
unit (GELU), and the class of radial basis activation functions.

Another crucial construction in modBIP is the {\em random} sub-interval
of $[-S_b,S_b]$ with a minimum size $S_c$,
on which the random target samples are generated.
This is the key that accounts for the high accuracy of the modBIP (combined
with ELM) results. If this sub-interval is fixed, instead of being random,
much of the high accuracy will be lost.
As demonstrated by ample numerical experiments, when combined with ELM,
the modBIP method typically produces much more accurate simulation results
than BIP.


In the current paper we have used partial differential equations to test
the combined ELM/modBIP method, and presented extensive numerical experiments
to evaluate its computational performance.
We have the following observations:
\begin{itemize}
\item
  The combined ELM/modBIP method produces highly accurate simulation results,
  and its accuracy is insensitive to the initial random coefficients of
  the neural network. More precisely, its accuracy is insensitive to the $R_m$
  for generating the initial random coefficients.
  In contrast, without pre-training the random coefficients,
  the accuracy of the ELM solution
  is strongly dependent on $R_m$, where the initial random coefficients are generated
  on $[-R_m,R_m]$.

\item
  The combined ELM/modBIP method works well with  non-monotonic activation
  functions, and produces highly accurate results. With the Gaussian
  activation function, the combined ELM/modBIP method appears to produce
  generally the most accurate results among the activation functions tested
  herein.
  In contrast, the BIP method~\cite{NeumannS2013} breaks down with
  non-monotonic activation functions.

\item
  Irrespective of the initial random coefficients, the errors of
  the combined ELM/modBIP solution decrease exponentially or nearly exponentially,
  as the number of training data points in the domain or
  the number of training parameters in the neural network increases.

\item
  The combined ELM/modBIP method works well with both shallow and deep
  neural networks. The favorable numerical properties,
  such as the error insensitivity
  to initial random coefficients and the exponential convergence with respect to
  the collocation points and the training parameters,
  are observed with both shallow and deep neural networks.

\item
  The computational cost of the modBIP pre-training of the random coefficients
  is low, and it is only a fraction of the ELM training cost of the neural network.
  In typical simulations the modBIP pre-training cost is within $10\%$
  of the ELM training cost for the neural network.
  Furthermore, the modBIP pre-training of the random coefficients logically
  only needs to be performed once for a given network architecture and the input data,
  and the pre-trained random coefficients can be saved and used later directly by
  ELM.

\end{itemize}


The numerical results demonstrate unequivocally that modBIP provides an efficient and
effective technique for pre-training the random coefficients to achieve
high accuracy with ELM. It significantly boosts the computational performance
of ELM. The combined ELM/modBIP method can produce accurate
simulation results, regardless of the initial random coefficients
in the neural network. This method is promising in terms of
both the accuracy and the computational cost. We
anticipate that the ELM/modBIP method will be useful to and instrumental in
neural network-based scientific computing and the computational
understanding of important physical processes and
phenomena~\cite{Dong2018,Dong2014obc}.
